\documentclass[twocolumn,twocolappendix]{aastex631}
\usepackage{natbib}
\usepackage{comment}
\usepackage{amsmath}
\usepackage{algorithm}
\usepackage{amsfonts}
\usepackage{mathrsfs}
\usepackage{amssymb}
\usepackage{color}
\usepackage{graphicx} 
\let\tablenum\relax
\usepackage{siunitx}

\newcommand{\cha}{Chandra}
\newcommand{\XMM}{{XMM-{{Newton}}}}
\newcommand{\NuSTAR}{{{NuSTAR}}}

\newcommand{\lns}{$\log N$--$\log S$}

\begin{document}
\defcitealias{Zhao2021}{Z21}

\title{PEARLS: \textbf{N\MakeLowercase{u}STAR} and XMM-\textbf{N\MakeLowercase{ewton}} Extragalactic Survey of the \textbf{JWST} North Ecliptic Pole Time-Domain Field II}

\author[0000-0002-7791-3671]{Xiurui Zhao}
\affiliation{Center for Astrophysics $|$ Harvard \& Smithsonian, 60 Garden Street, Cambridge, MA 02138, USA}

\author[0000-0002-2115-1137]{Francesca Civano}
\affiliation{NASA Goddard Space Flight Center, Greenbelt, MD 20771, USA}

\author[0000-0001-9262-9997]{Christopher~N.~A. Willmer}
\affiliation{Steward Observatory, University of Arizona, 933 North Cherry Avenue, Tucson, AZ 85721, USA}



\author[0000-0002-6381-2052]{Silvia Bonoli}
\affiliation{Donostia International Physics Center (DIPC), Manuel Lardizabal Ibilbidea, 4, San Sebastian, Spain}
\affiliation{Ikerbasque, Basque Foundation for Science, E-48013 Bilbao, Spain}


\author[0000-0002-4945-5079]{Chien-Ting Chen}
\affiliation{Science and Technology Institute, Universities Space Research Association, Huntsville, AL 35805, USA}
\affiliation{Astrophysics Office, NASA Marshall Space Flight Center, ST12, Huntsville, AL 35812, USA}



\author[0000-0002-9041-7437]{Samantha Creech}
\affiliation{Department of Physics \& Astronomy, University of Utah, 115 South 1400 East, Salt Lake City, UT 84112, USA}
\affiliation{Astrophysics Science Division, SURA/GSFC/CRESST II, Greenbelt, MD 20771}

\author[0000-0003-1477-3453]{Renato Dupke}
\affiliation{Observatório Nacional, Rua General José Cristino, 77, São Cristóvão, 20921-400, Rio de Janeiro, RJ, Brazil}
\affiliation{Department of Astronomy, University of Michigan, 311 West Hall, 1085 South University Ave., Ann Arbor, MI 48109, USA}
\affiliation{Department of Physics and Astronomy, University of Alabama, Box 870324, Tuscaloosa, AL 35487-0324, USA}


\author[0000-0002-9286-9963]{Francesca~M. Fornasini}
\affiliation{Department of Physics and Astronomy, Stonehill College, 320 Washington Street, Easton, MA 02357, USA}




\author[0000-0003-1268-5230]{Rolf~A. Jansen}
\affiliation{School of Earth and Space Exploration, Arizona State University, P.O. Box 871404, Tempe, AZ 85287-1404, USA}

\author[0000-0003-3214-9128]{Satoshi Kikuta}
\affiliation{National Astronomical Observatory of Japan, 2-21-1 Osawa, Mitaka, Tokyo 181-8588, Japan}

\author[0000-0002-6610-2048]{Anton~M. Koekemoer}
\affiliation{Space Telescope Science Institute, 3700 San Martin Drive, Baltimore, MD 21218, USA}

\author[0000-0003-2714-0487]{Sibasish Laha}
\affiliation{Center for Space Science and Technology, University of Maryland Baltimore County, 1000 Hilltop Circle, Baltimore, MD 21250, USA}
\affiliation{NASA Goddard Space Flight Center, Greenbelt, MD 20771, USA}



\author[0000-0002-2203-7889]{Stefano Marchesi}
\affiliation{Dipartimento di Fisica e Astronomia (DIFA), Università di Bologna, via Gobetti 93/2, I-40129 Bologna, Italy}
\affiliation{Department of Physics and Astronomy, Clemson University, Kinard Lab of Physics, Clemson, SC 29634, USA}
\affiliation{INAF, Osservatorio di Astrofisica e Scienza dello Spazio di Bologna, via P. Gobetti 93/3, 40129 Bologna, Italy}

\author[0000-0003-3351-0878]{Rosalia O'Brien}
\affiliation{School of Earth and Space Exploration, Arizona State University, Tempe, AZ 85287-1404, USA}



\author[0000-0001-6564-0517]{Ross Silver}
\affiliation{NASA Goddard Space Flight Center, Greenbelt, MD 20771, USA}


\author[0000-0002-9895-5758]{S.~P. Willner}
\affiliation{Center for Astrophysics $|$ Harvard \& Smithsonian, 60 Garden Street, Cambridge, MA 02138, USA}

\author[0000-0001-8156-6281]{Rogier~A. Windhorst}
\affiliation{School of Earth and Space Exploration, Arizona State University, P.O. Box 871404, Tempe, AZ 85287-1404, USA}

\author[0000-0001-7592-7714]{Haojing Yan}
\affiliation{Department of Physics and Astronomy, University of Missouri, Columbia, MO 65211, USA}

\author[0000-0003-2441-1413]{Jailson Alcaniz}
\affiliation{Observatório Nacional, Rua General José Cristino, 77, São Cristóvão, 20921-400, Rio de Janeiro, RJ, Brazil}

\author[0000-0002-0403-7455]{Narciso Benitez}

\author[0000-0001-7098-383X]{Saulo Carneiro}
\affiliation{Instituto de Física, Universidade Federal da Bahia, 40210-340, Salvador, BA, Brazil}

\author{Javier Cenarro}
\affiliation{Centro de Estudios de Física del Cosmos de Aragón (CEFCA), Plaza San Juan, 1, E-44001, Teruel, Spain}

\author{David Cristóbal-Hornillos}
\affiliation{Centro de Estudios de Física del Cosmos de Aragón (CEFCA), Plaza San Juan, 1, E-44001, Teruel, Spain}

\author[0000-0003-3656-5524]{Alessandro Ederoclite}
\affiliation{Centro de Estudios de Física del Cosmos de Aragón (CEFCA), Plaza San Juan, 1, E-44001, Teruel, Spain}

\author[0000-0002-4237-5500]{Antonio Hernán-Caballero}
\affiliation{Centro de Estudios de Física del Cosmos de Aragón (CEFCA), Plaza San Juan, 1, E-44001, Teruel, Spain}

\author[0000-0002-5743-3160]{Carlos López-Sanjuan}
\affiliation{Centro de Estudios de Física del Cosmos de Aragón (CEFCA), Plaza San Juan, 1, E-44001, Teruel, Spain}

\author[0000-0002-9026-3933]{Antonio Marín-Franch}
\affiliation{Centro de Estudios de Física del Cosmos de Aragón (CEFCA), Plaza San Juan, 1, E-44001, Teruel, Spain}

\author[0000-0002-7736-4297]{Claudia Mendes de Oliveira}
\affiliation{Departamento de Astronomia, Instituto de Astronomia, Geofísica e Ciências Atmosféricas, Universidade de São Paulo, São Paulo, Brazil}

\author{Mariano Moles}
\affiliation{Centro de Estudios de Física del Cosmos de Aragón (CEFCA), Plaza San Juan, 1, E-44001, Teruel, Spain}

\author[0000-0002-3876-268X]{Laerte Sodré Jr.}
\affiliation{IAG/USP - Departamento de Astronomia, Instituto de Astronomia, Geofísica e Ciências Atmosféricas, Universidade de São Paulo, São Paulo, Brazi}

\author{Keith Taylor}
\affiliation{Instruments4, 4121 Pembury Place, La Canada Flintridge, CA 91011, U.S.A.}

\author[0000-0003-0286-5940]{Jesús Varela}
\affiliation{Centro de Estudios de Física del Cosmos de Aragón (CEFCA), Plaza San Juan, 1, E-44001, Teruel, Spain}

\author[0000-0003-3135-2191]{Héctor Vázquez Ramió}
\affiliation{Centro de Estudios de Física del Cosmos de Aragón (CEFCA), Plaza San Juan, 1, E-44001, Teruel, Spain}

\begin{abstract}
We present the second \NuSTAR\ and \XMM\ extragalactic survey of the JWST North Ecliptic Pole (NEP) Time-Domain Field (TDF)\null. The first \NuSTAR\ NEP-TDF survey (Zhao et al.\ 2021) had 681\,ks total exposure time executed in \NuSTAR\ cycle 5, in 2019 and 2020. This second survey, acquired from 2020 to 2022 in cycle~6, adds 880\,ks of \NuSTAR\ exposure time. The overall \NuSTAR\ NEP-TDF survey is the most sensitive \NuSTAR\ extragalactic survey to date, and a total of 60 sources were detected above the 95\% reliability threshold. We constrain the hard X-ray number counts, $\log N$--$\log S$, down to $1.7\times10^{-14}$~erg~cm$^{-2}$~s$^{-1}$ at 8--24~keV and detect an excess of hard X-ray sources at the faint end. About 47\% of the \NuSTAR-detected sources are heavily obscured ($N\rm _{H}>10^{23}$\,cm$^{-2}$), and $18_{-8}^{+20}$\% of the \NuSTAR-detected sources are Compton-thick ($N\rm _{H}>10^{24}$\,cm$^{-2}$). These fractions are consistent with those measured in other \NuSTAR\ surveys. Four sources presented ${>}2\sigma$ variability in the 3-year survey.
In addition to \NuSTAR, a total of 62\,ks of \XMM\ observations were taken during \NuSTAR\ cycle 6. The \XMM\ observations provide soft X-ray (0.5--10\,keV) coverage in the same field and enable more robust identification of the visible and infrared counterparts of the \NuSTAR-detected sources. A total of 286 soft X-ray sources were detected, out of which 214 \XMM\ sources have secure counterparts from multiwavelength catalogs.
\end{abstract}


\section{Introduction} \label{sec:intro}

The Nuclear Spectroscopic Telescope Array (\NuSTAR) mission, launched in June 2012, is the first telescope focusing hard X-rays \citep[3--79\,keV;][]{harrison}. The 100 times deeper hard X-ray sensitivity of \NuSTAR\ compared with the previous collimated or coded mask instruments allows the peak (20--40\,keV) of the Cosmic X-ray background (CXB) to be resolved into individual objects. Indeed, about 35--60\% of the CXB at 8--24\,keV was resolved by previous \NuSTAR\ extragalactic surveys (Hickox et al., in prep). 

 \NuSTAR\ performed a series of extragalactic surveys in its first two-year baseline mission to probe AGN activity over cosmic time. The surveys followed a wedding cake strategy covering from small areas with deep exposures and broader surveys with shallow exposures \citep[see,][hereafter Z21, for an overview]{Zhao2021}.

JWST, successfully launched on Dec. 25, 2021, is a NASA/ESA/CSA Flagship mission focusing on near- and mid-infrared wavelengths (0.6--28.5~$\mu$m) with its 6.5~m aperture and state-of-the-art scientific instruments \citep{Gardner06,Gardner2023}. JWST Interdisciplinary Scientist (IDS) R.~Windhorst allocated $\sim$47 hours of his guaranteed time to the North Ecliptic Pole (NEP) time-domain field (TDF) is part of the ``Prime Extragalactic Areas for Reionization and Lensing Science" (PEARLS) project \citep[GTO-2738;][]{Windhorst_2023}. JWST has observed this field in four orthogonal spikes in cycle~1. Each observation includes eight filters of NIRCam observations and coordinated parallel observations with NIRISS/WFSS. This field was selected to be located within the JWST northern continuous viewing zone (CVZ) to enable time-domain studies. Furthermore, this NEP-TDF has the best combination of low foreground extinction and absence of AB$\le$16 mag stars \citep{Jansen_2018}. 
The NEP-TDF has become a comprehensive multiwavelength survey.\footnote{A comprehensive table can be found at \url{http://lambda.la.asu.edu/jwst/neptdf/}} The multiwavelength coverage of the NEP-TDF approved to date is presented in Table~\ref{Table:NEP-TDF}. 

\begingroup
\renewcommand*{\arraystretch}{1.1}
\begin{table*}
\caption{Approved NEP-TDF multiwavelength surveys.}
\centering
\label{Table:NEP-TDF}
  \begin{tabular}{llll}
       \hline
   \hline     
 	Telescope&PI&Exposure&Reference\\
	\hline
	{NuSTAR}&F.~Civano&3.3~Ms&\citet{Zhao2021}\\
	XMM-{Newton}&F.~Civano&120~ks&\citet{Zhao2021}\\
	{Chandra}&W.~P.~Maksym&1.8~Ms&Maksym et al., in prep.\\
	{AstroSat}/UVIT&K.~Saha&98~hrs&\\
	HST/WFC3+ACS& R.~Jansen \& N.~Grogin&173~hrs&\citet{OBrien2024}, Jansen et al., in prep.\\
	{LBT}/LBC&R.~Jansen&11~hrs&\\
	{Subaru}/HSC&G.~Hasinger \& E.~Hu&5~hrs&\citet{Taylor2023}\\
	{GTC}/HiPERCAM&V.~Dhillon&16~hrs&\\
	{TESS}&G.~Berriman \& B.~Holwerda&357~days&\\
	{MMT}/MMIRS&C.~N.~A.~Willmer&68~hrs&\citet{Willmer2023}\\
	{JWST}/NIRCam+NIRISS&R.~A.~Windhorst \& H.~B.~Hammel&49~hrs&\citet{Windhorst_2023,Adams2023}\\
	{JCMT}/SCUBA-2&I.~Smail \& M.~Im&63~hrs&\citet{Hyun2023}\\
	{IRAM}/NIKA 2&S.~H.~Cohen&30~hrs&\\
	 {SMA} &G.~Fazio&112~hrs&\\
	{VLA}&R.~A.~Windhorst \& W.~Cotton&47~hrs&\citet{Hyun2023,Willner2023}\\
	{VLBA}&W.~Brisken&137~hrs&\\
	{eMERLIN} &A.~Thomson&140~hrs&\\
	 {LOFAR} &R.~Van~Weeren&72~hrs&\\      
	 \hline
	 Spectroscopic\\
	{J-PAS} (56 filters)&S.~Bonoli \& R.~Dupke&29~hrs&\citet{Caballero2023}\\
	MMT/Binospec&C.~N.~A.Willmer&26~hrs&Willmer et al., in prep.\\
	MMT/MMIRS&C.~N.~A.~Willmer&11~hrs&Willmer et al., in prep.\\	
	  \hline
   \hline   	
	
\end{tabular}
\end{table*}
\endgroup

This paper presents the multi-year \NuSTAR\ and \XMM\ extragalactic survey in the NEP-TDF. The paper's focus is the two X-ray source catalogs. 
The paper is organized as follows. Section~\ref{sec:data} describes the \NuSTAR\ data reduction. Section~\ref{sec:simul} describes the construction of simulated data and the resulting reliability, completeness, sensitivity, positional uncertainty, and input-measured relation of the \NuSTAR\ NEP-TDF survey. Section~\ref{sec:NuSTAR_catalog} presents the \NuSTAR\ source catalog. Section~\ref{sec:XMM_catalog} describes our \XMM\ source detection and astrometric corrections, the sensitivity, and the \XMM\ catalog including matching with \NuSTAR. Section~\ref{sec:multi} describes matching \XMM\ sources to existing visible-wavelength and infrared (IR) catalogs using a maximum-likelihood method. We present the X-ray to optical properties of these sources. Section~\ref{sec:discuss} discusses the number counts of the sources as a function of flux, the X-ray hardness ratios, and the Compton-thick (CT) fraction. The Appendixes present the source catalogs of both the \NuSTAR\ and \XMM\ NEP-TDF surveys, spectroscopic redshifts of some VLA and Chandra detected sources in the NEP-TDF, and a newly developed pipeline to analyze source variability of \NuSTAR\ observations. 

Uncertainties are quoted at a 90\% confidence level throughout the paper unless otherwise stated. Magnitudes used here are in the AB system, and standard cosmological parameters are adopted as follows: $H_0 = 70$~km s$^{-1}$ Mpc$^{-1}$, $\Omega_M = 0.30$, and $\Omega_\Lambda = 0.70$.

\section{\NuSTAR\ data processing} \label{sec:data}
\NuSTAR\ (3--24~keV) surveyed the NEP-TDF in both cycle~5 (PI: Civano, ID: 5192) and cycle~6 (PI: Civano, ID: 6218, two-year program). The cycle~5 results were published \citepalias{Zhao2021}. This work focused on the cycle~6 and the combined cycles 5 and 6 data. The \NuSTAR\ cycle 6 NEP-TDF survey comprises 12 observations taken in four epochs spanning from 2020 Oct. to 2022 Jan. with a total of 880\,ks exposure time. The cycle 6 survey was designed with a primary focus on variability, and therefore each epoch's observations pointed to the same area with similar effective position angles.\footnote{\NuSTAR\ reaches similar effective position angles every 3 month due to its square CCD, but the source might lie on different sub-detectors in different epochs.} Table~\ref{Table:obs} presents the details of the individual \NuSTAR\ observations in cycle 5 and cycle 6. This work, following previous \NuSTAR\ extragalactic surveys, focuses on the 3--24~keV band because only the brightest sources can be detected at $>$24~keV owing to the decrease of the effective area and significant increase of the background at $>$24~keV \citep[e.g.,][]{Masini_2018}.

\subsection{Data Reduction}
\begingroup
\renewcommand*{\arraystretch}{1.1}
\begin{table}
\caption{List of \NuSTAR\ and \XMM\ observations.}
\centering
\label{Table:obs}

  \begin{tabular}{ccccc}
       \hline
       \hline     
 	ObsID&Date&R.A.&Dec.&Exp.\\
	&&(deg)&(deg)&(ks)\\
	\hline
	Cycle 5 \NuSTAR\\
	60511001002&2019-09-30&260.8664&65.8298&73.5\\
	60511002002&2019-10-02&260.7643&65.8305&77.6\\
	60511003002&2019-10-04&260.6297&65.8316&68.7\\
	\hline
	60511004002&2020-01-03&260.4992&65.9184&89.8\\
	60511005002&2020-01-04&260.7289&65.8757&84.7\\
	60511006002&2020-01-05&260.8676&65.9159&83.0\\
	\hline
	60511007002&2020-03-01&260.5070&65.7602&65.2\\
	60511008001&2020-03-02&260.7292&65.7406&70.2\\
	60511009001&2020-03-03&260.9369&65.7399&68.3\\
	\hline
	Total&&&&681\\
	\hline
	Cycle 6 \NuSTAR\\
    60666001002&2020-10-18&260.8664&65.8298&72.8\\
	60666002002&2020-10-13&260.6219&65.8075&71.9\\
	60666003002&2020-10-15&260.8534&65.8463&72.7\\
	\hline
	60666004002&2021-01-14&260.6508&65.8776&76.9\\
	60666005002&2021-01-17&260.6541&65.7791&77.5\\
	60666006002&2021-01-18&260.8448&65.8204&80.4\\
	\hline
	60666007002&2021-10-12&260.5930&65.9025&78.1\\
	60666008002&2021-10-14&260.5868&65.7996&49.9\\
	60666009002&2021-10-15&260.8223&65.8393&52.7\\
	\hline
	60666010002&2022-01-19&260.6164&65.8882&80.9\\
	60666011002&2022-01-22&260.5980&65.7878&82.9\\
	60666012002&2022-01-23&260.8546&65.8173&82.9\\
	\hline	
	Total&&&&880\\
	\hline
	\hline
	Cycle 6 XMM\\
	0870860101&2020-10-14&260.6917&65.8711&17.0\\
	0870860201&2021-01-16&260.6917&65.8711&23.2\\
	0870860301\rlap{\tablenotemark{a}}&2021-10-14&260.6917&65.8711&0$^*$\\	
	0870860401&2022-01-24&260.6917&65.8711&21.9\\	
	\hline
	Total&&&&62\\
	\hline
	\end{tabular}
\raggedright
\tablenotetext{a}{This observation was entirely lost to high particle background during the observation.}
\end{table}
\endgroup

The reduction of the cycle 6 observations used the same method as for the cycle 5 data (\citetalias{Zhao2021}). In brief, the \NuSTAR\ data were processed using HEASoft v.6.29c and \NuSTAR\ Data Analysis Software (NuSTARDAS) v.2.1.1 with the updated calibration and response files CALDB v.20211115. The level 1 raw data were calibrated, cleaned, and screened by running the \texttt{nupipeline} tool. Following \citetalias{Zhao2021}, we removed the high-background time intervals (when the count rate in the 3.5--9.5~keV band was at least double the average count rate of the entire observation) in each observation. The total exposure losses of the two \NuSTAR\ focal plane modules (FPMs) due to the high background were 6.5\,ks and 8\,ks for FPMA and FPMB, respectively, corresponding to 0.8\% of the entire cycle 6 \NuSTAR\ NEP-TDF survey exposures.

\subsection{Exposure Map Production} \label{sec:exposure}
\begin{figure} 
\centering
\includegraphics[width=.48\textwidth]{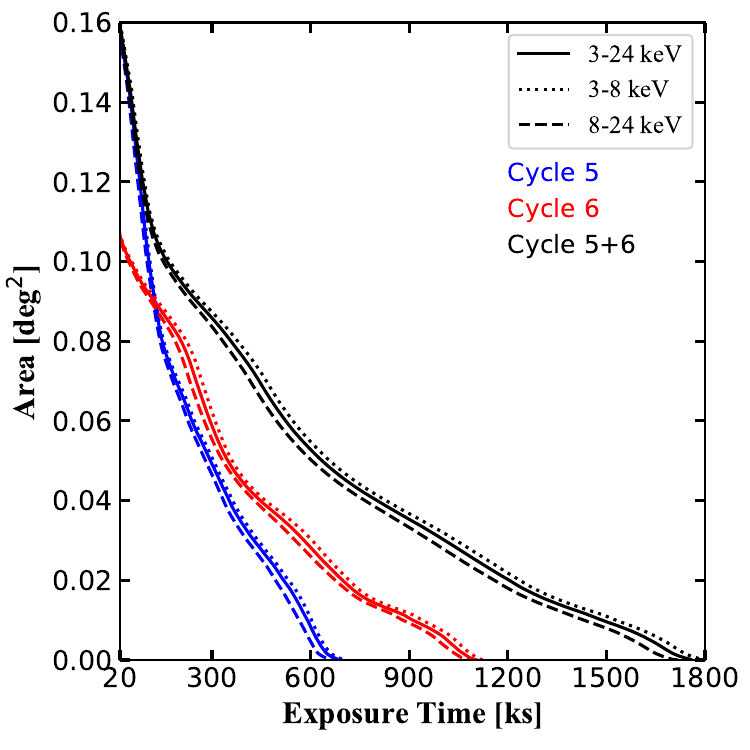}
\caption{Cumulative survey area as a function of the FPMA+B vignetting-corrected exposure time. Colors distinguish the \NuSTAR\ NEP-TDF surveys in cycle 5 (blue), cycle 6 (red), and combined cycles 5+6 (black), and line types distinguish energy bands as shown in the legend.} 
\label{fig:expomap}
\end{figure}

We generated the vignetting-corrected exposure map of each \NuSTAR\ observation in three energy bands: 3--24\,keV, 3--8\,keV, and 8--24\,keV using the NuSTARDAS tool \texttt{nuexpomap}. The exposure map of the entire cycle 6 survey was produced by merging the 12 individual maps into a mosaic. That included merging the two FPMs' observations as FPMA+B. Figure~\ref{fig:expomap} shows the cumulative areas as a function of the vignetting-corrected exposure in the three energy bands. The cycle~6 survey repeatedly observed the same region, so its exposure is $\sim$60\% deeper than the cycle~5 survey (Figure~\ref{fig:expomap}) but with $\sim$35\% smaller area coverage ($\sim$0.107~deg$^2$ in cycle~6 compared with $\sim$0.16~deg$^2$ in cycle~5). To achieve the deepest exposure, we also merged the cycle~5 and cycle~6 observations to achieve a central exposure time of $\approx$1.7~Ms (FPMA+B). We used the 8--24 keV exposure map to present the 8--16 keV, and 16--24 keV exposure maps as they show only marginal differences. 

\begin{figure*} 
\begin{minipage}[b]{.488\textwidth}
\centering
\includegraphics[width=\textwidth]{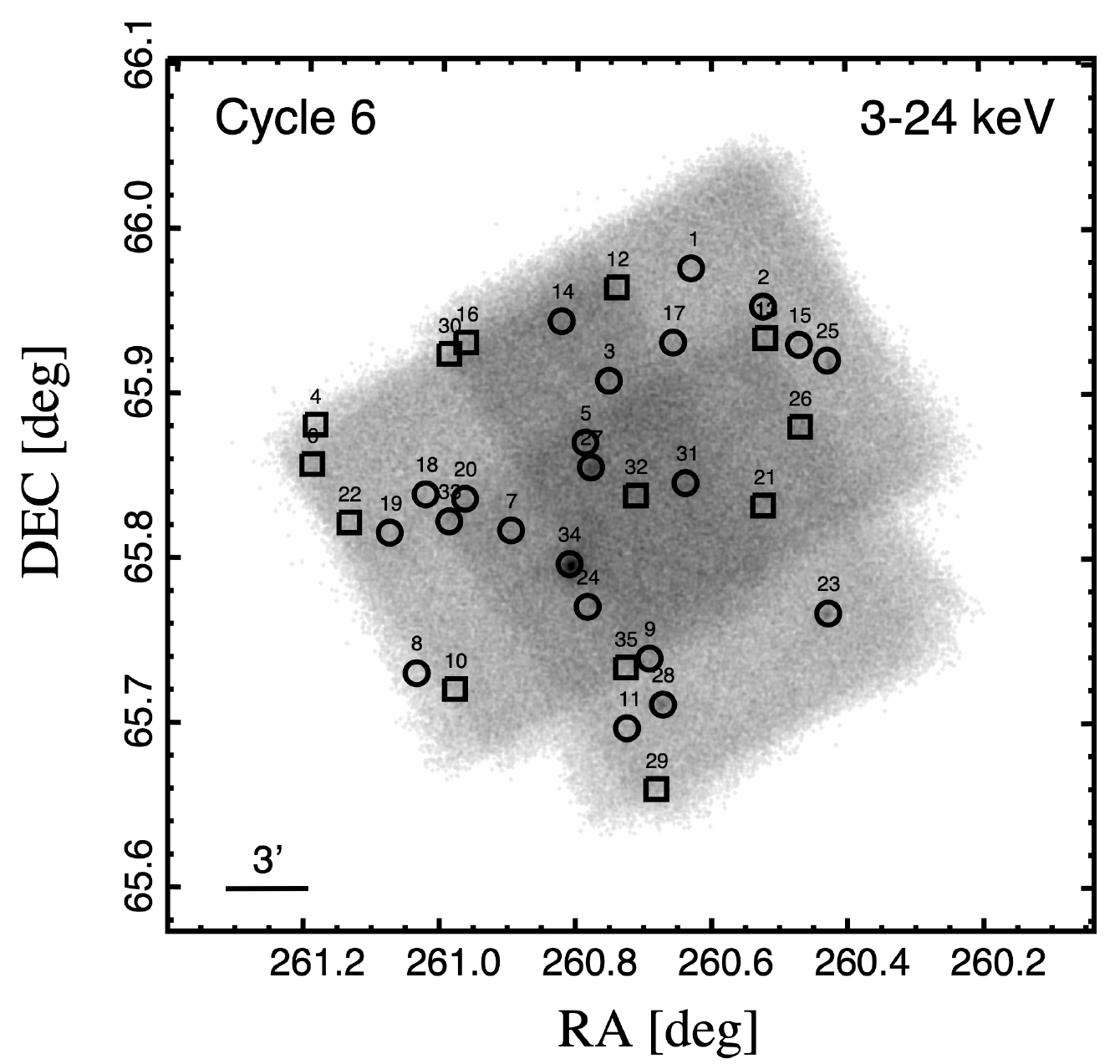}
\end{minipage}
\begin{minipage}[b]{.49\textwidth}
\centering
\includegraphics[width=\textwidth]{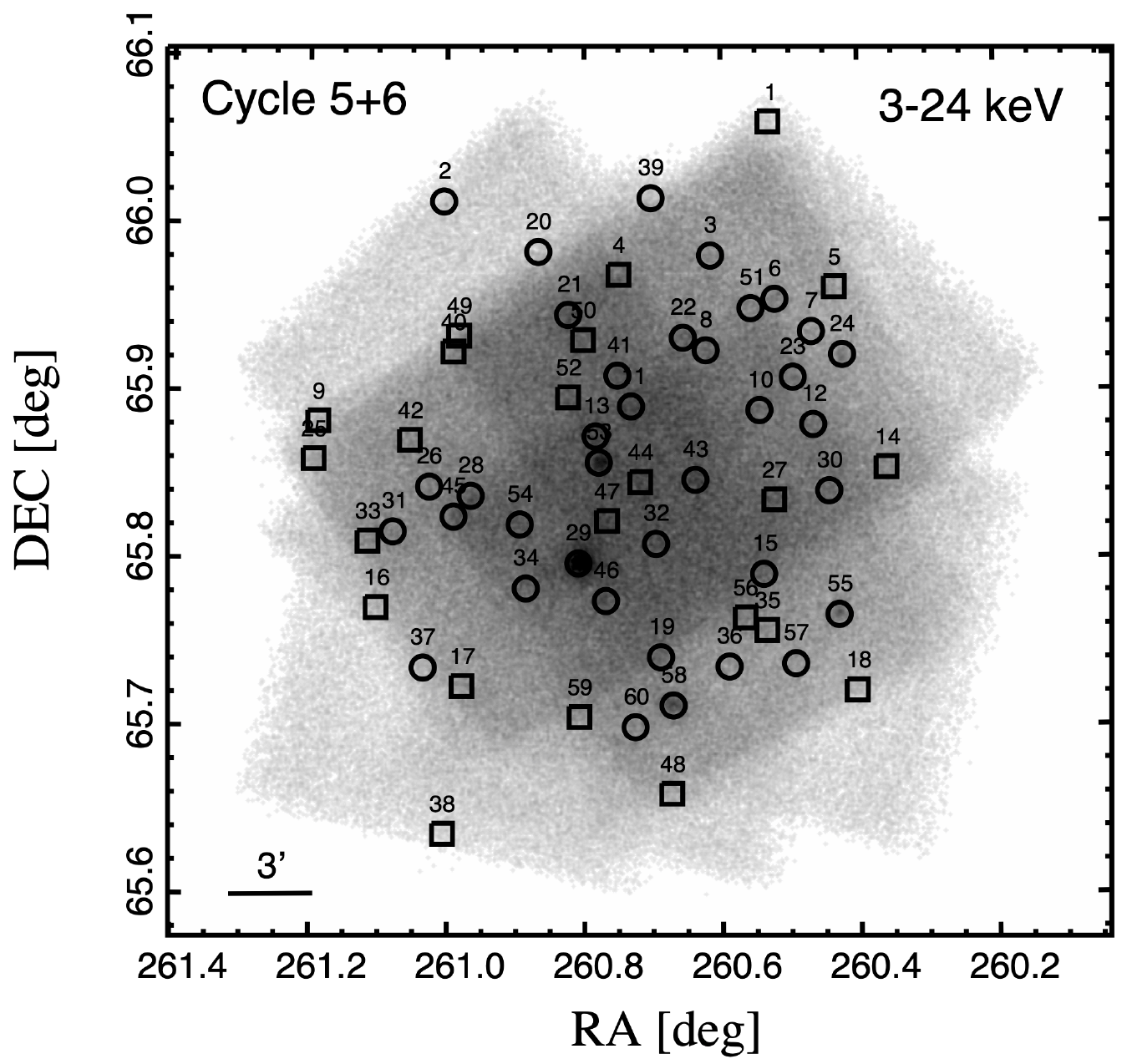}
\end{minipage}
\caption{\NuSTAR\ FPMA+B 3--24\,keV mosaics of cycle 6 (left) and cycles 5+6 (right) observations. The \NuSTAR-detected sources with (black circles, 25\arcsec\ radius) and without (black squares, 45\arcsec\ width) soft X-ray counterparts are marked.
Labels are source IDs in the respective catalogs.}
\label{fig:mosaic}
\end{figure*}   

\subsection{Observation Mosaic Creation} \label{sec:mosaic}

For each of the 12 cycle~6 observations, summed FPMA+B mosaics were created in five energy bands: 3--24\,keV, 3--8\,keV, 8--24\,keV, 8--16\,keV, and 16--24\,keV. The five bands were separated using the HEASoft \texttt{Xselect} tool. Each band was then merged into a full-exposure mosaic using the \texttt{Ximage} tool. Loss of spatial resolution due to the resampling is negligible because the pixel angular resolution (\ang[angle-symbol-over-decimal]{;;2.45}) of the standard \NuSTAR\ sky binning by NuSTARDAS is much smaller than the $\sim$18\arcsec\ full width at half-maximum (FWHM) of \NuSTAR. The cycle~6 mosaic was also merged with the cycle~5 mosaic to achieve the deepest sensitivity. The astrometric offsets of the \NuSTAR\ NEP-TDF survey could not be measured reliably due to the limited number of bright sources in the field, but the previous \NuSTAR\ COSMOS survey \citep{Civano_2015} found a \NuSTAR\ astrometric offset of 1\arcsec--7\arcsec, small compared to the \NuSTAR\ FWHM\null. Therefore, astrometric offsets should only marginally affect our results \citep{Civano_2015}, and we did not apply any astrometric correction when merging the observations. (In any case, there is only one bright source in the FoV that could have been used for astrometric correction.) Figure~\ref{fig:mosaic} shows the 3--24\,keV FPMA+B merged mosaics.

\subsection{Background Map Production} \label{section:background}
The background map was used for both source detection and simulation. The background of \NuSTAR\ is spatially non-uniform across the field of view (FoV) and is variable among different observations, adding complexities when producing the background maps. We used the {\tt nuskybgd}\footnote{\url{https://github.com/NuSTAR/nuskybgd}} package \citep{Wik_2014}, which was used for all previous \NuSTAR\ extragalactic surveys, to produce background maps of each cycle~6 observation following the same method as \citetalias{Zhao2021}. We merged the cycle~6 background maps into an FPMA+B mosaic, which was then merged with the cycle~5 mosaic into a cycles 5+6 mosaic. 

To test the accuracy of the generated background maps, we compared the number of counts in the observed images and the corresponding background maps. As the background dominates the observation in most areas, the observed numbers of counts should be consistent except for the regions where (bright) X-ray sources exist. The comparison was based on 64 circular, 45\arcsec-radius regions across the FoV in each observation, and the mean difference between the observed images (Data) and the background maps (Bgd) is $\rm (Data - Bgd)/Bgd = -0.7\%$ and 2.0\% for FPMA and FPMB, respectively. The standard deviations of the differences are 12.2\% and 13.8\% for FPMA and FPMB, respectively. These suggest good modeling of the \NuSTAR\ background and are consistent with the accuracy obtained in cycle~5 (\citetalias{Zhao2021}) and by \citet{Civano_2015}. 

\section{\NuSTAR\ Simulations} \label{sec:simul}
Comprehensive simulations in each energy band (3--24\,keV, 3--8\,keV, 8--24\,keV, 8--16\,keV, and 16--24\,keV) were used to (1) determine the reliability and completeness of our source detection technique and the resulting source catalogs, (2) measure the sensitivity of the surveys, and (3) demonstrate the quality of our source detection technique by comparing the input and measured source properties (e.g., positions and fluxes).
\begingroup
\renewcommand*{\arraystretch}{1.1}
\begin{table*}
\caption{Source detections in simulated and real data}
\label{Table:number_cycle6}
  \begin{tabular}{lcccccc}
       \hline
       \hline   
 	&3--24\,keV&3--8\,keV&8--24\,keV&8--16\,keV&16--24\,keV\\
	\hline
       &&Cycle 6\\
       \hline
 	Simulations&&&&&\\
	Detections in each simulated map&53&51&49&49&43\\
	Detections matched to input catalog&38&36&29&29&16\\
	DET\underline{\;\;}ML$>$95\% reliability threshold&20.9&17.4&8.2&9.0&0.6\\
	\hline 
	Real Data&&&&&&Total\\
	DET\underline{\;\;}ML$>$95\% reliability threshold&28&24&13&15&3&35\\
	\hline 
 	&&Cycles 5+6\\
	\hline
	Simulations&&&&&\\
	Detections in each simulated map&80&78&74&72&63\\
	Detections matched to input catalog&57&53&43&44&24\\
	DET\underline{\;\;}ML$>$95\% reliability threshold&32.9&27.7&14.1&15.9&1.3\\
	\hline 
	Real Data&&&&&&Total\\
	DET\underline{\;\;}ML$>$95\% reliability threshold&45&32&24&26&2&60\\
	\hline\end{tabular}
\tablecomments{The mean number of the detected sources using \texttt{SExtractor} in each simulated map (line 1). The mean number of the detected sources matched the input source catalog within 30\arcsec\ (line 2). The mean number of the detected sources with DET\_ML above 
95\% reliability threshold in the simulated maps (lines 3) and the real data (lines 5).}
\end{table*}
\endgroup

\subsection{Generating Simulated Observations} \label{sec:simulation}
We generated simulated observations in the five energy bands following \citetalias{Zhao2021}'s procedure. To summarize, each iteration randomly places mock sources on the background maps described in Section~\ref{section:background}. The fluxes of the mock sources were randomly assigned following the X-ray source flux distribution ($\log N$--$\log S$) measured by \citet{Treister09}. The minimum fluxes in the 3--24\,keV bands were $3 \times 10^{-15}$\,erg\,cm$^{-2}$\,s$^{-1}$ for cycle~6 and $2 \times 10^{-15}$\,erg\,cm$^{-2}$\,s$^{-1}$ for cycle~5+6, about 10 times fainter than the expected limits of each survey. Adopting a much fainter flux limit than the sensitivity of the survey would result in many mismatched measured and input sources and produce an incorrect reliability curve of source detection. The fluxes of the input sources in each energy band were extrapolated from their 3--24\,keV fluxes assuming an absorbed power-law model with photon index $\Gamma = 1.80$ and a Galactic absorption $N\rm_H = 3.4 \times 10^{20}$\,cm$^{-2}$ \citep{nh}. The fluxes were converted to count rates using conversion factors (CF) of 4.86, 3.39, 7.08, 5.17, and $16.2 \times 10^{-11}$\,erg\,cm$^{-2}$ count$^{-1}$ in the 3--24, 3--8, 8--24, 8--16, and 16--24\,keV bands, respectively. The CF was computed using WebPIMMS\footnote{\url{https://heasarc.gsfc.nasa.gov/cgi-bin/Tools/w3pimms/w3pimms.pl}} assuming the above spectral model. The simulated observations of each exposure were then merged into mosaics in five energy bands for both FPMA and FPMB, which were then combined into FPMA+B mosaics. We used 1200 simulations for the \NuSTAR\ NEP-TDF cycle 6 survey and 2400 for the combined cycles 5+6 survey. 

\begin{figure*} 
\begin{minipage}[b]{.48\textwidth}
\centering
\includegraphics[width=\textwidth]{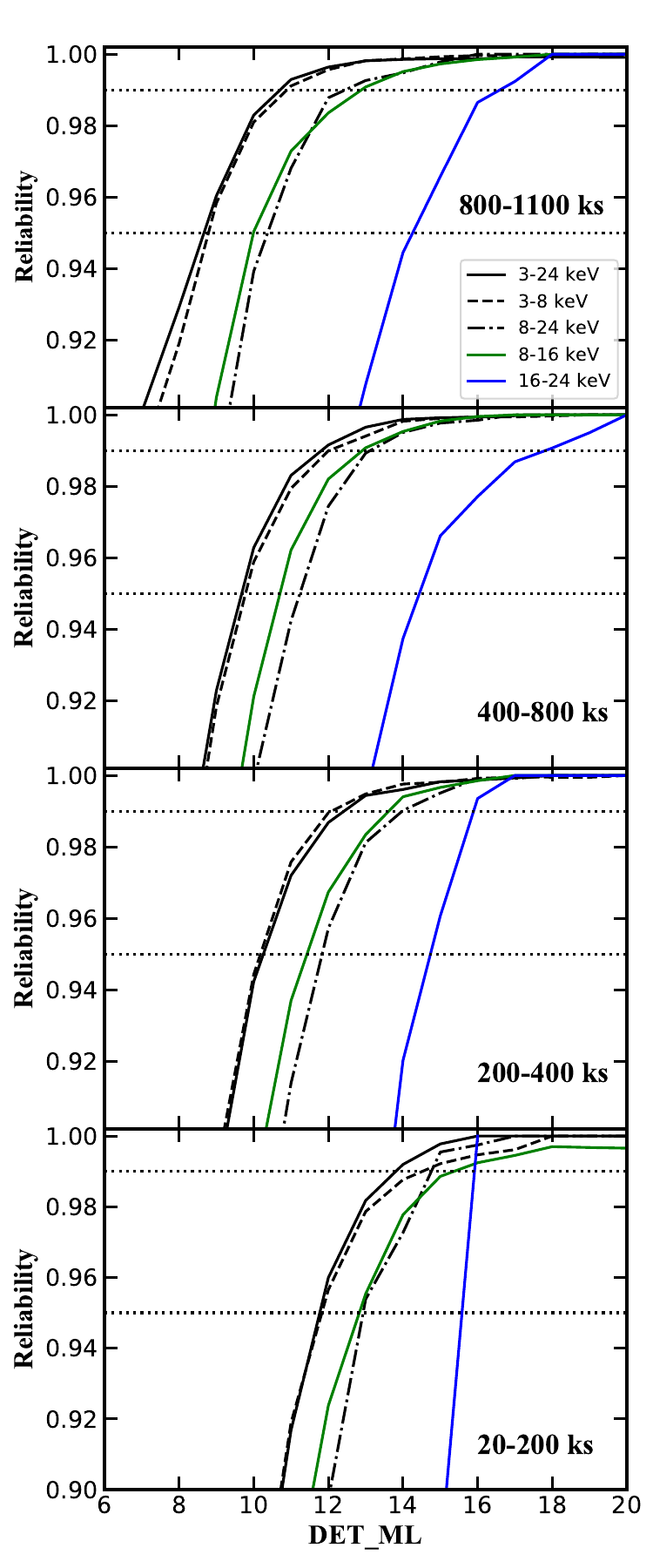}
\end{minipage}
\begin{minipage}[b]{.48\textwidth}
\centering
\includegraphics[width=\textwidth]{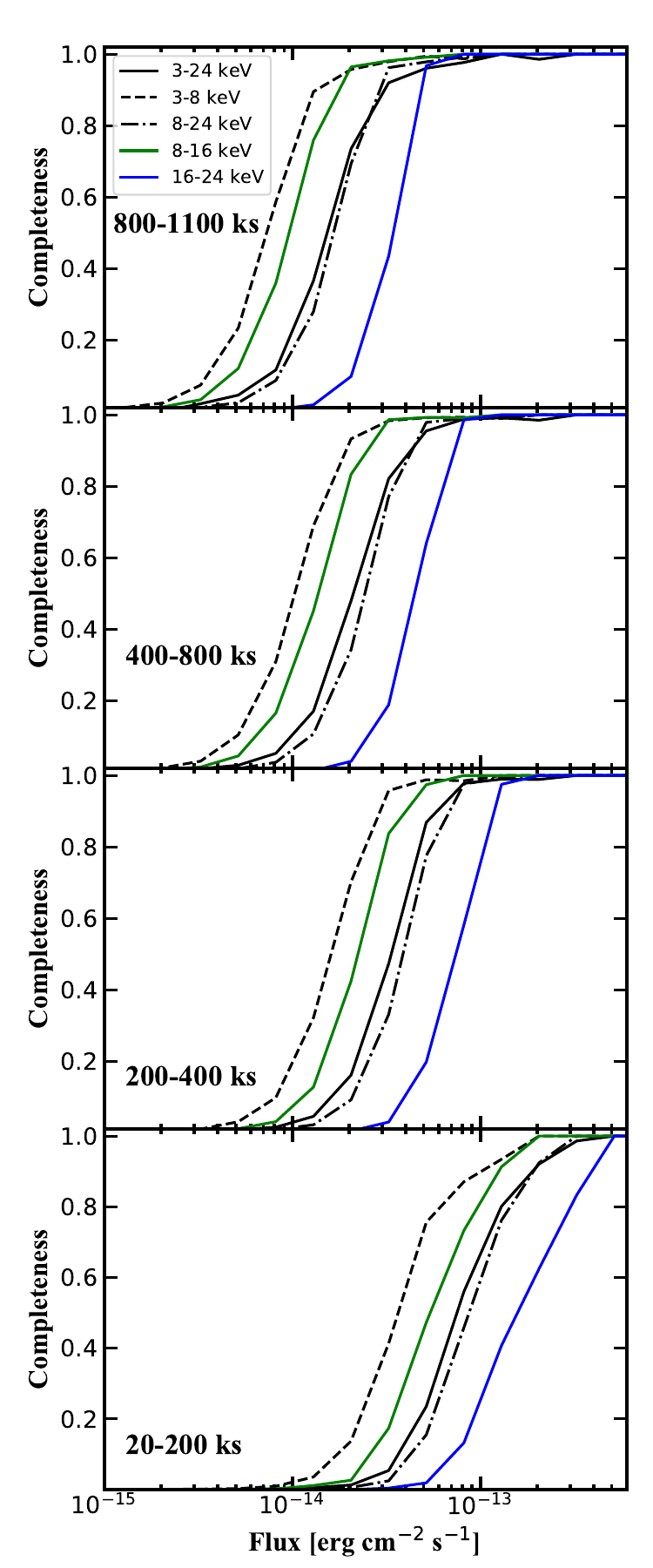}
\end{minipage}
\caption{Left column: Reliability (equation~\ref{eq:reliablity}) as a function of DET\_ML\null. Right column: Completeness (equation~\ref{eq:completeness}) at 95\% reliability as a function of flux. 
All values come from the cycle~6 simulations described in Section~\ref{sec:simul}.
Panels from top to bottom show four different exposure-time ranges, as labeled. Line types indicate energy ranges as indicated in the legends in the top panels. Dotted horizontal lines in the left panels show 95\% and 99\% reliability.}
\label{fig:relia_comple_cycle6}
\end{figure*}

\begin{figure*} 
\begin{minipage}[b]{.48\textwidth}
\centering
\includegraphics[width=\textwidth]{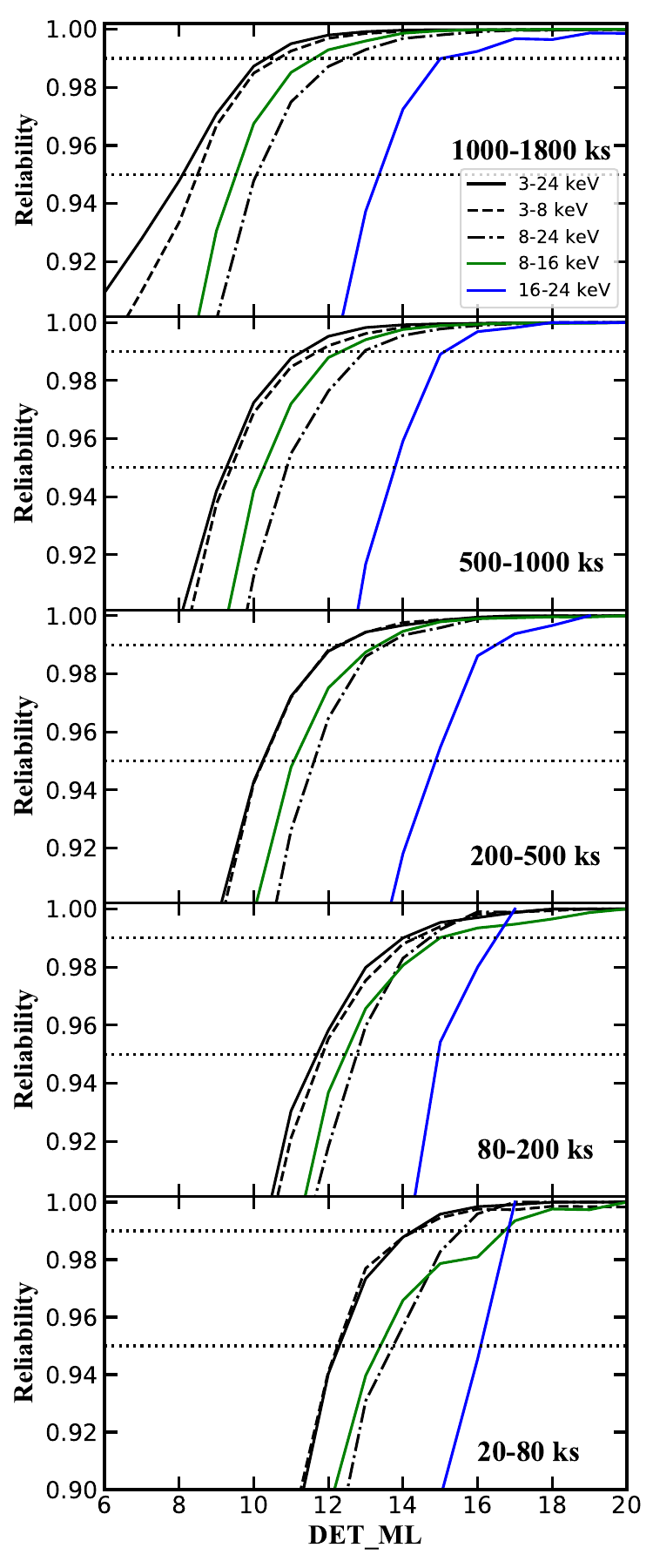}
\end{minipage}
\begin{minipage}[b]{.48\textwidth}
\centering
\includegraphics[width=\textwidth]{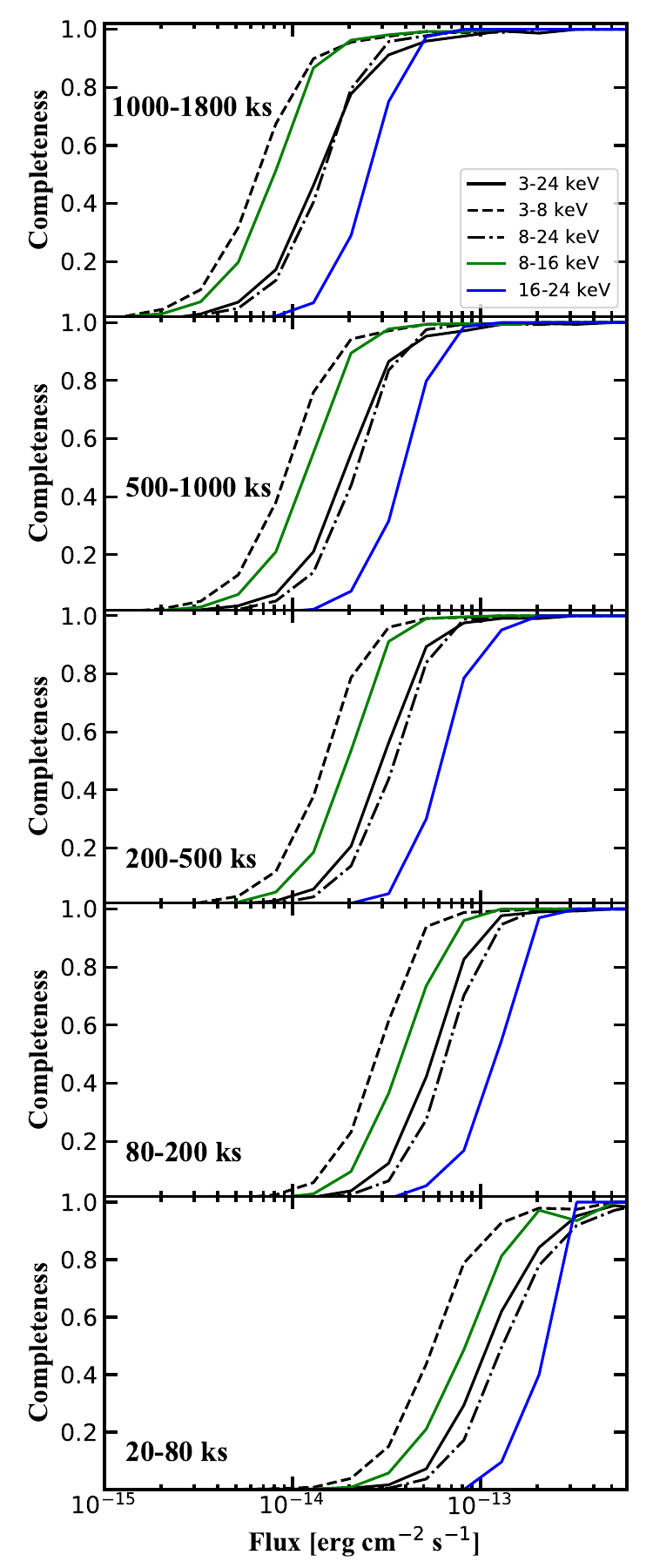}
\end{minipage}
\caption{Left column: Reliability (equation~\ref{eq:reliablity}) as a function of DET\_ML\null. Right column: Completeness (equation~\ref{eq:completeness}) at 95\% reliability as a function of flux. 
All values come from the cycle~5+6 simulations described in Section~\ref{sec:simul}.
Panels from top to bottom show four different exposure-time ranges, as labeled. Line types indicate energy ranges as indicated in the legends in the top panels. Dotted horizontal lines in the left panels show 95\% and 99\% reliability.}
\label{fig:relia_comple_cycle56}
\end{figure*}

\subsection{Source Detection on Simulated Observations} \label{sec:detection}
We performed source detection on the simulated cycle 6 and cycles 5+6 FPMA+B mosaics using the technique developed by \citet{Mullaney15}. In summary, source detection used \texttt{SExtractor} \citep{Bertin1996} on the false-probability maps produced by the simulated observations and background maps. We defined the maximum likelihood (DET\underline{\;\;}ML) of each detection, which measured the chance that the detection is from the background fluctuation rather than from a real source. A higher DET\underline{\;\;}ML suggests a lower chance that the detection is from the background fluctuation. Further details were described by \citetalias{Zhao2021}. 


\begingroup
\renewcommand*{\arraystretch}{1.25}
\begin{table*}
\centering
\caption{DET\_ML values required for 95\% reliability}
\label{Table:DETML_cycle6}
  \begin{tabular}{lcccccc}
       \hline
       \hline   
 	&3--24\,keV&3--8\,keV&8--24\,keV&8--16\,keV&16--24\,keV\\
	\hline
       &&Cycle 6\\
       
	DET\underline{\;\;}ML(20--200\,ks) threshold&11.77&11.83&12.93&12.83&15.58\\
	DET\underline{\;\;}ML(200--400\,ks) threshold&10.26&10.18&11.83&11.43&14.74\\
	DET\underline{\;\;}ML(400--800\,ks) threshold&9.68&9.78&11.24&10.70&14.44\\
	DET\underline{\;\;}ML(800-1100\,ks) threshold&8.67&8.79&10.37&9.99&14.26\\
	\hline
 	&&Cycles 5+6\\
	DET\underline{\;\;}ML(20--80\,ks) threshold&12.29&12.25&13.74&13.40&16.08\\
	DET\underline{\;\;}ML(80--200\,ks) threshold&11.70&11.85&12.77&12.46&14.95\\
	DET\underline{\;\;}ML(200--500\,ks) threshold&10.24&10.26&11.62&11.08&14.87\\
	DET\underline{\;\;}ML(500--1000\,ks) threshold&9.28&9.40&10.88&10.26&13.78\\
	DET\underline{\;\;}ML(1000-1800\,ks) threshold&8.10&8.50&10.08&9.53&13.36\\	
	\hline
	\end{tabular}
\end{table*}
\endgroup

The measured counts associated with a detected source might be contaminated by other sources within 90\arcsec, corresponding to 85--90\% encircled energy fraction (EEF) of the \NuSTAR\ point spread function (PSF). Therefore, we applied a deblending process to the detected sources in each simulation following \citet{Mullaney15}. The deblended source counts and background counts were then used to update DET\underline{\;\;}ML values for each detection. 

The detections with the updated DET\underline{\;\;}ML of each simulation were then matched with the input catalog using a 30\arcsec\ search radius. The average numbers of the sources detected and matched to the input catalogs in each simulation are listed in Table~\ref{Table:number_cycle6}.

\subsection{Reliability and Completeness} \label{sec:relia}
To evaluate the accuracy and efficiency of the source detection in the real observations, we used the statistics of the simulated sources described in Section~\ref{sec:detection}. Reliability is the ratio of true detections, i.e., matching input sources, to the total number of detected sources at or above a particular DET\_ML threshold:
\begin{equation}\label{eq:reliablity}
{\rm Rel (DET\_ML)}= \frac{N_{\rm matched}({\ge} \rm DET\_ML)}{N_{\rm detected}({\ge}\rm DET\_ML)}
\end{equation}
\noindent Therefore, if 95 out of 100 detected sources with $\rm DET\_ML\ge15$ were matched to input sources, then the reliability of the detection at $\rm DET\_ML\ge15$ is 95\%. 
Completeness is defined as the ratio of the number of detected true sources to the number in the input catalog at a particular flux assuming a particular realiability:
\begin{equation}\label{eq:completeness}
{\rm Completeness (flux)} = \frac{N_{\rm matched \& \ge Rel\_({\rm flux})}}{N_{\rm input(flux)}}
\end{equation}
\noindent Therefore, if 90 out of 100 input sources at flux $1 \times 10^{-13}$\,erg~cm$^{-2}$s$^{-1}$ were detected above the 95\% reliability level, then the completeness of the survey at this particular flux is 90\% at the 95\% reliability level. 
A higher reliability level requires a higher DET\_ML threshold, but that leads to lower completeness. 
Reliability and completeness curves obtained from the cycle 6 and cycles 5+6 simulations are plotted in Figures~\ref{fig:relia_comple_cycle6} and~\ref{fig:relia_comple_cycle56}, respectively.

The reliability and completeness curves heavily depend on the effective exposure time because the spurious detection rate at a given threshold decreases exponentially with exposure. As in cycle 5, the effective exposure time across the entire NEP-TDF cycle 6 and cycles 5+6 survey area is nonuniform (Figure~\ref{fig:expomap}) because of the observing strategy. We therefore analyzed the reliability function in different exposure intervals as given in Table~\ref{Table:DETML_cycle6}. The exposure intervals were selected to keep a similar number of detected sources in each interval to achieve similar statistics, and we chose an exposure cutoff at 20\,ks to avoid potentially spurious detections on the edge of the observations. 
Table~\ref{Table:number_cycle6} reports the average number of sources detected at $>$95\% reliability level in each simulation.

\subsection{Sensitivity Curves}
\begin{figure} 
\centering
\includegraphics[width=.5\textwidth]{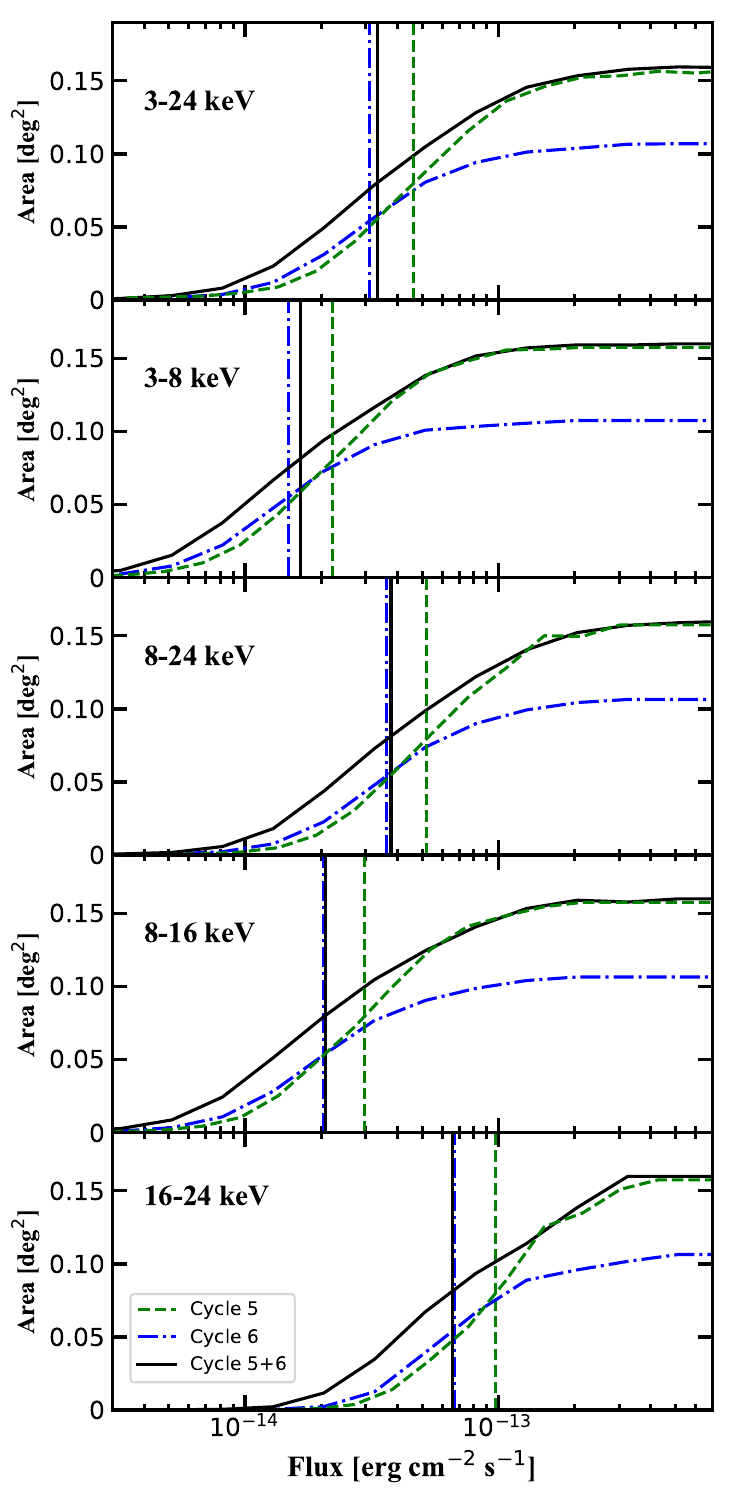}
\caption{Sky coverage as a function of flux at the 95\% reliability level. Panels from top to bottom are for the five energy bands as labeled. As indicated in the legend, different line types show different survey components. The vertical dashed line shows the half-area flux of each survey component.  (At the highest energies, the cycle~5 and cycle~6 lines overlap.)}
\label{fig:coverage}
\end{figure}

The effective sky coverage of the survey at a particular flux is the completeness at that flux multiplied by the maximum covered area. For instance, if a survey covering 0.1\,deg$^2$ has 80\% completeness at a particular flux, the survey's effective area at that flux is 0.08\,deg$^2$. 
As sensitivity depends on the exposure time, the effective sky coverage of the NEP-TDF survey was calculated by adding up the sky-coverage curves of all different exposure intervals (Table~\ref{Table:exposure_area}). The effective sky-coverage curves in the five energy bands of cycle 6 and cycles 5+6 surveys are plotted in Figure~\ref{fig:coverage}. The half-area and 20\%-area sensitivity of the three surveys in different energy bands are reported in Table~\ref{Table:sensitivity}. 

\begingroup
\begin{table}
\centering
\caption{Areas covered with different exposure times}
\label{Table:exposure_area}
  \begin{tabular}{cccccc}
       \hline
       \hline   
\multicolumn{2}{c}{Cycle 5}&
\multicolumn{2}{c}{Cycle 6}&
\multicolumn{2}{c}{Cycles 5+6}\\
       \hline
 	Exposure&Area&Exposure&Area&Exposure&Area\\
	ks&deg$^2$&ks&deg$^2$&ks&deg$^2$\\
	\hline
	20--200&0.091&20--200&0.026&20--80&0.034\\
	200--500&0.047&200--400&0.037&80--200&0.030\\
	500--700&0.019&400--800&0.030&200--500&0.033\\
	&&800--1100&0.014&500--1000&0.033\\
	&&&&1000--1800&0.031\\
	\hline
	\end{tabular}
\raggedright
\tablecomments{Exposure times are vignetting-corrected times for FPMA+B in the 3--24~keV band.}
\end{table}
\endgroup

\begin{figure} 
\centering
\includegraphics[width=.49\textwidth]{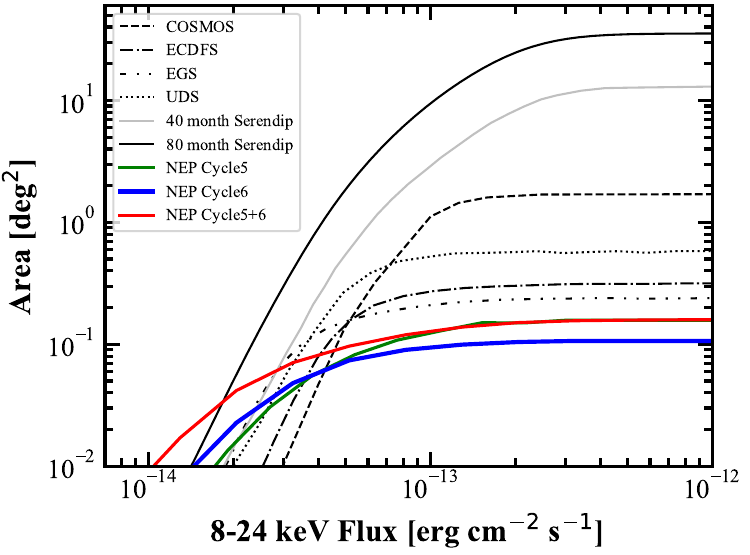}
\caption{8--24\,keV sky coverage as a function of flux in \NuSTAR\ surveys. Different line types represent different \NuSTAR\ extragalactic surveys as indicated in the legend. 
}
\label{fig:sensitivity}
\end{figure}

Observations in the 8--24\,keV band are unique to \NuSTAR. Therefore, Figure~\ref{fig:sensitivity} compares the 8--24~keV sensitivities of the cycle 5, cycle 6, and cycles 5+6 NEP-TDF surveys with previous \NuSTAR\ surveys. NEP-TDF currently reaches the deepest flux in a contiguous \NuSTAR\ survey.

\begingroup
\renewcommand*{\arraystretch}{1.2}
\begin{table}
\caption{\NuSTAR\ Survey Sensitivities}
\centering
\label{Table:sensitivity}
  \begin{tabular}{ccc}
       \hline
       \hline     
 	Energy (keV)&Half-area&20\%-area\\
	\hline
	\NuSTAR	&Cycle 5/6/5+6&Cycle 5/6/5+6\\
	&10$^{-14}$\,erg\,cm$^{-2}$\,s$^{-1}$&10$^{-14}$\,erg\,cm$^{-2}$\,s$^{-1}$\\
	3--24&4.6/3.1/3.3&2.4/1.7/1.6\\
	3--8&2.2/1.5/1.7&1.1/0.80/0.74\\
	8--24&5.2/3.6/3.8&2.7/2.0/1.7\\
	8--16&3.0/2.1/2.1&1.5/1.1/0.95\\
	16--24&9.8/6.7/6.6&5.2/3.9/3.1\\
	\hline
	\XMM& Cycle 6&Cycle 6\\
	&10$^{-15}$\,erg\,cm$^{-2}$\,s$^{-1}$&10$^{-15}$\,erg\,cm$^{-2}$\,s$^{-1}$\\
	0.5--2&0.87&0.63\\
	2--10&6.3&4.0\\
	\hline
\end{tabular}
\raggedright
\tablecomments{Sensitivities are shown for half and for 20\% of the maximum survey area in all relevant energy bands.}
\end{table}
\endgroup

\subsection{Positional Uncertainty} \label{sec:position}
\begin{figure} 
\centering
\includegraphics[width=.5\textwidth]{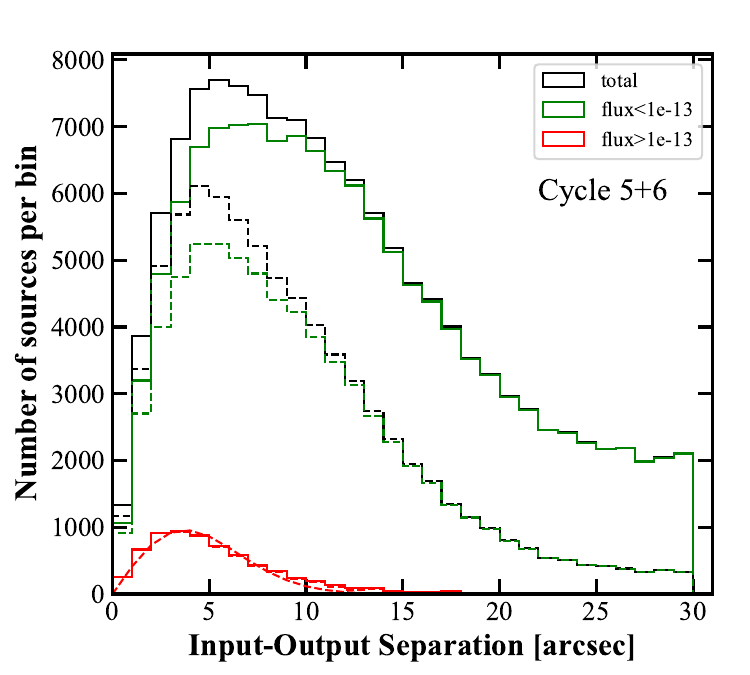}
\caption{Distributions of position offsets from the \NuSTAR\ simulations of the 3--24\,keV cycles 5+6 survey. Solid lines refer to the whole sample and dashed lines refer only to sources above the 95\% reliability level. The dashed red curve shows the best Rayleigh fit to the offsets of sources above the 95\% reliability level. 
}
\label{fig:separation}
\end{figure}

The simulations described in Section~\ref{sec:simulation} can quantify the positional uncertainties of the sources detected. Figure~\ref{fig:separation} shows the separations between the detected and input-catalog positions in the cycles 5+6 survey simulations as an example. The separation histograms follow a Rayleigh distribution \citep{Pineau17}. The best-fit Rayleigh scale parameter for all matched sources are $\sigma_{\rm all, C6}$ = \ang[angle-symbol-over-decimal]{;;9.5} and $\sigma_{\rm all, C56}$ = \ang[angle-symbol-over-decimal]{;;9.2} for cycle 6 and cycles 5+6 surveys, respectively. Eliminating the faintest sources by limiting the sample to sources detected above the 95\% reliability level gives smaller separations $\sigma_{\rm 95\%, C6} = 6\farcs6$ and $\sigma_{\rm 95\%, C56} = 6\farcs5$. The separations are even smaller for sources with 3--24\,keV flux ${>}10^{-13}$\,erg\,cm$^{-2}$\,s$^{-1}$, $\sigma_{\rm 95\%,bright,C6} = 3\farcs8$ and $\sigma_{\rm 95\%,bright,C56} = 3\farcs7$. 
The measured separations are consistent with the previous cycle~5 survey and other \NuSTAR\ extragalactic surveys, and therefore we used these distributions as the expected positional uncertainty of real detections. The simulations did not include the astrometric offsets, and therefore they do not reflect the full positional uncertainty, but the effect is likely minimal. 

\begin{figure} 
\centering
\includegraphics[width=.49\textwidth]{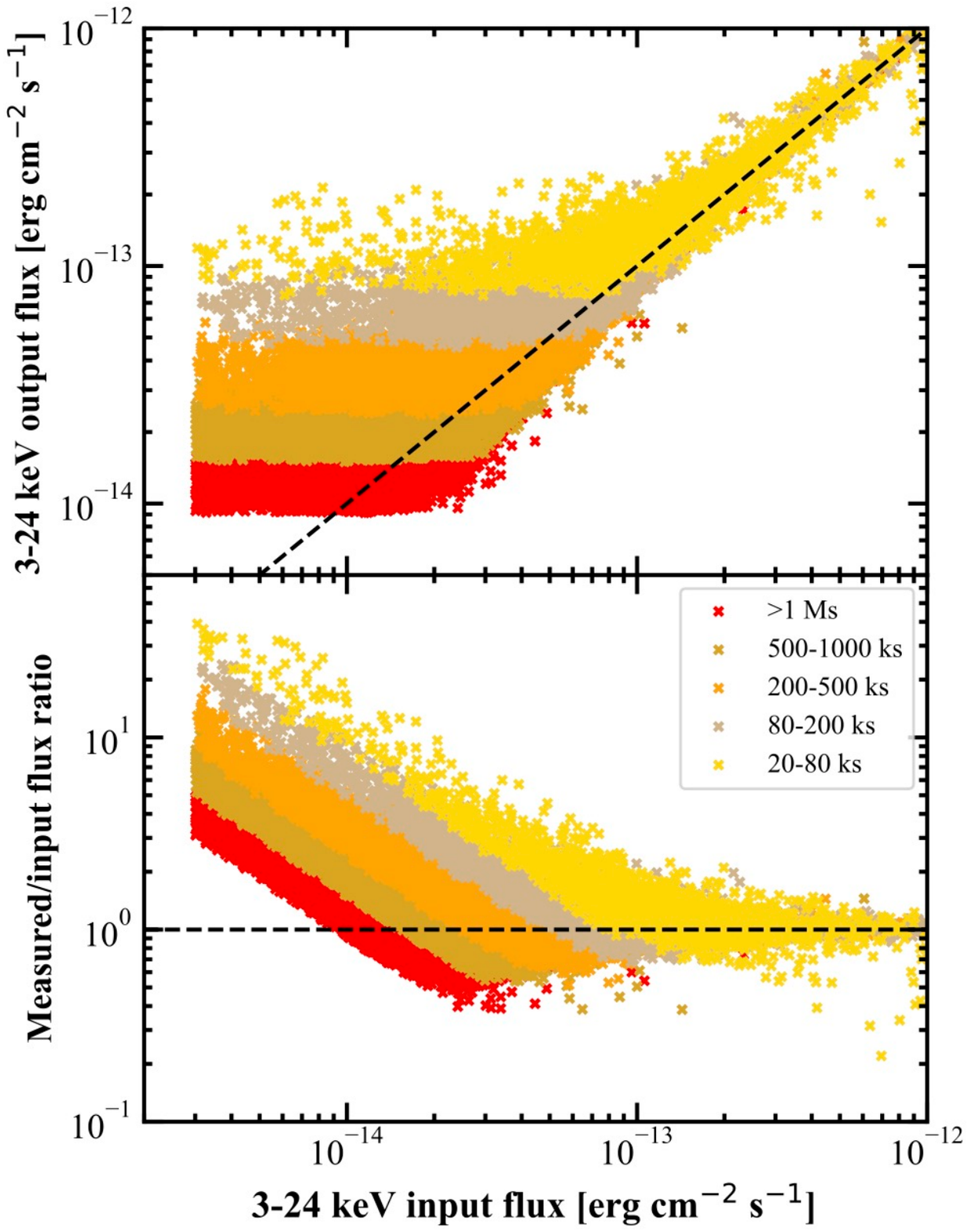}
\caption{Upper: Measured versus input 3--24 keV fluxes for simulated sources. Lower: ratio of measured to input fluxes for the same sources. Both panels show only sources above the 95\% reliability level of the cycles 5+6 survey. Sources in different exposure intervals are plotted in different colors as indicated in the legend.  Dashed lines in both panels show equality. The excess at lower fluxes is due to the Eddington bias.}
\label{fig:flux}
\end{figure}   

\subsection{Fluxes}\label{sec:Fluxes}
The fluxes of the simulated sources detected in each energy band were measured and compared with the input fluxes to quantify the accuracy of the flux measurement technique. We extracted the source counts and deblended background counts of each matched source using the CIAO \citep{CIAO} tool {\tt dmextract}. The effective exposure of each source was measured from the exposure maps (Section~\ref{sec:exposure}). The net counts were then converted to in-band fluxes using the CF (Section~\ref{sec:simulation}). The counts were extracted in a 20\arcsec\ circular region, and we converted this aperture flux to total flux using an aperture correction factor of $F({20\arcsec})/F_{\rm tot} = 0.32$, as calculated from the \NuSTAR\ PSF.\footnote{\url{https://heasarc.gsfc.nasa.gov/docs/nustar/NuSTAR_observatory_guide-v1.0.pdf}} 
Figure~\ref{fig:flux} shows the ratio of measured to input 3--24\,keV fluxes. Flux measurements for faint sources are over-estimated, as expected from Eddington bias which favors the detection of faint sources with positive noise deflections. This excess corresponds to the detection limits of the survey and is also exposure-dependent (Figure~\ref{fig:relia_comple_cycle56}). Therefore, the fluxes of the fainter sources can be better measured with deeper exposures. 

\citetalias{Zhao2021} found an under-estimate of the measured fluxes for bright ($F_{3-24}>10^{-12}$\,erg\,cm$^{-2}$\,s$^{-1}$) sources. This was due to a computing error when generating false probability maps where the false probability in the center pixel of the bright sources was saturated. The bug led to an incorrect measurement of the source position and therefore underestimating its flux. This error did not affect the flux measurement of the real observations in the NEP-TDF because all detected sources are much fainter than 10$^{-12}$\,erg\,cm$^{-2}$\,s$^{-1}$. The error is fixed in this work.

\section{\NuSTAR\ Source Catalog} \label{sec:NuSTAR_catalog}
To maximize the signal-to-noise ratio (SNR), we performed source detections on the FPMA+B mosaics of the actual \NuSTAR\ observations in the cycle 6 and cycles 5+6 surveys using the same detection strategy as in the simulations (Section~\ref{sec:detection}). Source detection, requiring DET\_ML above the 95\% reliability level, was performed separately in each of the five energy bands. The resulting coordinates, source counts, background counts, DET\_MLs, and vignetting-corrected exposure times of the detected sources were then merged into a master catalog by using a 30\arcsec\ matching radius among the five energy bands. The master catalog therefore includes all sources detected in at least one energy band above the 95\% reliability level. The coordinates of the sources reported in the master catalog are taken from the detections that have the highest DET\_ML among the five energy bands. The positions of the sources detected in cycle 6 and cycles 5+6 surveys are plotted in Figure~\ref{fig:mosaic}.

The master catalog includes 35 and 60 sources for cycle 6 and cycles 5+6 surveys, respectively. The number of the sources detected above 95\% reliability level in each energy band and the merged master catalog are listed in Table~\ref{Table:number_cycle6}. Statistically, we expect about two to three spurious detections in the 95\% reliability master catalogs. Table~\ref{Table:FSH} reports the number of sources detected in each combination of energy bands.

\begin{figure*} 
\begin{minipage}[b]{.33\textwidth}
\centering
\includegraphics[width=\textwidth]{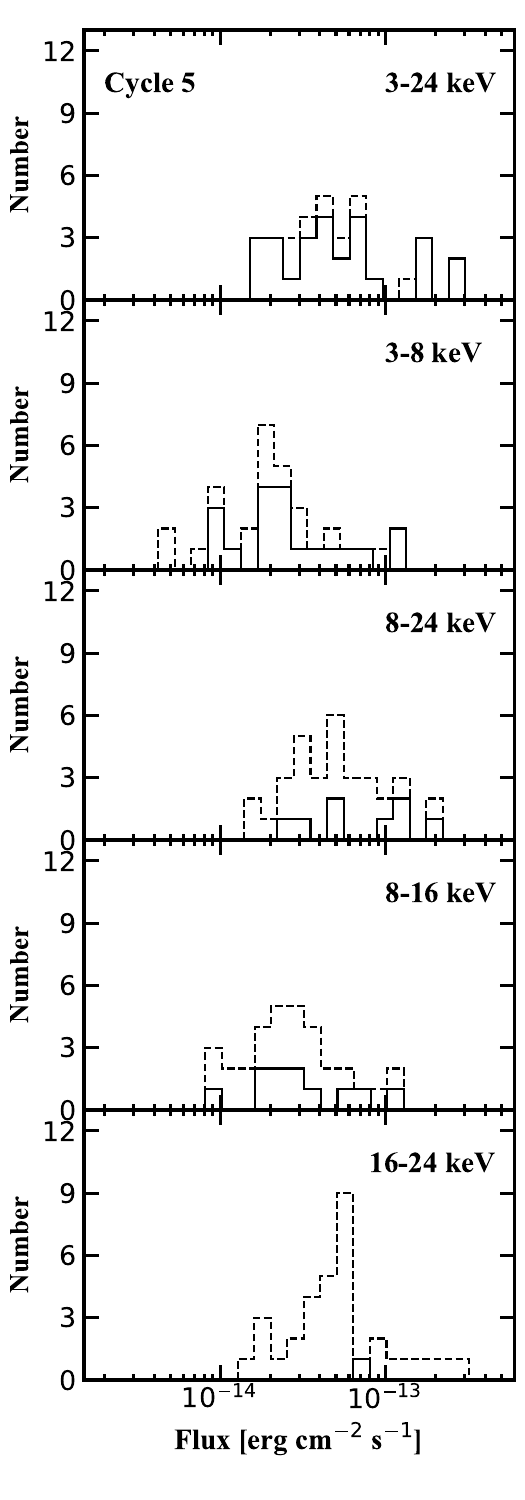}
\end{minipage}
\begin{minipage}[b]{.33\textwidth}
\centering
\includegraphics[width=\textwidth]{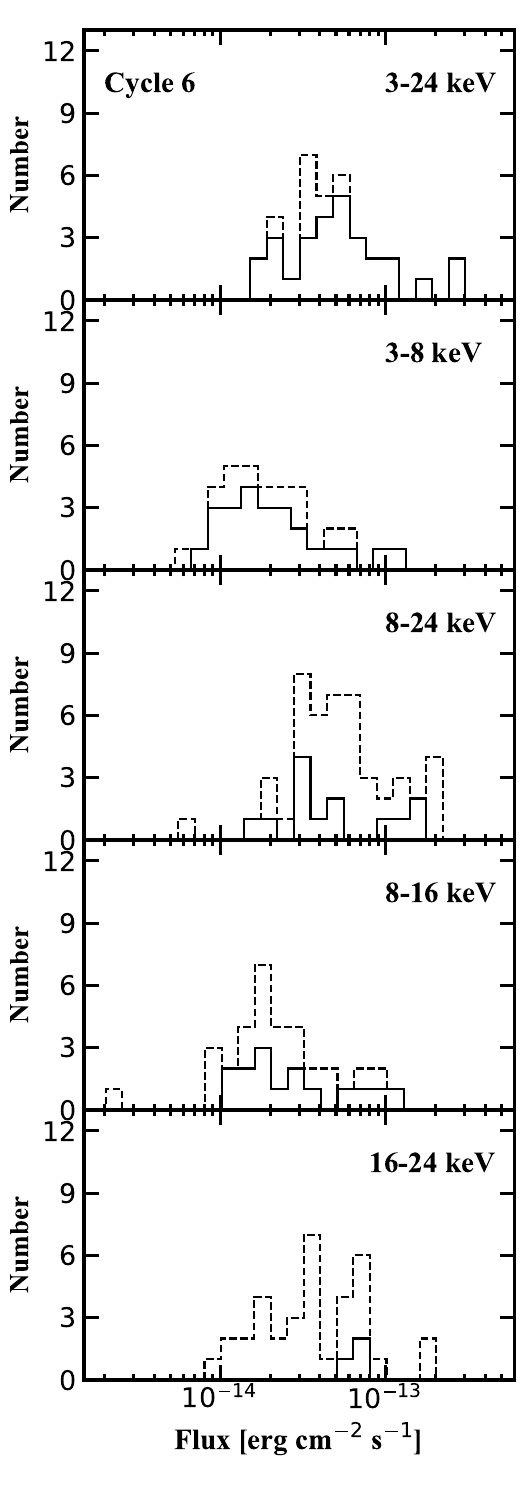}
\end{minipage}
\begin{minipage}[b]{.33\textwidth}
\centering
\includegraphics[width=\textwidth]{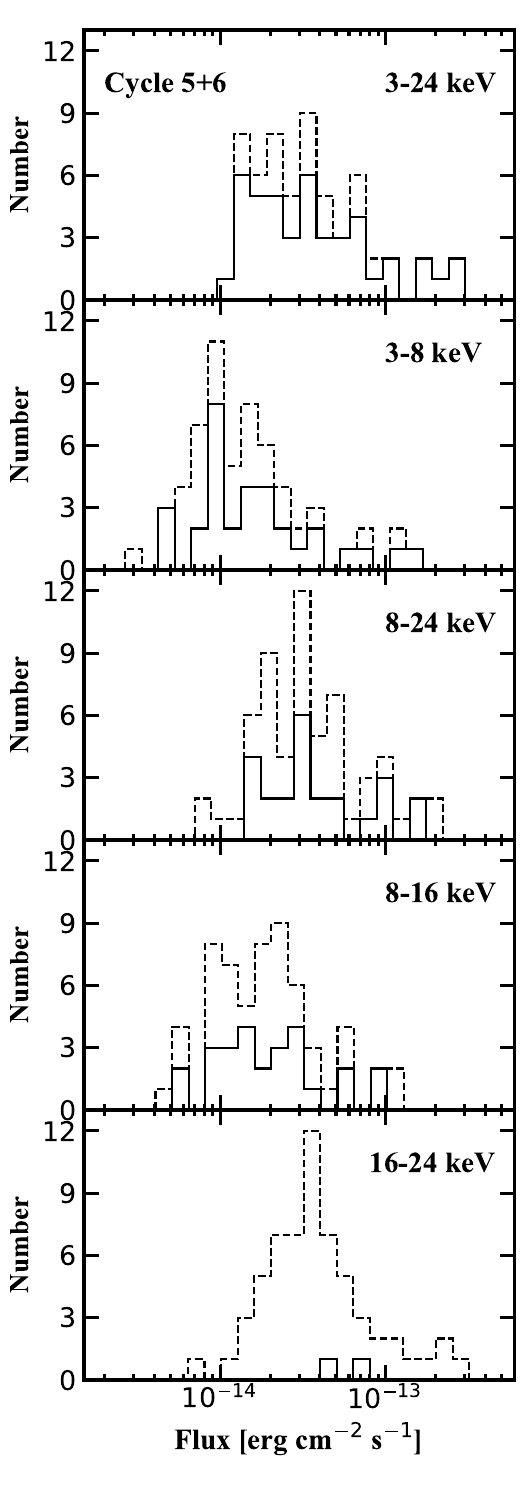}
\end{minipage}
\caption{Flux distributions of \NuSTAR\ sources. Panels from left to right show results for the cycle 5 (\citetalias{Zhao2021}), cycle 6, and cycles 5+6 surveys, respectively. Panels from top to bottom show different energy bands as labeled in each panel. Solid lines represent sources detected above the 95\% reliability level in the specific survey and band, and dashed lines represent all 60 sources detected by \NuSTAR\null. For sources not detected in a particular survey and energy range, the flux plotted is the 90\%-confidence upper limit.}
\label{fig:flux_all}
\end{figure*}

Figure~\ref{fig:flux_all} shows the distribution of the source fluxes in different energy bands from the various master catalogs. The source fluxes were calculated with the method described in Section~\ref{sec:Fluxes}. For sources not detected in an energy band, background counts were not deblended. We calculated 1$\sigma$ net count-rate and flux uncertainties for the sources that were detected above the 95\% reliability level in a given energy band using Equations (9) and (12) of \citet{Gehrels1986} with $S = 1$. For sources not detected above the 95\% reliability level, we calculated the 90\% confidence level upper limits of net count-rates and fluxes using Equation (9) of \citet{Gehrels1986} with $S = 1.645$. 

The master catalogs of the \NuSTAR\ detected sources in cycle 6 and cycles 5+6 surveys are made public with this paper. Table~\ref{Table:catalog} explains each column in the catalogs. 

Variability detection is the prime goal of the \NuSTAR\ NEP-TDF survey. We developed a dedicated pipeline, briefly introduced in Appendix~\ref{sec:variability}, using a Bayesian method to analyze the source variability in \NuSTAR\ observations. As a preliminary result, four sources showed variability in cycles 5+6 at $p<$0.05 (${\sim}2\sigma$) in at least one energy band in the 26-months of observations. Systematic discussion of source variability in the \NuSTAR\ and \XMM\ NEP-TDF will be presented in future work.

\begingroup
\renewcommand*{\arraystretch}{1.}
\begin{table}
\centering
\caption{Energy bands of detected sources}
\label{Table:FSH}
  \begin{tabular}{lclclcl}
       \hline
       \hline
       Energy&&Cycle 5&&Cycle 6&&Cycles 5+6\\
       \hline
        F+S+H&\quad\quad&9 (27\%)&\quad\quad&13 (37\%)&\quad\quad&17 (28\%)\\
	F+S+h&&6 (18\%)&&4 (11\%)&&5 (8\%)\\
	F+S&&1 (3\%)&&2 (6\%)&&4 (7\%)\\
	F+s+H&&0 (0\%)&&2 (6\%)&&2 (3\%)\\
	F+s+h&&5 (15\%)&&5 (14\%)&&12 (20\%)\\
	F+s&&3 (9\%)&&0 (0\%)&&0 (0\%)\\
	F+H&&0 (0\%)&&2 (6\%)&&2 (3\%)\\
	F+h&&2 (6\%)&&0 (0\%)&&1 (2\%)\\
	F&&0 (0\%)&&0 (0\%)&&1 (2\%)\\
	f+S+h&&1 (3\%)&&1 (3\%)&&1 (2\%)\\
	f+S&&1 (3\%)&&2 (6\%)&&3 (5\%)\\
	f+s+H&&1 (3\%)&&0 (0\%)&&3 (5\%)\\
	f+H&&2 (6\%)&&1 (3\%)&&2 (3\%)\\
	S+h&&0 (0\%)&&1 (3\%)&&1 (2\%)\\
	S&&1 (3\%)&&1 (3\%)&&1 (2\%)\\
	H&&1 (3\%)&&1 (3\%)&&5 (8\%)\\
	\hline
	Total&&33&&35&&60\\
	\hline
\end{tabular}
\raggedright
\tablecomments{Numbers are for the master catalogs for cycle 5 (\citetalias{Zhao2021}), cycle 6, and cycles 5+6 surveys. F(f), S(s), and H(h) represent the full (3--24\,keV), soft (3--8\,keV), hard (8--24\,keV and/or 8--16\,keV and/or 16--24\,keV) energy bands. F, S, and H represent sources detected above the 95\% reliability threshold in the given energy band, while f, s, and h refer to the sources detected below the 95\% reliability threshold.}
\end{table}
\endgroup

\section{XMM-Newton NEP-TDF Survey} \label{sec:XMM_catalog}
\begin{figure} 
\centering
\includegraphics[width=\linewidth]{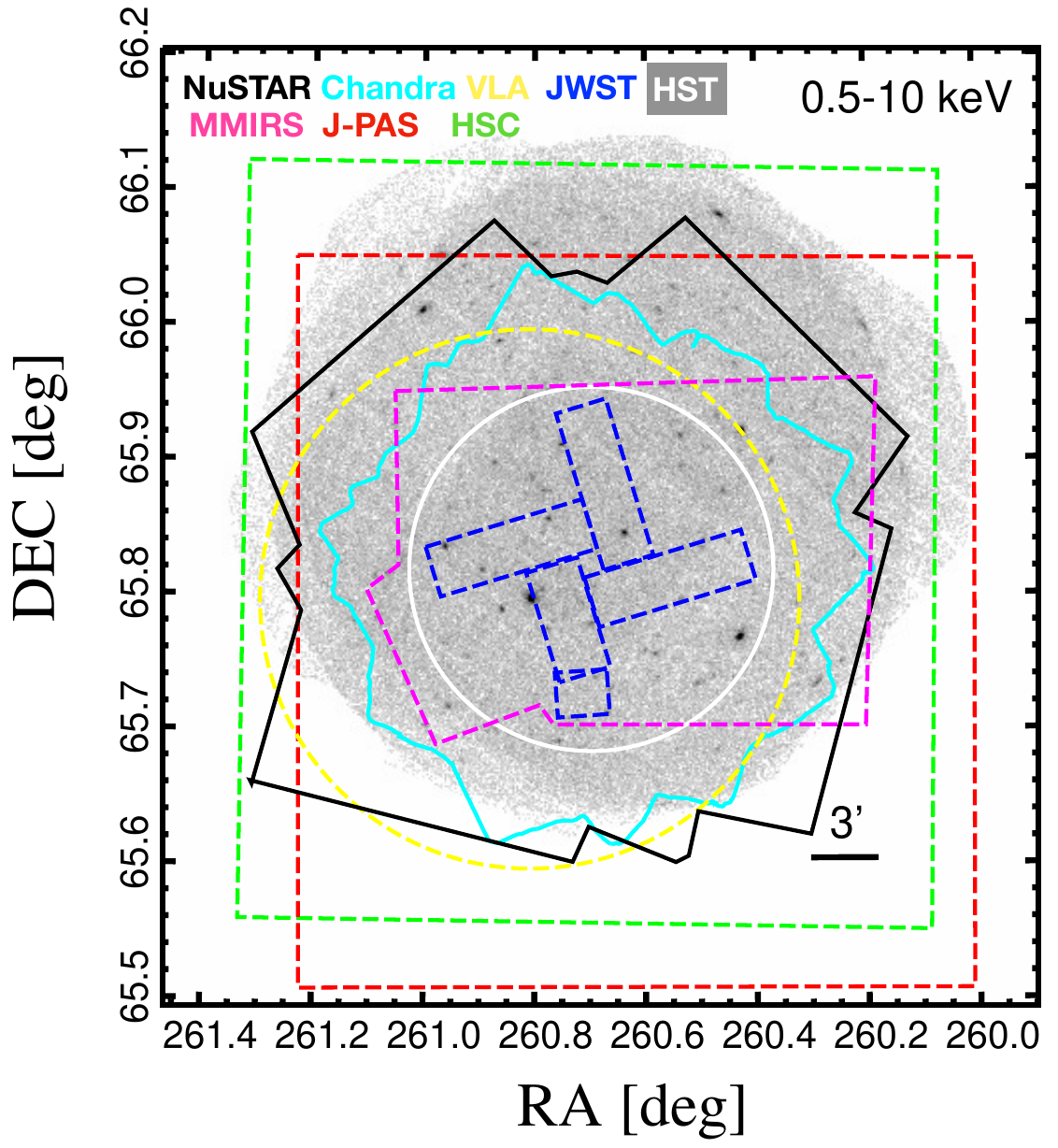}
\caption{\XMM\ MOS+pn mosaics combining 0.5--10 keV observations from all three epochs. The footprints of other surveys in the NEP-TDF field are plotted in different colors as indicated in the key. SDSS and WISE catalogs cover the entire XMM and NuSTAR regions.}
\label{fig:XMM}
\end{figure}

To provide lower-energy (0.5--10\,keV) information, 
\XMM\ observed the NEP-TDF field simultaneously with the four cycle~6 \NuSTAR\ epochs. The observations utilized all three \XMM\ cameras, i.e., MOS1, MOS2, and pn. Unfortunately, the entire 16\,ks of data in the third epoch were lost to high particle background. Otherwise, each \XMM\ epoch had exposure time $\sim$20\,ks, and the total effective exposure time is 62\,ks. Details are in Table~\ref{Table:obs}. The \XMM\ observations cover a field of 0.21~deg$^2$, about 90\% of the \NuSTAR\ NEP-TDF field. The two missing bottom corners of the field (Figure~\ref{fig:XMM}) contain one \NuSTAR\ source (ID~38).

\subsection{Data Reduction}
The \XMM\ data were reduced following \citet{Brunner08,Cappelluti09,LaMassa_2016}; and \citetalias{Zhao2021}. Details of the \XMM\ Science Analysis System (SAS) packages are described in the \XMM\ data analysis threads.\footnote{\url{https://www.cosmos.esa.int/web/xmm-newton/sas-thread-src-find-stepbystep}} The observational data files (ODFs) of MOS1, MOS2, and pn were generated using the SAS version 20.0.0 tasks {\tt emproc} and {\tt epproc}. High-background time intervals of the three instruments were excluded using $>$10\,keV count-rate thresholds of 0.2 and 0.3~counts~s$^{-1}$ for MOS and pn, respectively. We also excluded data in energy bands that might be contaminated by fluorescent emission lines, i.e., the Al~K$\alpha$ line at 1.48\,keV in both MOS and pn and two Cu lines at $\sim$7.4\,keV and 8.0\,keV only in pn. The specific energy intervals removed were 1.45-- 1.54\,keV in both MOS and pn data and 7.2--7.6\,keV and 7.8--8.2\,keV in pn. We then used the clean event files to generate the images of MOS 1,2 and pn in the 0.5--2\,keV and 2--10\,keV bands. 

To generate exposure maps, we used the SAS task {\tt eexpmap}. The maps were weighted by each instrument's energy conversion factor (ECF), which converts count rate to flux. ECFs were calculated using WebPIMMs assuming an absorbed power-law model with $\Gamma = 1.80$ and Galactic column density $N\rm _H = 3.4 \times 10^{20}$~cm$^{-2}$, as for the \NuSTAR\ observations (Section~\ref{sec:simulation}). The ECFs used for MOS and pn are 0.54 and $0.15 \times 10^{-11}$~erg\,cm$^{-2}$\,count$^{-1}$ in the 0.5--2\,keV band, and 2.22 and $0.85 \times 10^{-11}$~erg\,cm$^{-2}$\,count$^{-1}$ in the 2--10\,keV band, respectively. 

Background maps were generated for each instrument after masking detected sources. Preliminary source detection used a sliding-box method with SAS package {\tt eboxdetect} and detection likelihood {\tt LIKE} set to $>$4 to avoid any possible sources. (The detection likelihood is defined as ${\tt LIKE} \equiv -\ln p$, where $p$ is the probability of a Poissonian random fluctuation of the counts in the detection box which would have resulted in at least the observed number of source counts.) We generated the background map using SAS package {\tt esplinemap} assuming a 2-component model of the \XMM\ background by setting {\tt fitmethod} = model. This 2-component model considers background from both the detector (particles) and the CXB.

\subsection{Source Detection}
To maximize sensitivity, we co-added the cleaned images, exposure maps, and background maps of the three instruments into mosaic images for the two energy bands using the SAS {\tt emosaic} task. Figure~\ref{fig:XMM} shows the merged mosaic. Source detection was performed using the SAS {\tt eboxdetect} and {\tt emldetect} tasks, the latter to optimize detection of the center of the source. 
Source detection was performed in the 0.5--2\,keV and 2--10\,keV bands simultaneously to minimize uncertainties in source positions and fluxes. A detection required {\tt mlmin}$>$6 in either of the two bands. This threshold corresponds to a reliability of 97.3\% in the 0.5--2\,keV band and 99.5\% in the 2--10\,keV band based on simulations of the XMM-COSMOS survey \citep{Cappelluti_2007}, which has a $\sim$60\,ks depth similar to the \XMM\ NEP-DTF survey. Source detection excluded the margin of the FoV where the exposure time is $<$1\,ks. 

\subsection{Astrometric Correction and Uncertainty}
Before merging the three epochs of observations into mosaics to maximize the sensitivity of the survey, we estimated the astrometric offsets of the three observations. 
The astrometric offset of an \XMM\ observation is typically less than 3\arcsec\ and on average is 1\farcs0--1\farcs5 \citep[e.g.,][]{Cappelluti_2007,Ni_2021}. To determine the astrometric offset of our three epochs, we matched $>$6$\sigma$ ({\tt mlmin}$>$20) \XMM\ sources to optical sources from the Sloan Digital Sky Survey (SDSS) DR16.\footnote{\url{https://www.sdss4.org/dr16/}} Only {\tt Type} = Star sources were used, and the matching radius was 4\farcs5. \XMM\ epochs 1, 2, and 4 had 13, 22, and 15 matched SDSS counterparts, respectively. The median offsets in R.A.\ and Decl.\ are $(\Delta\alpha, \Delta\delta) = (3\farcs97, 0\farcs95$) for epoch 1, (0\farcs88, 1\farcs13) for epoch 2, and (0\farcs06, 1\farcs46) for epoch 4. We applied these offsets to the event and attitude files in each observation and remade the images, background maps, and exposure maps with the corrected files. 

To test the astrometric corrections, we performed the source detection again and measured the offsets of the same sources from their optical counterparts. The average \XMM\ to SDSS separations were reduced by 30\% for epoch 1, 45\% for epoch 2, and 41\% for epoch 4 after the astrometric correction. Furthermore, the median \XMM\ offsets among different epochs decreased by 84\%--96\%, suggesting that the three epochs became better aligned. The resulting images, background maps, and exposure maps of the three epochs were then merged into mosaics. The new images have more high-count pixels than the old images, again suggesting that the alignment is better corrected. The new average X-ray-to-optical offset is 1\farcs22, and we take this to be the systematic position uncertainty of the \XMM\ NEP-TDF survey.

\subsection{Sensitivity}
\begin{figure} 
\centering
\includegraphics[width=.49\textwidth]{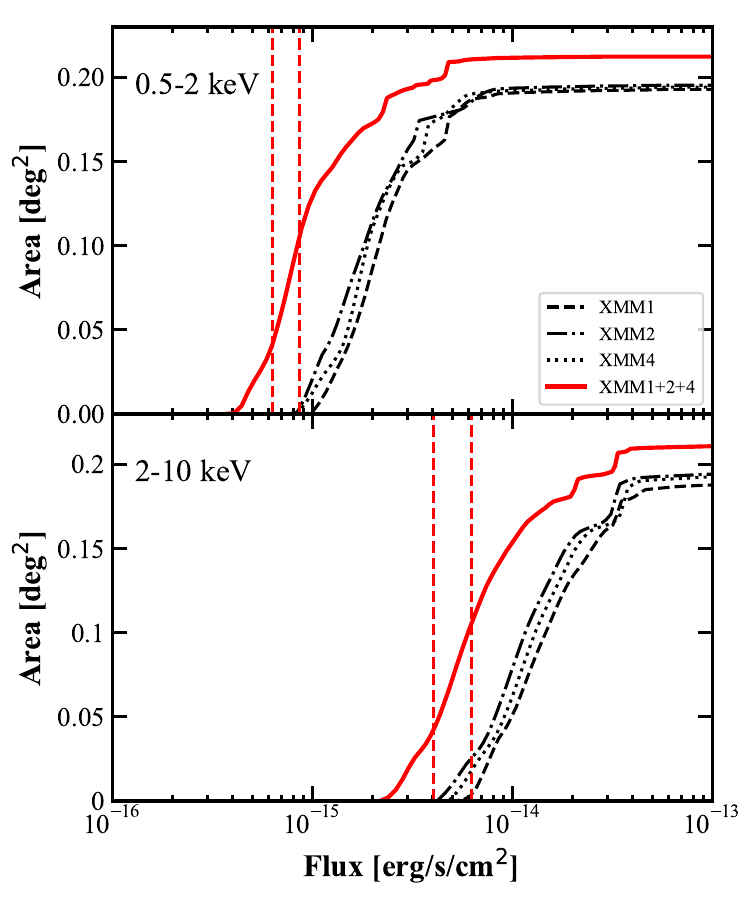}
\caption{\XMM\ NEP-TDF sensitivity maps. The upper panel shows 0.5--2\,keV, and the lower shows 2--10\,keV. Different line types show different epochs as indicated in the legend. Vertical dashed lines show the half-area and 20\%-area sensitivities for the three epochs combined.}
\label{fig:XMM_sen}
\end{figure}

The sensitivity curves of the \XMM\ NEP-TDF survey are plotted in Figure~\ref{fig:XMM_sen}. The sensitivity maps were generated using the SAS {\tt esensmap} package assuming a maximum likelihood ${\tt mlmin}>6$. The half-area and 20\%-area sensitivities are reported in Table~\ref{Table:sensitivity}.

\subsection{\XMM\ Source Catalog}
\begin{figure} 
\centering
\includegraphics[width=.49\textwidth]{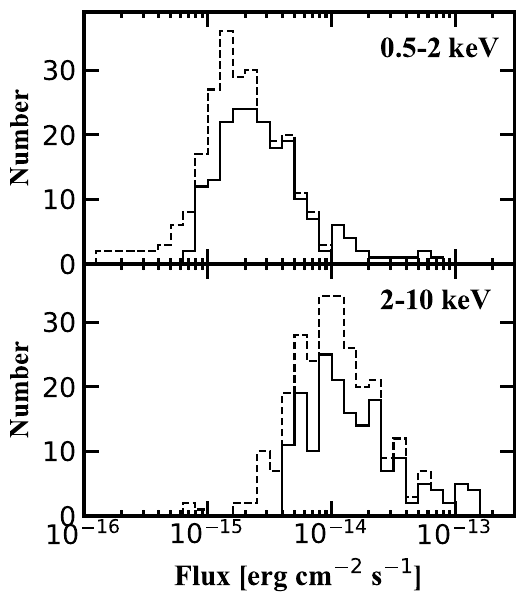}
\caption{Flux distribution of sources detected in the 0.5--2\,keV (top) and 2--10\,keV (bottom) bands of the three-epoch combined \XMM\ mosaic. Solid lines represent the flux distribution of detected sources with ${\tt mlmin}>6$ in a given band. Dashed lines represent the flux distributions of all 286 \XMM\ sources with those not detected or with ${\tt mlmin}\le6$ in the given band plotted at their 90\% confidence upper limits.}
\label{fig:XMM_flux_dis}
\end{figure}

The final \XMM\ NEP-TDF catalog includes 194 sources in the 0.5--2\,keV band and 172 sources in the 2--10\,keV band at ${\tt mlmin}>6$. There were only 80 sources in common in the two bands giving a total of 286 individual sources detected in at least one band. The source properties are listed in the \XMM\ source catalog, and descriptions of each column of the catalog are in Table~\ref{Table:XMM_catalog}. The source flux distributions are plotted in Figure~\ref{fig:XMM_flux_dis}.

\subsection{Cross-match with NuSTAR} \label{sec:match_NuSTAR}
We cross-matched the \XMM\ sources with the 60 sources detected in the \NuSTAR\ cycles 5+6 survey using a simple position match. The match radius was the 20\arcsec\ \NuSTAR\ position uncertainty combined in quadrature with the position uncertainty of individual \XMM\ sources ($\sigma_{\rm XMM}$ from the {\tt emldetect} best-fit results) and the 1\farcs22 \XMM\ systematic uncertainty. The 20\arcsec\ \NuSTAR\ uncertainty is three times the best-fit Rayleigh scale parameter ($\rm \sigma_{95\%, C56} = 6\farcs5$) of the simulated position errors (Section~\ref{sec:position}). In all, 36 \NuSTAR\ sources match at least one \XMM\ counterpart. Thirty of these have a single \XMM\ counterpart, and six (ID 11/13/23/31/41/46) have two \XMM\ counterparts within the search radius. 
In all six cases, one of the \XMM\ sources is both brighter and closer to the \NuSTAR\ position than the other, and we took this to be the primary counterpart. 
One \NuSTAR\ source (ID~51) has two \XMM\ sources (ID 134/181) just outside the search radius (22\farcs8 and 23\farcs3, respectively). The two are in opposite directions and were both detected in the 2--10\,keV band with similar flux $F({\rm \hbox{2--10\,keV}})\sim 9 \times 10^{-15}$\,erg\,cm$^{-2}$\,s$^{-1}$. This looks like a case of source confusion where both \XMM\ sources contribute to the \NuSTAR\ detection.

\begin{figure} 
\centering
\includegraphics[width=\linewidth]{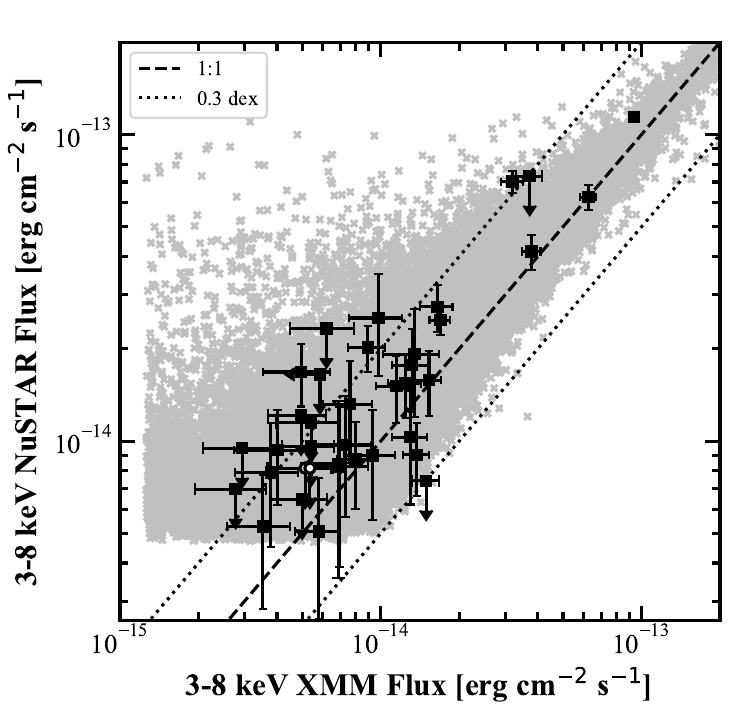}
\caption{Comparison between \NuSTAR\ and \XMM\ fluxes for the 37 \NuSTAR\ sources with \XMM\ counterparts (black filled squares). The dashed line represents the 1:1 relation, and the dotted lines show a factor of two difference. The black open circles show the two \XMM\ candidate counterparts of \NuSTAR\ ID 51. For undetected sources, the indicated upper limits are 90\% confidence. The gray crosses in the background are from the \NuSTAR\ simulations in the 3--8\,keV band.}
\label{fig:flux_compare}
\end{figure}   

In all, 37 out of 60 \NuSTAR\ sources have at least one \XMM\ association. Of the remaining 23, 17 (ID 1/4/5/9/14/16/17/18/40/42/44/47/48/49/50/56/59) were undetected or below the 95\% reliability level in the \NuSTAR\ soft 3--8\,keV band, suggesting that they might be heavily obscured and therefore detectable only in hard X-rays. The other six sources (ID 25/27/33/35/38/52) were above the 95\% reliability level in the 3--8\,keV band but do not have \XMM\ counterparts. They might be variable sources that were bright only in \NuSTAR\ cycle~5, or some of them could be spurious \NuSTAR\ sources. Statistically, only 2--3 spurious detections are expected in the \NuSTAR\ 95\% reliability catalog, and therefore variability is likely to be a factor.


For comparison between \NuSTAR\ and \XMM, the \XMM\ 2--10\,keV fluxes were converted to 3--8\,keV assuming an absorbed power-law intrinsic SED with photon index $\Gamma = 1.80$ and Galactic absorption $N\rm _H = 3.4 \times 10^{20}$\,cm$^{-2}$. This gives a conversion factor of 0.62, and Figure~\ref{fig:flux_compare} shows the comparison. 
Most sources have comparable fluxes measured by the two observatories. The tendency for \NuSTAR\ fluxes to be higher than \XMM\ fluxes at the faint end is due to the Eddington Bias as demonstrated by Figure~\ref{fig:flux}. Other offsets could arise from variability in the last three years or different spectral shapes of the sources than assumed for converting the \XMM\ fluxes to the \NuSTAR\ energy band.

\section{Multiwavelength Counterparts}\label{sec:multi}
The {JWST} NEP-TDF has extensive multiwavelength coverage \citep{Windhorst_2023}. Because \XMM\ has a better PSF than \NuSTAR, we first matched \XMM\ sources with the visible-wavelength and infrared (IR) catalogs.

\subsection{Visible-wavelength Catalogs}
We matched X-ray positions to three visible-wavelength catalogs covering the NEP-TDF: SDSS DR17 \citep{Abdurrouf2022}, the HEROES catalog \citep{Taylor2023} made from {Subaru} Hyper Suprime-Cam \citep[HSC;][]{Aihara2018} images, and the NEP portion \citep[J-NEP;][]{Caballero2023} of the Javalambre-Physics of the Accelerating Universe Astrophysical Survey \citep[J-PAS;][]{Benitez2014}. We used the $i$-band catalogs as they include the largest number of detected sources.

The SDSS catalog covers the entire field, and data were downloaded from the public database.\footnote{\url{https://www.sdss4.org/dr17/}} The HSC images were reduced by S.\ Kikuta, and the HSC NEP-TDF catalog was generated by C.~N.~A. Willmer \citep{Willmer2023}. Sources with $m_i<17.5$ are saturated in the HSC observations, and we replaced their magnitudes with the magnitudes in the SDSS catalog. The J-NEP was performed with the single-CCD Pathfinder camera on the 2.55 m Javalambre Survey Telescope (JST) at the Javalambre Astrophysical Observatory with 56 narrow filters and used the SDSS $u$, $g$, $r$, and $i$ filters. J-NEP covered about 80\% area of the \XMM\ field. We applied magnitude cuts at $ S/N >3$ (corresponding to $i$-band magnitudes $m_i\lesssim22.5$, 25.8, and 24.5 for SDSS, HSC, and J-PAS, respectively) to the three catalogs to ensure reliable detections and accurate measurements of the source fluxes.

\subsection{Infrared Catalogs}
We used two near-IR (NIR) catalogs covering the NEP-TDF: the \textit{YJHK} catalog \citep{Willmer2023} made with the MMT--Magellan Infrared Imager and Spectrometer \citep[MMIRS;][]{McLeod2012} on the {MMT}, and the unWISE catalog \citep{Schlafly2019} made from five years of the {Wide-field Infrared Survey Explorer} \citep[{WISE};][]{wright10} observations. The unWISE wavelengths are 3.4 (W1) and 4.6~\micron\ (W2).

The MMIRS catalog covers 30--40\% of the \XMM-observed field, and 
the S/N $>$3 sensitivity cuts correspond to $m\lesssim24.6$, 24.5, 24.1, and 23.5 (in AB magnitudes) in the $Y$, $J$, $H$, and $K$ bands, respectively. The unWISE catalog covers the entire \XMM\ field, and the $S/N >3$ sensitivity cuts correspond to $m\lesssim21.5$ in the W1 band and 20.5~AB in the W2 band.

\begin{figure*} 
\includegraphics[width=\textwidth]{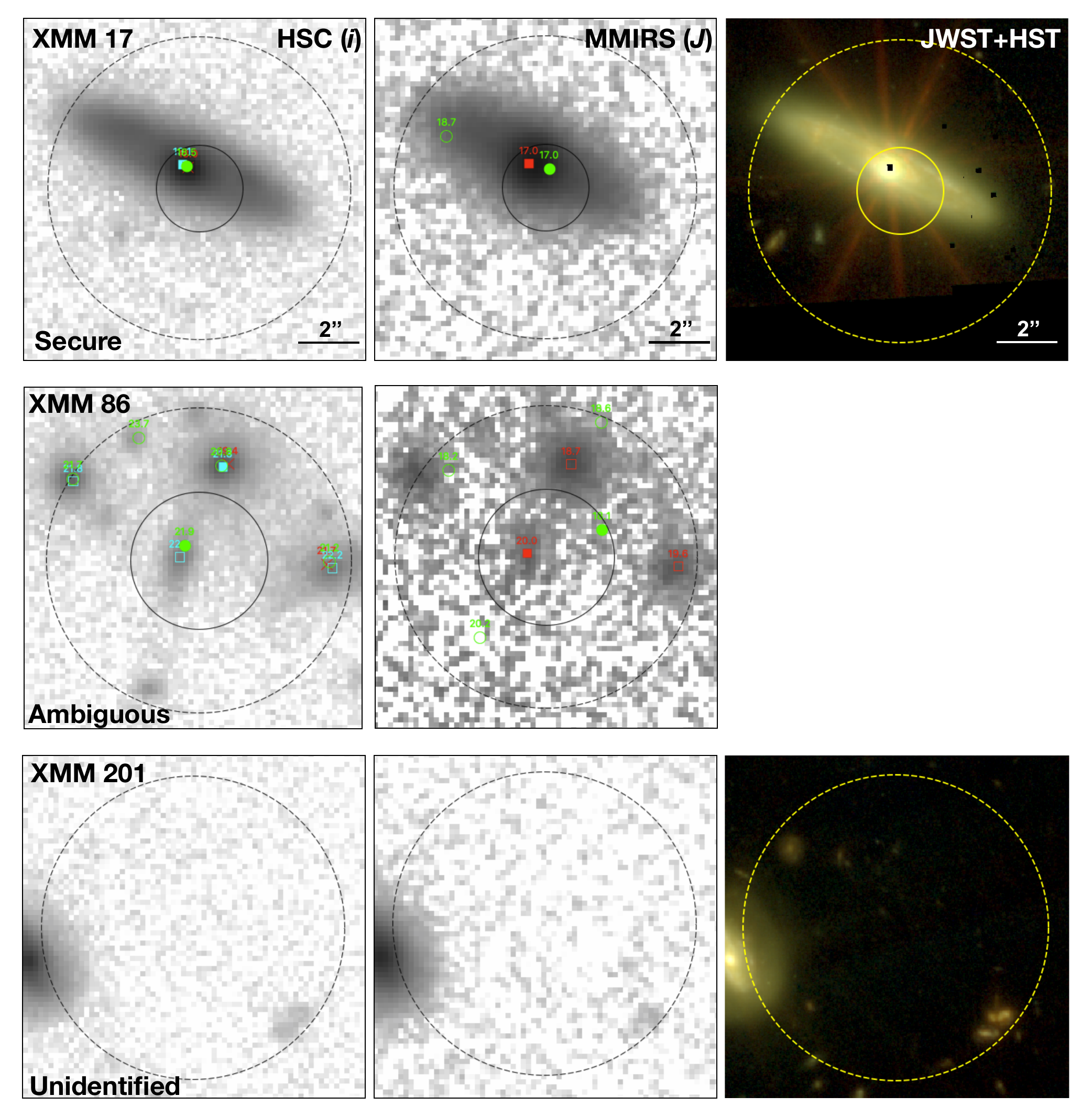}
\caption{Illustrations of secure (top), ambiguous (middle), and unidentified (bottom) associations of \XMM\ sources. Negative images left to right are HSC $i$, MMIRS $J$, and HST$+$JWST (11-filter mosaic: HST F275W, F435W, and F606W; JWST F090W, F115W, F150W, F200W, F277W, F356W, F410M, and F444W)\null. All images are oriented north up, east left and centered at the centroid of the \XMM\ detections. The image scale is indicated in the top row. Solid circles show the \XMM\ position uncertainty, and dashed circles show the 5\arcsec\ matching radius. Labels show $i$-band magnitudes measured from HSC (cyan square), J-PAS (green circle), or SDSS (red cross) or IR magnitudes measured by MMIRS ($J$-band, red square) or {WISE} (W1, green circle). Only ancillary counterparts within the matching radius and with $\rm LR>LR_{\rm th}$ are plotted. The ancillary counterparts with the highest $\rm LR$ are plotted as filled symbols. XMM~17 is the bright Seyfert galaxy discussed by \citet[][their Figure~3]{Willner2023}. XMM~86 was not covered by JWST.}
\label{Fig:visual_check}
\end{figure*}  

\subsection{Multiwavelength Matching} \label{sec:multiwavelength_match}
We used a 5\arcsec\ matching radius to identify candidate counterparts of the \XMM\ sources. (More than 95\% of \XMM\ sources are detected within this radius based on the simulations made in the STRIPE 82 \XMM\ survey: \citealt{LaMassa_2016}.) To choose among possible counterparts within the match radius, we used a maximum likelihood estimator \citep[MLE;][]{Sutherland1992} as applied for X-ray sources detected in previous \XMM\ and \cha\ extragalactic surveys \citep[e.g.,][]{Brusa_2007,Civano2012,Marchesi_2016,LaMassa_2016}. These previous results showed $>$80\% reliability. The MLE method considers both the flux and the offset of the candidate counterparts in the context of the position uncertainties of the surveys and the flux distribution of survey sources. The likelihood ratio ($\rm LR$) that a candidate is the real counterpart is:
\begin{equation}
{\rm LR} = \frac{q(m)f(r)}{n(m)}\quad,
\end{equation}
where $m$ is the catalog magnitude of the candidate, $n(m)$ is the local magnitude distribution of background sources, $q(m)$ is the expected magnitude distribution of the real multiwavelength counterparts, and $r$ is the position offset between the X-ray source and the candidate. In practice, $n(m)$ was measured in an annulus between 5\arcsec\ and 30\arcsec\ from the X-ray source. The function $q(m)$ is the normalization of $q'(m)$, where $q'(m)$ is the magnitude distribution of catalog objects within 5\arcsec\ of the X-ray source after subtracting $n(m)$ and rescaling to the 5\arcsec\ circular area. \citet[][their Figure~1]{Civano2012} gave an example of $q(m)$. The function $f(r)$ is the probability distribution of the positional uncertainties, assumed to be a two-dimensional Gaussian $f(r) = 1/(2\pi\sigma^2)\times \exp(-r^2/2\sigma^2)$, where $\sigma$ is the quadrature combination of the position uncertainty of the \XMM\ source (Section~\ref{sec:match_NuSTAR}) and the ancillary object (0\farcs2). The choice of 0\farcs2 was validated by cross-matching our ancillary table with the extragalactic sources (proper motion {\tt pm}$<$10~mas~yr$^{-1}$) in the GAIA DR3 catalog \citep{Prusti+16,Brown+21}. Table~\ref{t:matches} lists the number of sources matched by each survey.

\begin{table}
\centering
\caption{\XMM\ Match Statistics}
\label{t:matches}
\begin{tabular}{lccccc}
\hline\hline
Survey & band & XMM  & matched & candidates&CP\\
\hline
HSC   & $i$ & 285 & 251 & 514&197\\
J-PAS & $i$ & 261 & 141 & 178&131\\
SDSS  & $i$ & 286 &  93 &  97&93\\ 
MMIRS & $J$ & 132 & 125 & 222&117\\
WISE  & W1  & 286 & 210 & 272&210\\
\hline
\end{tabular}
\raggedright
\tablecomments{The four numbers in each row are respectively the number of \XMM\ sources in the survey's footprint, the number of those sources with at least one candidate within 5\arcsec, the total number of candidates within that area for all sources, and the number of XMM sources with at least one ancillary counterpart (CP) above the chosen $\rm LR_{th}$.}
\end{table}

We used the LR threshold ($\rm LR_{th}$) to distinguish whether an ancillary object is the true counterpart of the \XMM\ detection or is a background source within the search radius. The $\rm LR_{th}$ was determined by balancing the reliability and the completeness of the final selected sample. 
The reliability and completeness of the matching can be estimated from the survey statistics
\citep{Civano2012}. The reliability $R_i$ for an individual candidate $j$ is
\begin{equation}
R_i = \frac{{\rm LR}_i}{\sum_i({\rm LR})_i+(1-Q)}\quad,
\end{equation}
where $Q$ is the fraction of \XMM\ sources having at least one candidate counterpart (i.e., the ratio of Table~\ref{t:matches} column 4 to column 3). The LR was summed over all potential counterparts within the search radius of a given \XMM\ source
for all ancillary objects within the search radius. The reliability ($R$) of the entire sample is defined as the ratio between the sum of the reliabilities of all the candidate counterparts and the total number of sources with $\rm LR>LR_{th}$. The completeness ($C$) of the sample is defined as the ratio between the sum of the reliability of all the candidate counterparts and the number of the X-ray sources that have ancillary objects within the search radius.

A higher $\rm LR_{th}$ suggests a higher reliability of the matching but lower sample completeness, while a lower $\rm LR_{th}$ suggests a lower reliability of the matching but higher sample completeness. We selected $\rm LR_{th}$ following \citet{Brusa_2007} by maximizing ($R+C$)/2. We applied this criterion to the five ancillary catalogs (HSC, J-PAS, SDSS, MMIRS, WISE) and the resulting $\rm LR_{th}$ are 0.3, 0.2, 0.2, 0.3, and 0.1, respectively. The corresponding numbers of the \XMM\ sources that have at least one counterpart above the chosen $\rm LR_{th}$ are given in Table~\ref{t:matches}. We used the HSC catalog as the primary reference for visible-wavelength counterparts because it is the deepest and covers the most area. Similarly, we used the MMIRS catalog as the primary for the infrared counterparts. Other catalogs were checked if no counterpart was found in the primary catalog.

The identified counterparts of \XMM\ sources can be separated into three classes:
\begin{itemize}
\item {\textbf{Secure}}: these are sources with a single counterpart with $\rm LR>LR_{th}$, or it has more than one candidate counterpart but the $\rm LR$ of the primary counterpart is four times higher than the $\rm LR$ of the secondary counterpart. (For there to be more than one candidate, both primary and secondary counterparts must have $\rm LR>LR_{th}$).

\item {\textbf{Ambiguous}}: these are sources with multiple candidate counterparts with the $\rm LR$ of the primary counterpart being less than four times higher than the $\rm LR$ of the secondary counterpart or the secure optical counterpart being different from the IR counterpart. 

\item {\textbf{Unidentified}}: these are sources with no optical or IR counterpart with $\rm LR>LR_{th}$ within the search radius. 

\end{itemize}

\begin{figure} 
\includegraphics[width=.47\textwidth]{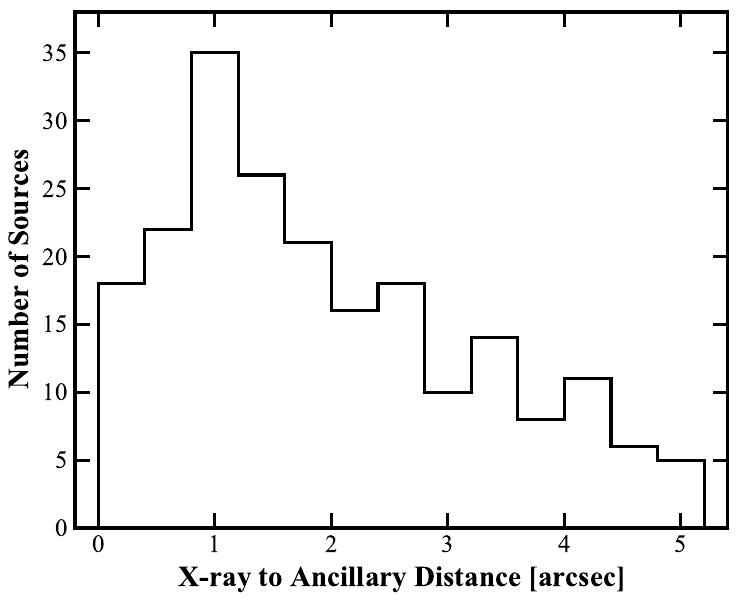}
\caption{Distribution of separations between \XMM\ sources and their secure ancillary counterparts. The median separation is 1\farcs69.}
\label{fig:XMM_distance}
\end{figure}   

The NEP-TDF also has \cha\ coverage, which provides better localization of some of the X-ray sources. Therefore, we utilized the \cha\ NEP-TDF source catalog (Maksym et al., in prep.) to help identify the ancillary counterparts of the \XMM\ sources with ambiguous counterparts. In ambiguous cases, we selected the ancillary counterpart that is closer to the \cha\ measured position (which is also the primary counterpart of the \XMM\ source in most cases) as the secure ancillary counterpart of the \XMM\ source. We also visually inspected all identifications on the \XMM, HSC, and MMIRS images. Figure~\ref{Fig:visual_check} shows three \XMM\ sources with secure, ambiguous, and unidentified ancillary counterparts. 

The source XMM ID 17 (=\NuSTAR\ ID 58 = J172241+6542.6, $z$ = 0.1791) shown in Figure~\ref{Fig:visual_check} is a heavily obscured Seyfert galaxy with column density $N_{\rm H}\sim10^{23}$~cm$^{-2}$ measured by \NuSTAR\ and $N_{\rm H}\ge10^{23}$~cm$^{-2}$ by \XMM. (See Section~\ref{sec:HR} for more discussion.) This heavily obscured scenario is also supported by the JWST and HST imaging. The strong red spikes seen on the JWST images suggest that the nucleus is a dusty point source. The clear host-galaxy feature in yellow shown on the HST image also implies a significantly obscured core. 

A total of 214 \XMM\ sources have secure ancillary counterparts. 19 \XMM\ sources have ambiguous ancillary counterparts, out of which 14, 3, 1, and 1 have two, three, four, and five candidate ancillary counterparts within the search radius and with $\rm LR>LR_{th}$. The coordinates and fluxes of the optical and IR counterparts of the \XMM\ sources are listed in Tables~\ref{Table:catalog} and~\ref{Table:XMM_catalog}. The distribution of the separations between \XMM\ sources and their secure ancillary counterparts is shown in Figure~\ref{fig:XMM_distance}; the median separation is 1\farcs69.

\subsection{Radio Counterparts}
The NEP-TDF was observed in the radio ``$S$ band'' ($\nu = 3$~GHz) by the Karl G.\ Jansky Very Large Array (VLA; PIs: R.~A.~Windhorst \& W.~Cotton). The 48-hour VLA survey \citep{Hyun2023} covered an area of $\sim$0.126~deg$^2$ (24\arcmin\ in diameter), centered at the bright (in both radio and X-ray) blazar (\NuSTAR\ ID~29, $z = 1.441$). Therefore, only about 55\% of the \XMM\ area was covered by VLA\null. The \citet{Hyun2023} source list comprises 756 sources at S/N $>$ 5. The 1$\sigma$ noise is 1\,$\mu$Jy beam$^{-1}$ at the primary-beam center. As the angular resolution is FWHM 0\farcs7, we matched the VLA catalog with the ancillary counterparts of the \XMM-detected sources using 0\farcs7 as the matching radius and did not consider the unidentified sources. With this procedure, 55 out of the 171 \XMM\ sources covered by the VLA have VLA counterparts. This fraction is consistent with what was discovered in COSMOS, where $\sim$40\% of the X-ray sources have VLA counterparts \citep{Smolcic2017,Marchesi_2016}. The radio-brightest source in the \XMM\ sample is the blazar (\NuSTAR\ ID 29) with a 3~GHz flux of 0.2~Jy. The median flux of the \XMM\ matched VLA sample is 23~$\mu$Jy. Both the \NuSTAR\ and \XMM\ catalogs report the VLA ID and fluxes from \citet{Hyun2023}.

\citet{Willner2023} reported the 62 VLA sources that have JWST counterparts. Six sources were detected by \XMM\ (ID 17/29/49/65/179/197). \citet{Hyun2023} also reported the James Clerk Maxwell Telescope (JCMT) SCUBA-2 850~$\mu$m survey of the JWST NEP-TDF with 114 sources detected at S/N$>$3.5. Four \XMM\ sources (ID 1/42/70/91) have JCMT counterparts.

\subsection{HST and JWST counterparts}
HST observations of the JWST NEP-TDF (GO15278, PI: R.~Jansen and GO16252/16793, PIs: R.~Jansen \& N.~Grogin) were taken between 2020 September 25 and 2022 October 31. These observations include imaging with WFC3/UVIS in the F275W (272~nm) filter and with ACS/WFC in the F435W (433~nm) and F606W (592~nm) filters (\citealt{OBrien2024}, R.\ A.\ Jansen et al., in prep.). The 2$\sigma$ limiting depths are mag$\rm _{AB}$ $\simeq$ 28.0, 28.6, and 29.5 mag in F275W, F435W, and F606W, respectively, and the ACS/WFC observations cover a total area of $\sim$194 arcmin$^2$. 

The JWST observations of the NEP-TDF (PI: R.~A.~Windhorst \& H.~B.~Hammel, PID 2738) were taken in four epochs between 2022 August 26 and 2023 May 21 \citep{Windhorst_2023}. The survey includes eight NIRCam filters with 5$\sigma$ point-source AB limits for each epoch of observation $\simeq$ 28.6, 28.8, 28.9, 29.1, 28.8, 28.8, 28.1, and 28.3 mag in F090W, F115W, F150W, F200W, F277W, F356W, F410W, and F444W, respectively. Each NIRCam epoch of observation covers an area of 2\farcm15 $\times$ 6\farcm36 with the four epochs together (Figure~\ref{fig:XMM}) covering $\sim$55 arcmin$^2$. The survey also includes NIRISS grism data with 1$\sigma$ continuum sensitivity 25.9. Each NIRISS epoch covers an area of 2\farcm22 $\times$ 4\farcm90.

The HST and JWST NEP-TDF surveys are much deeper than the HSC and MMIRS/WISE catalogs, but they cover only the center 26\% and 7\% of the \XMM\ survey area, respectively. Therefore, we did not use the HST and JWST catalogs to identify the multiwavelength counterparts of the \XMM\ sources. Instead we used the coordinates of the counterparts of the \XMM\ sources (Section~\ref{sec:multiwavelength_match}) to match the HST (Jansen et al., in prep) and JWST catalogs \citep[][Windhorst et al., in prep]{Windhorst_2023}. We used the F606W HST catalog and F444W JWST catalog when matching, as they have the deepest sensitivities. In all, 102 \XMM\ sources have HST counterparts, and 32 \XMM\ sources have JWST counterparts. The \NuSTAR\ and \XMM\ catalogs report the F606W and F444W fluxes.

\subsection{Redshifts}
Some of the NEP-TDF X-ray sources have redshifts measured from optical spectra. Spectra came from Hectospec\footnote{\url{https://lweb.cfa.harvard.edu/mmti/hectospec.html}} \citep{Fabricant2005} and Binospec\footnote{\url{https://lweb.cfa.harvard.edu/mmti/binospec.html}} \citep{Fabricant2019}, both of which are mounted on the 6.5 m {MMT}. 

Hectospec is multi-object spectrograph with 300 optical fibers. Its 1\arcdeg-diameter field of view makes it an efficient instrument to survey the NEP-TDF X-ray sources because of their relatively low areal density. Therefore, we observed (PI: Zhao) the \XMM-selected sources with Hectospec on 2022 Sep~1 and with a different fiber configuration on 2023 May 20. Each exposure was 2 hours split into six 1200~s exposures to avoid saturation of bright targets, remove cosmic rays, and improve pipeline reduction. The 270 line mm$^{-1}$ grating provided spectral resolution $R\sim1000$--2000) over the wavelength range 3800--9200~\AA, and each exposure allows measuring redshifts of sources with $m_i\le22$~AB\null. A limitation of Hectospec is that adjacent fibers cannot be placed within 20\arcsec\ of each other. In all, we obtained 41 spectra with good enough S/N to measure the redshifts of \XMM\ targets in the 2022 run and 37 spectra in the 2023 run. Those include spectra of both possible counterparts of \XMM\ ID 187.
Besides the \XMM\ targets, we also observed 40 ($i\le21$~mag) targets selected from the VLA \citep{Hyun2023} and Chandra-detected (Maksym et al., in prep) sources. Table~\ref{Table:Hectospec_catalog} in Appendix~\ref{sec:hectospec} reports their redshifts and spectral types.

The Hectospec data were reduced using the IDL script HSRED\footnote{\url{http://www.mmto.org/hsred-reduction-pipeline/}} v2.0 (originally written by Richard Cool) developed by the Telescope Data Center at SAO \citep{Mink07}. This pipeline provides fine-tuned wavelength calibrated, improved cosmic-ray rejected, and sky-subtracted 1D spectra. The redshifts were measured using a semi-automated and interactive Java toolkit, A Spectrum Eye Recognition Assistant\footnote{\url{https://gitee.com/yuanhl1984/asera_pub/}} \citep[ASERA;][]{Yuan2013}, which was developed to classify the spectra observed by the Large Sky Area Multi-object Fiber Spectroscopic Telescope \citep[LAMOST;][]{Wang1996}. The spectroscopic redshifts were measured by cross-correlating the observed Hectospec spectra against a library of quasar, galaxy, and star template spectra\footnote{\url{http://www.sdss.org/dr5/algorithms/spectemplates/}} from SDSS integrated into ASERA.

\begin{figure} 
\centering
\includegraphics[width=.48\textwidth]{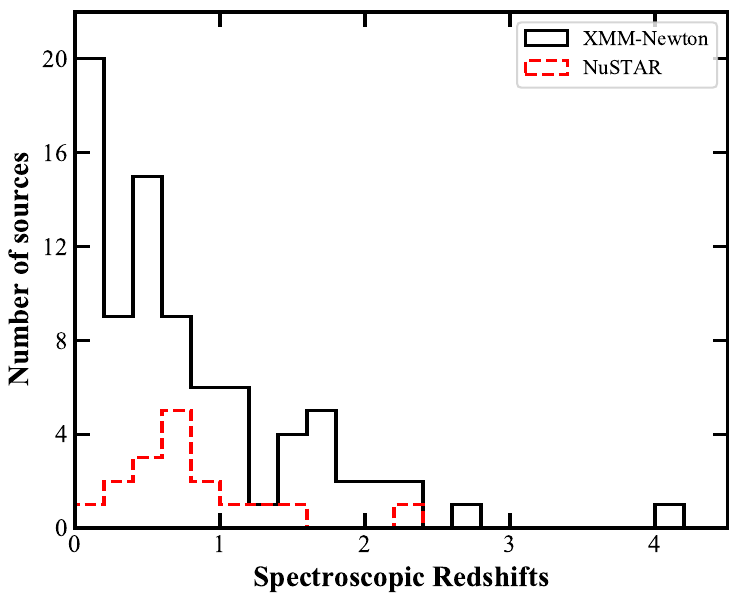}
\caption{Spectroscopic redshifts of the \XMM\ (black solid line) and \NuSTAR\ (red dashed line) sources.}
\label{fig:z_disribution}
\end{figure}

\begin{figure*} 
\begin{minipage}[b]{.49\textwidth}
\centering
\includegraphics[width=\textwidth]{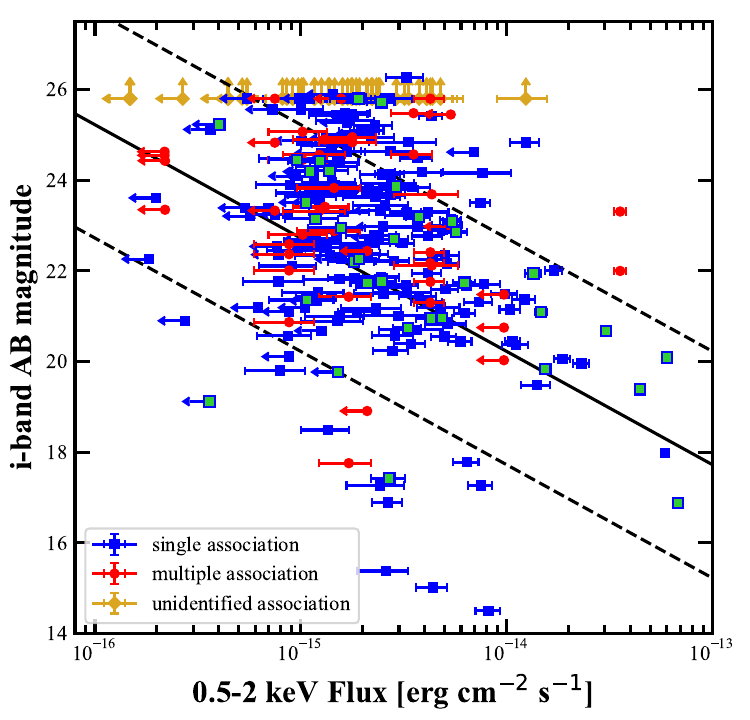}
\includegraphics[width=\textwidth]{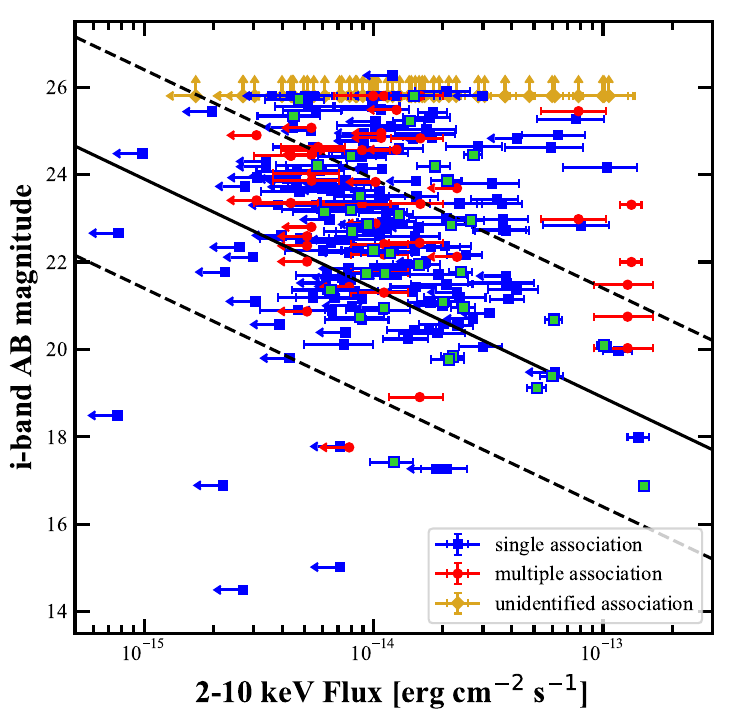}
\end{minipage}
\begin{minipage}[b]{.49\textwidth}
\centering
\includegraphics[width=\textwidth]{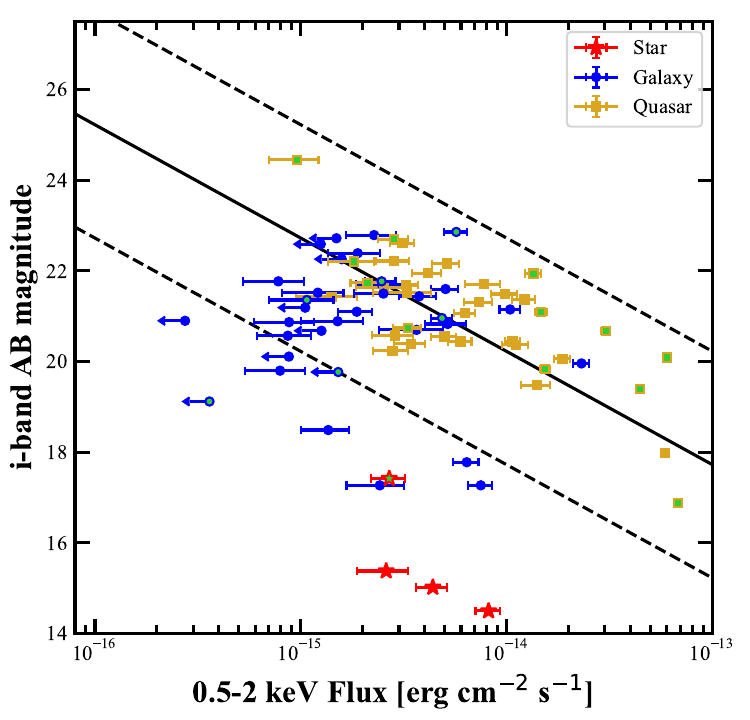}
\includegraphics[width=\textwidth]{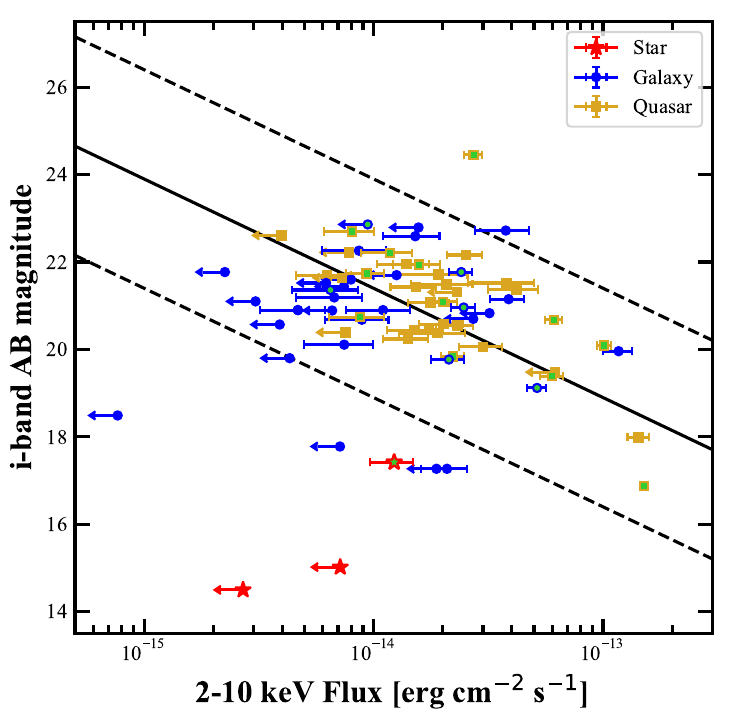}
\end{minipage}
\caption{The $i$-band magnitudes of \XMM\ counterparts as a function of X-ray fluxes.
\XMM\ sources with \NuSTAR\ counterparts are plotted as open symbols filled with green. The solid and dashed lines represent the classical AGN locus, $X/O = 0 \pm 1$ \citep{Maccacaro1988}. Upper panels show the distribution of sources in soft X-rays and lower panels those in hard X-rays. The left panels show the entire sample of \XMM\ sources. Symbols identify sources that have single, multiple, or no candidate counterparts as indicated in the legend. 
For the 18 \XMM\ sources with ambiguous optical counterparts, all candidate optical counterparts are plotted.
The 66 \XMM\ sources without optical counterparts are plotted with 3$\sigma$ upper limits $m_i> 25.8$.
The right panels show only sources that have secure counterparts with spectroscopic identifications. Symbols indicate the type of source.
}
\label{fig:f_ox}
\end{figure*}   

Binospec is an imaging spectrograph covering 3900--10,000~\AA\ \citep{Fabricant2019}. C.~N.~A. Willmer obtained more than 1,378 optical spectra with Binospec and successfully measured more than 1,000 redshifts of the sources in NEP-TDF (Willmer et al., in prep). These include five additional \XMM\ sources. Therefore, a total of 82 \XMM\ sources have spectroscopic redshifts. We categorized the sources into quasars (presenting broad emission lines), galaxies (including Type~2 AGN, which have only narrow emission lines), and stars. Future efforts to identify type 2 AGN can include methods such as the BPT diagram \citep[e.g.,][]{Kewley2001,Kauffmann2003}.

Photometric redshifts of the X-ray source counterparts were adopted from the SDSS DR17 catalog \citep{Abdurrouf2022}. Only photometric redshifts with low root-mean-square (RMS) uncertainties, specifically the {\tt photoErrorClass} flag = $-1$, 1, 2, or 3 were considered. That added two \XMM\ sources without spectroscopic redshifts for a total of 84 \XMM\ sources with redshifts, reported in Tables~\ref{Table:catalog} and~\ref{Table:XMM_catalog}.

\subsection{X-ray to Optical Properties}\label{sec:X_O}
The X-ray to optical flux ($X/O$) ratio has been historically used to identify the nature of X-ray sources \citep[e.g.,][]{Maccacaro1988}. The ratio is defined as:
\begin{equation}
X/O \equiv \log(f_{\rm X}/f_{\rm opt}) = \log(f_{\rm X}) + m_{\rm opt}/2.5 + C
\end{equation}
where $f_{\rm X}$ is the X-ray flux in a given band in units of erg~cm$^{-2}$~s$^{-1}$, $m_{\rm opt}$ is the optical AB magnitude in a given filter, and $C$ is a constant depending on the bands chosen in X-ray and optical. Figure~\ref{fig:f_ox} shows the $i$-band (HSC) magnitudes as a function of the soft (0.5--2~keV) and hard (2--10~keV) X-ray fluxes of the \XMM-detected sources. The constants used to calculate $X/O$ are $C_{0.5-2}$ = 5.91 and $C_{0.5-2}$ = 5.44 for the soft and hard bands, respectively \citep{Marchesi_2016}.

The majority of X-ray (both \XMM\ and \NuSTAR) detected sources are AGN with $-1<X/O<1$, as found in previous surveys \citep[e.g.,][]{Stocke1991, Schmidt1998, Akiyama_2000,Marchesi_2016}. Previous \cha\ and \XMM\ surveys \citep{Hornschemeier_2001, Fiore03, Civano05, Brusa_2007, Laird_2008, Xue_2011} and \NuSTAR\ surveys \citep{Civano_2015, Lansbury_2017}
detected sources having $X/O>1$. These sources were associated with high redshifts or large obscurations. Some 
sources with $X/O<-1$ are stars, as shown in Figure~\ref{fig:f_ox}. Many galaxies are in the AGN locus. Most of these sources are likely to be type 2 AGN rather than quiescent or star-forming galaxies because their 2--10\,keV luminosities (Figure~\ref{Fig:Lx_z_XMM}) are mostly $>$10$^{42}$~erg~s$^{-1}$ \citep[the conventional threshold when separating AGN from galaxies; e.g.,][]{Basu-Zych2013}. Furthermore, only about five galaxies are expected in the FoV considering the galaxy number density \citep{Ranalli2005,Luo2017,Marchesi2020}.

\subsection{Luminosity--Redshift Distribution}
\label{s:lz}
\begin{figure} 
\includegraphics[clip=true,trim=0 30 0 0,width=\linewidth]{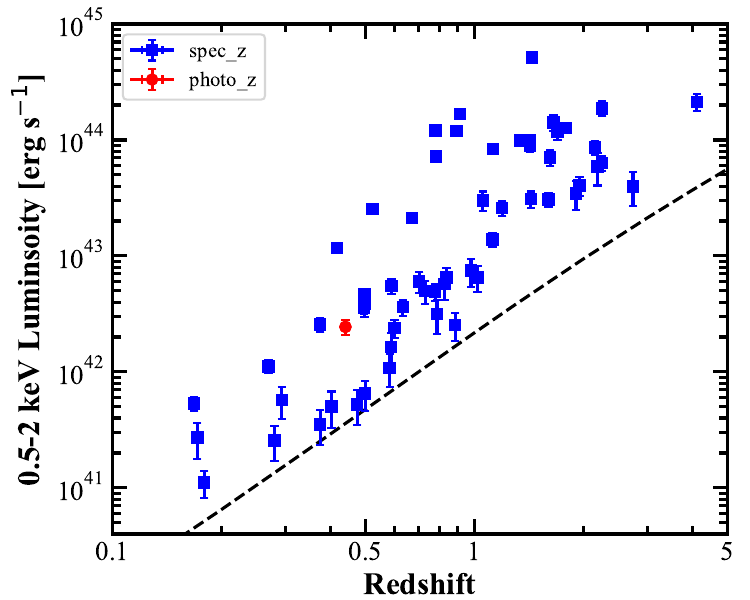}
\includegraphics[clip=true,trim=0 0 0 10,width=\linewidth]{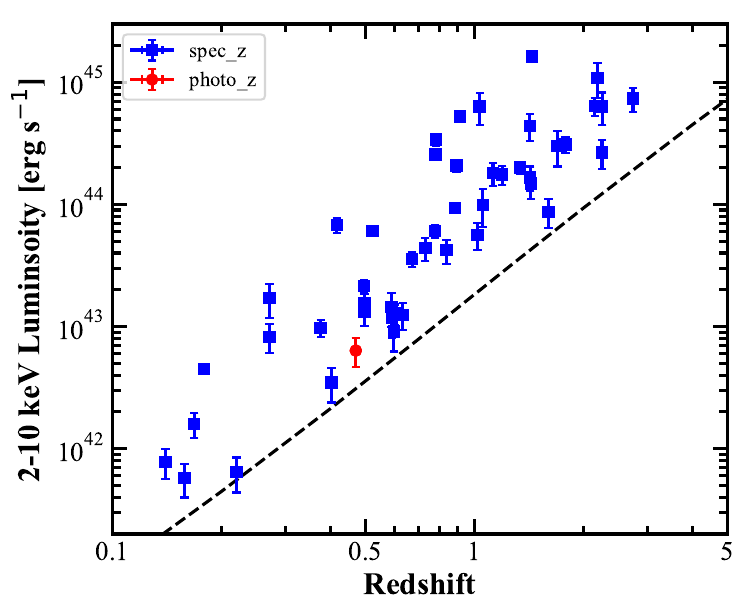}
\caption{X-ray rest-frame luminosity versus redshift for the 84 \XMM\ NEP-TDF sources with redshifts. The upper panel shows soft X-rays, and the lower panel shows hard X-rays. X-ray luminosities are as observed, not corrected for absorption.
Sources with spectroscopic redshifts are plotted using blue squares, and a source with only a photometric redshift is plotted as a red circle. The 20\%-area sensitivities are plotted as black dashed lines.}
\label{Fig:Lx_z_XMM}
\end{figure}

Figure~\ref{Fig:Lx_z_XMM} shows the 0.5--2\,keV and 2--10\,keV rest-frame luminosities of the 85 \XMM\ sources as a function of redshift. The rest-frame luminosities were calculated by converting their observed 0.5--2\,keV and 2--10\,keV fluxes with a K-correction assuming X-ray spectral indices $\Gamma(\hbox{0.5--2\,keV}) = 1.40$ and $\Gamma(\hbox{2--10\,keV}) = 1.80$. The calculated luminosities were not corrected for absorption, although the absorbing effect is partly mitigated by the choice of $\Gamma$ = 1.40 in the 0.5--2~keV band.

\begin{figure} 
\centering
\includegraphics[width=.5\textwidth]{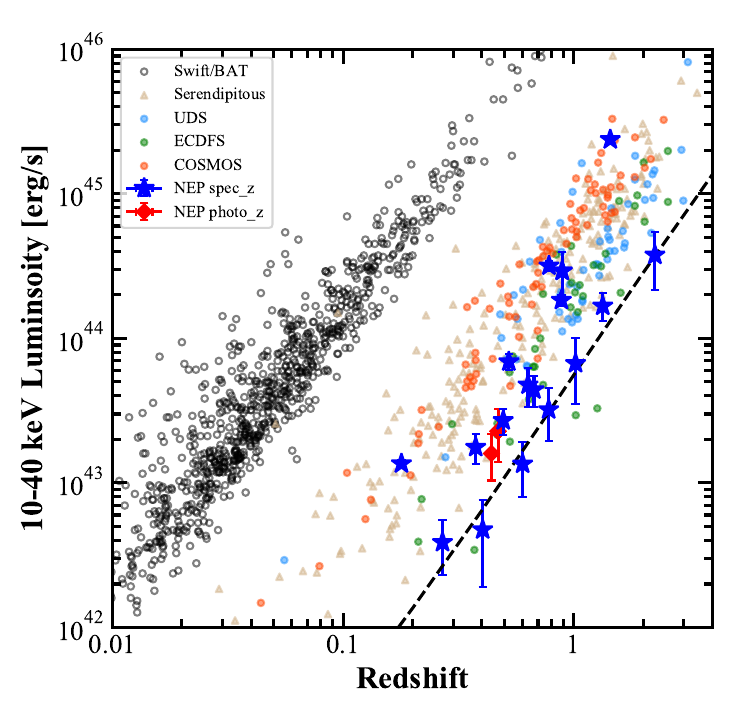}
\caption{10--40\,keV rest-frame luminosity versus redshift for the \NuSTAR\ sources with redshifts. Sources with spectroscopic (photometric) redshifts are plotted as blue stars (red diamonds). The sensitivity of the NEP-TDF cycles 5+6 survey at 20\% sky coverage is plotted as a dashed line. \NuSTAR\ COSMOS \citep[red circles;][]{Civano_2015}, ECDFS \citep[green circles;][]{Mullaney15}, UDS \citep[blue circles;][]{Masini_2018}, 40-month serendipitous \citep[brown triangles;][]{Lansbury_2017}, and {{Swift}}-BAT 105-month \citep[black open circles;][]{Oh_2018} surveys are shown as well. The luminosities were not corrected for intrinsic absorption.}
\label{fig:Lx_z_NuS}
\end{figure}

Figure~\ref{fig:Lx_z_NuS} shows the 10--40\,keV rest-frame luminosities of the 22 \NuSTAR\ sources that have redshift measurements. The 10--40\,keV rest-frame luminosities were calculated by converting the observed 3--24\,keV fluxes with a K-correction assuming $\Gamma = 1.80$. 
The brightest source in the FoV is a flat-spectrum radio quasar (FSRQ) blazar (ID 29, $z = 1.441$). 
Figure~\ref{fig:Lx_z_NuS} also shows sources detected in previous \NuSTAR\ surveys.
Most of the detected sources from previous \NuSTAR\ extragalactic surveys are well above the NEP-TDF sensitivity line, consistent with the \NuSTAR\ survey being the deepest. 
The all-sky {{Swift}}-BAT survey is also shown in Figure~\ref{fig:Lx_z_NuS}. Its measured 14--195~keV luminosities \citep{Gehrels04,Barthelmy05} 
from the 105-month {{Swift}}-BAT catalog \citep{Oh_2018} were converted to 10--40~keV luminosities (assuming a $\Gamma = 1.8$ power law model) for plotting. {{Swift}}-BAT samples sources mostly in the local Universe with median redshift of $\left<z_{\rm BAT} \right> = 0.044$, while \NuSTAR\ samples sources at $\left<z_{\rm NuS} \right> = 0.734$.

\section{Discussion}\label{sec:discuss}
\subsection{\lns}\label{sec:logN_logS}
The cumulative hard X-ray source number-counts distributions (log$N$-log$S$) in three energy bands (3--8\,keV, 8--24\,keV, and 8--16\,keV) were calculated using the NEP-TDF cycle 5 (681\,ks) observations \citetalias{Zhao2021}. Here, we update the log$N$-log$S$ distributions by using the combined cycles 5+6 (1.56\,Ms) data, which can provide constraints of log$N$-log$S$ at fainter hard X-ray fluxes. 

The \lns\ distribution is defined following \citet{Cappelluti09} as:
\begin{equation}
N({>}S)\equiv\sum_{i=1}^{N_S}\frac{1}{\Omega_i} \rm ~deg^{-2}\quad,
\end{equation}
where $N({>}S)$ is the surface density of sources detected above 95\% reliability level in a given energy band with flux greater than $S$, and $\Omega_i$ is the sky coverage associated with the flux of the $i$th source (Figure~\ref{fig:coverage}). The variance of $N({>}S)$ is 
\begin{equation}
\sigma^2_S = \sum_{i=1}^{N_S}(\frac{1}{\Omega_i})^2\quad.
\end{equation}
The \lns\ distribution depends on the minimum flux limit and the S/N limit of the sources \citep{Cappelluti09, Puccetti_2009}. We selected a flux limit equal to 1/3 of the flux corresponding to the half-area coverage sensitivity reported in Table~\ref{Table:sensitivity} in each band (\citealt{Masini_2018} ,\citetalias{Zhao2021}). This reduces the effect of Eddington bias (Figure~\ref{fig:flux}). To reduce the large uncertainties in the flux of low-S/N sources (Appendix~\ref{Sec:flux_SN} and Figure~\ref{fig:SN_flux}), we kept only sources detected with $\rm S/N>2.5$ \citep[following][]{Puccetti_2009}. Here S/N is defined as $C_{\rm net}/(C_{\rm tot}+C_{\rm bk})^{0.5}$, where $C_{\rm net}$ is the source net counts, $C_{\rm tot}$ is the total counts, and $C_{\rm bk}$ is the background counts. For the maximum flux, we adopted 10$^{-13}$\,erg\,cm$^{-2}$\,s$^{-1}$ for the 3--8\,keV and 8--24\,keV bands and $6 \times 10^{-14}$\,erg\,cm$^{-2}$\,s$^{-1}$ for the 8--16\,keV band to provide enough statistics at the high-flux end. 

To validate the selection of the minimum flux limit and the S/N limit, we calculated the \lns\ distributions in different energy bands using the selected minimum flux limits and the S/N limit from the 2400 simulations described in Section~\ref{section:background}. The calculated \lns\ distributions reproduce the input \lns\ distribution \citep{Treister09} in the simulations, suggesting the selected minimum flux limits and the S/N limits are reasonable for the real observations. Other choices of minimum flux (e.g., 20\%--area sensitivity) and S/N limits (e.g., S/N$_{\rm lim}$ = 2 or S/N$_{\rm lim}$ = 3) were unable to reproduce the input \lns\ distribution. A minimum flux limit at 20\%--area sensitivity leads to an $\sim$30\% over-estimation of $N({>}S)$ at the faint end, S/N$_{\rm lim} = 2$ leads to an over-estimation of $N({>}S)$ by $\sim$35\%, and S/N$_{\rm lim} = 3$ leads to an under-estimate of $N({>}S)$ by $\sim$40\% at the faint end. 
\begin{figure} 
\centering
\includegraphics[width=.48\textwidth]{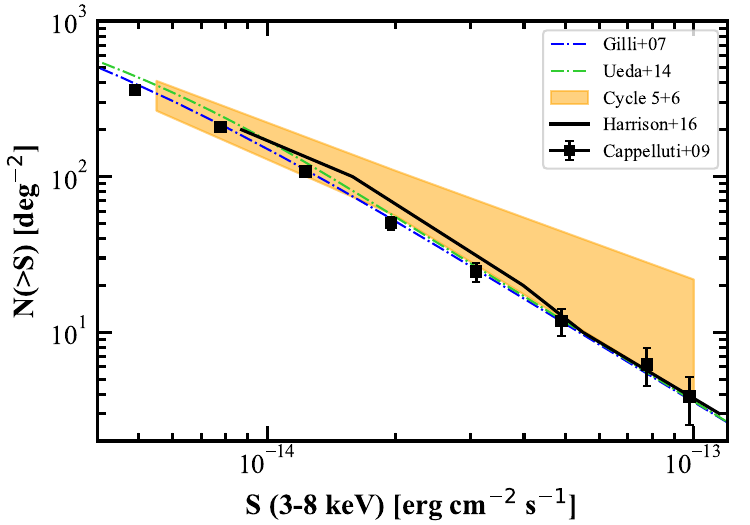}
\includegraphics[width=.48\textwidth]{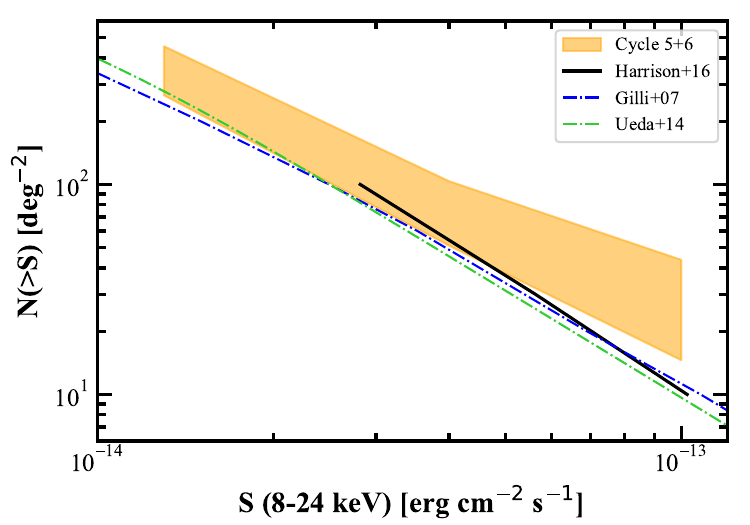}
\includegraphics[width=.48\textwidth]{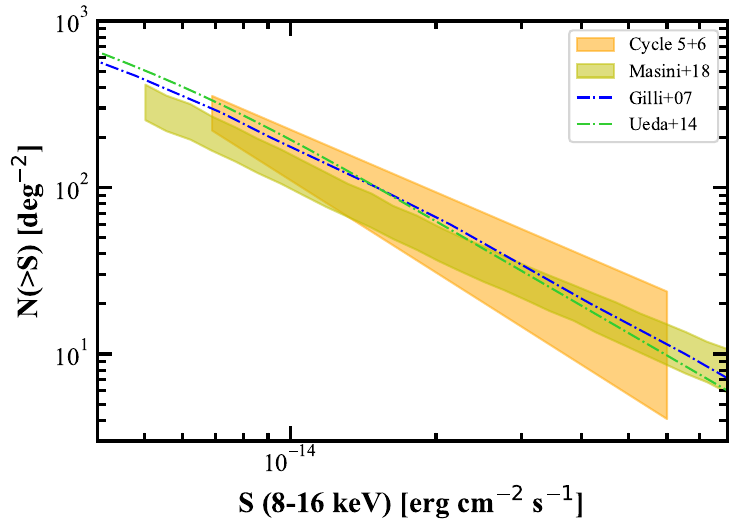}
\caption{Cumulative source number counts as a function of X-ray flux. Panels show three energy ranges as labeled. The orange-shaded areas represents the 68\%-confidence region at each energy. Black solid lines show results from \citet{harrison15}, and yellow shaded areas show those of \citet{Masini_2018b} using \NuSTAR. Black points in the top panel show \XMM\ results from \citet{Cappelluti09}. Expectations from population-synthesis models \citep{gilli07,Ueda14} are shown by dot-dash lines as indicated in the legend.}
\label{fig:logN-logS}
\end{figure}   

Figure~\ref{fig:logN-logS} shows the calculated \lns\ distributions from the actual cycle 5+6 observations. The NEP-TDF survey reaches fainter 8--24\,keV fluxes than previous \NuSTAR\ extragalactic surveys (i.e., COSMOS, EGS, and ECDFS; \citealt{harrison15}). The number of sources at the bright end in the 3--8\,keV and 8--24~keV bands is a little high but (at $\sim$1\,$\sigma$) consistent with previous measurements, especially given cosmic variance in the $\sim$0.16\,deg$^2$ area of the NEP-TDF survey. This excess cannot be explained solely by the bright blazar in the FoV\null.
The observed \lns\ distributions are also generally consistent with CXB population-synthesis models \citep[e.g.,][]{gilli07,Ueda14}. However, there may be an excess of hard X-ray sources at the faint end of the 8--24\,keV distribution, although again only at the $\sim$1\,$\sigma$) level. Extrapolating the \citet{harrison15} \lns\ distribution shows a possible excess, but \citet{Masini_2018} found no such excess at 8--16\,keV in the UDS field. If this excess is real, more heavily obscured sources exist than predicted by the population-synthesis models.  

\subsection{Hardness Ratio} \label{sec:HR}
Hardness ratio (HR) is useful to characterize the spectral shape of the \XMM\ and \NuSTAR\ NEP-TDF sources. 
The HRs of different column densities were estimated using a physical model which is typically used for modeling AGN X-ray spectra. The model includes a line-of-sight component (modeled by an absorbed power law), reflection component \citep[modeled by {\tt{borus02}},][]{Borus}, and scattered emission of soft X-rays (modeled by a fractional power law). The model was calculated with XSPEC as {\tt phabs$\times$(zphabs$\times$powerlw+borus+constant$\times$powerlw)}. {\tt phabs} models the Galactic absorption. We assumed a photon index $\Gamma = 1.8$ in both {\tt powerlw} and {\tt{borus02}} and a torus column density $N\rm _{H,Tor} = 1.4 \times 10^{24}$~cm$^{-2}$, a covering factor of $f_c = 0.67$ in {\tt{borus02}}, and an inclination angle $\theta_{\rm inc} = 60\arcdeg$ following the torus properties measured by \citet{Zhao2021a}. We assumed a {\tt constant} = 1\% fraction of the intrinsic emission being scattered \citep{Ricci2017}.

\begin{figure} 
\centering
\includegraphics[width=.49\textwidth]{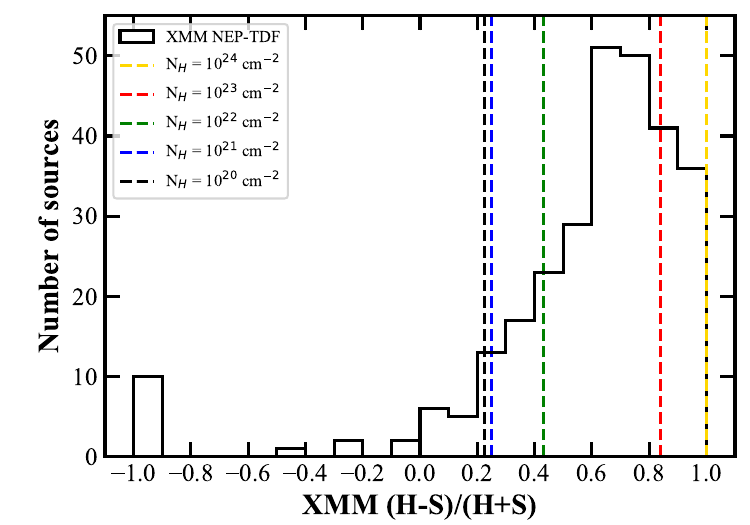}
\includegraphics[width=.49\textwidth]{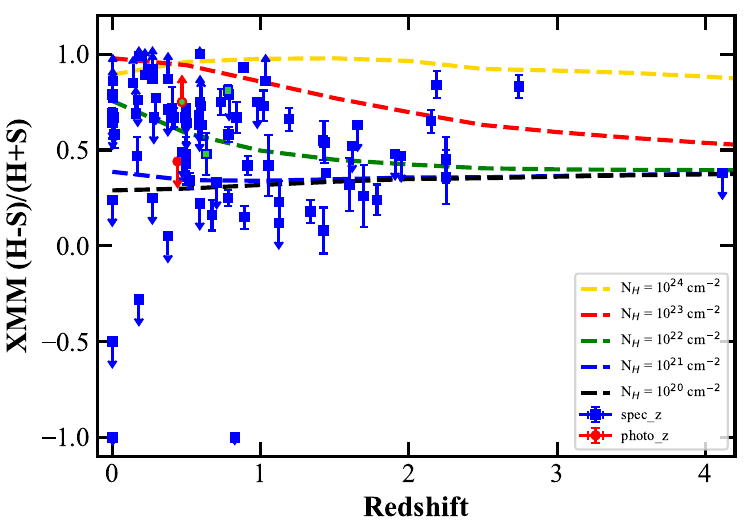}
\caption{Top: log(HR) distribution of the 286 \XMM-detected sources. Dashed lines show the expected log(HR) for a source at $z=0.60$ (the mean redshift of the \XMM\ sources with spectroscopic redshifts) and different obscuring column densities $N_{\rm H}$ as labeled. Bottom: log(HR) of \XMM\ sources versus redshift. Blue squares represent sources with spectroscopic redshifts and red circles those with photometric redshifts. Dashed lines show the expected HR $N_{\rm H}$ for different column densities as labeled. The three CT-AGN candidates detected by \NuSTAR\ (Figure~\ref{fig:HR_NuSTAR}) are shown as green-filled symbols.}
\label{fig:HR_XMM}
\end{figure}

Table~\ref{Table:XMM_catalog} reports the HRs of \XMM\ sources. They are defined as $(H-S)/(H+S)$ with 0.5--2\,keV flux as the soft band flux $S$ and 2--10\,keV flux as the hard band flux $H$. Table~\ref{Table:XMM_catalog} also reports $S$ and $H$, which were calculated with the SAS {\tt emldetect} tool. 
Figure~\ref{fig:HR_XMM} shows the HR distribution of the 286 \XMM\ sources. The expected HR for a given obscuration depends on the source redshift as shown in Figure~\ref{fig:HR_XMM}.
About half (48\%) of the \XMM-detected sources have HR larger than expected for column density $N\rm _{H} = 10^{22}$\,cm$^{-2}$ or have a lower limit of HR that implies an obscured source. For the 85 \XMM\ sources that have redshift measurements, 38\% are obscured. That lower percentage might be due to the bias from $m_i\le22$ mag selection for Hectospec observation: obscured sources are typically fainter in the visible light than unobscured ones.

\begin{figure} 
\centering
\includegraphics[width=.49\textwidth]{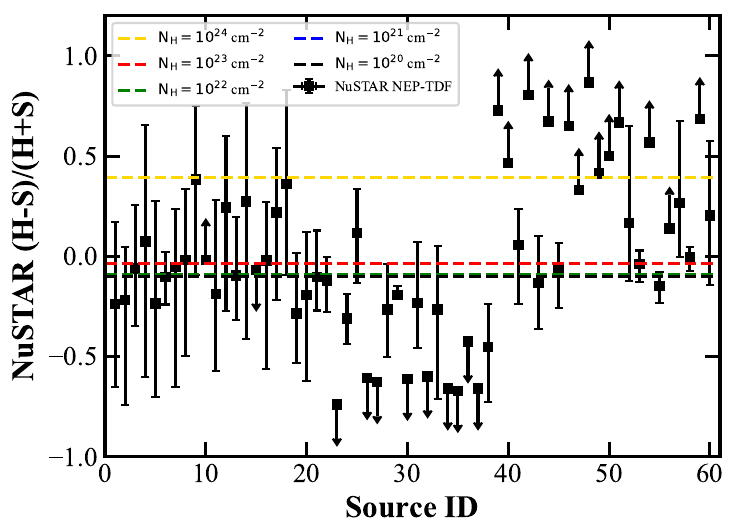}
\includegraphics[width=.49\textwidth]{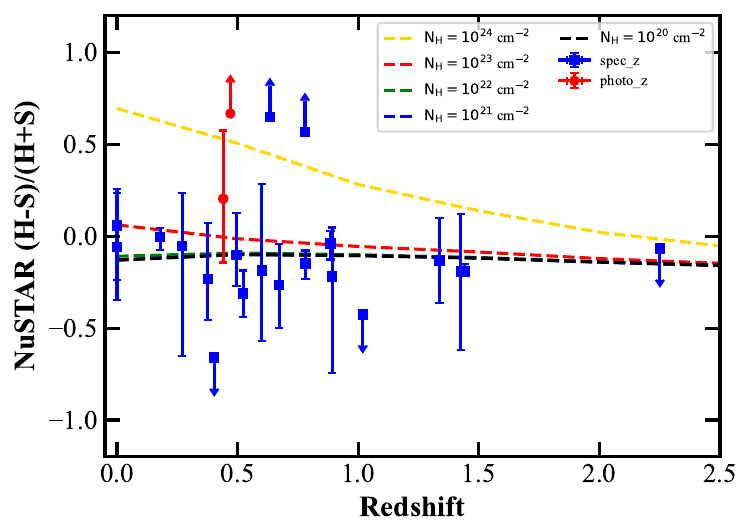}
\caption{Top: log(HR) distribution of all 60 \NuSTAR-detected sources versus source ID. Dashed lines show expected HRs for $z =0.734$, the median redshift of the \NuSTAR\ sources having measured redshifts, and for values of $N_{\rm H}$ as labeled.
Bottom: log(HR) of \NuSTAR\ sources with spectroscopic (blue square) or photometric (red circle) redshifts as a function of their redshifts. Dashed lines are the expected HR for different $N_{\rm H}$ values as labeled. (The green, blue, and black lines overlap.)}
\label{fig:HR_NuSTAR}
\end{figure}   

The HRs of the \NuSTAR\ sources detected in the cycles 5+6 survey were calculated using a Bayesian method \citep[BEHR;][]{Park_2006} following \citetalias{Zhao2021}. BEHR can estimate HR even for sources in the Poisson regime with a limited number of counts. BEHR also calculates the mode and uncertainty of the HR distribution of each source based on the source's total and background counts. Here, $S$ and $H$ were defined as net counts in the 3--8\,keV and 8--24\,keV bands, respectively. 
The 1\,$\sigma$ uncertainty was calculated by Gaussian-quadrature numerical integration when the number of the net counts of either energy band was less than 15 or by the Gibbs sampler method when the number of net counts was larger. The differences in the effective exposure times between the two bands were considered. The upper panel of Figure~\ref{fig:HR_NuSTAR} shows the HR of the 60 \NuSTAR\ sources. We converted the soft and hard band fluxes to count rate using the CF listed in Section~\ref{sec:simulation} when calculating the model predicted HR to directly compare with the BEHR calculated HR.

Unlike \XMM, \NuSTAR\ is sensitive to obscurations $N_{\rm H}>10^{23}$\,cm$^{-2}$ \citep[e.g.,][]{Masini_2018}. 47\% of the \NuSTAR\ detected sources are obscured above that level, and 23\% are Compton-thick (CT, $N_{\rm H}>10^{24}$\,cm$^{-2}$) assuming $z$ = 0.734 (median redshift of the \NuSTAR\ sources). CT candidates include sources that have lower limits on their HR\null. Figure~\ref{fig:HR_NuSTAR} shows HR as a function of redshift for the 22 \NuSTAR\ sources with redshift measurements. Of these, 23\% are heavily obscured, and 14\% are CT.

The HRs of the \XMM\ sources and the model predictions used to compare with the \NuSTAR\ HRs were calculated assuming particular spectral shapes when converting the count rates into fluxes. Different assumed column densities and photon indices lead to different ECFs. The ECF changes by about 70\% in the 0.5--2~keV band and about 20\% in the 8--24~keV band assuming no intrinsic absorption compared to a CT absorption. This might explain the discrepancy of the column densities estimated using \NuSTAR\ and \XMM. Changing the photon index has little effect: only $\sim$5\% for photon indices $\Gamma = 1.40$ or 2.20 rather than the assumed $\Gamma = 1.80$. 
Therefore, a broadband spectral analysis is needed to accurately measure the obscuration of these sources. Full spectral analysis of the \NuSTAR\ and \XMM\ sources will be presented by S.~Creech et al. in prep.


\subsection{CT Fraction}
As shown in Section~\ref{sec:HR}, \XMM\ is more sensitive to distinguishing between obscured and unobscured sources, while \NuSTAR\ is more powerful in determining whether a source is CT\null. 
Three CT sources (ID 46/51/54) with redshift measurements are shown in Figure~\ref{fig:HR_NuSTAR}. Three additional sources (ID 39/42/48) lack redshift measurements but have $\rm HR > 0.736$, the CT threshold at $z = 0$. Therefore, at least six sources are CT based on HR\null. For the rest of the sources without redshift measurements, five are CT candidates if $z\ge 0.734$.
Another 12 sources have HR uncertainty ranges that include the $z = 0.734$ CT threshold. A reasonable estimate is that $(3+3+5)/60 = 18\%$ of sources are CT with limits of 6--23 sources or 10--38\%. Additional redshift measurements are needed to tighten the constraints on the CT fraction.

\begin{figure} 
\centering
\includegraphics[width=.5\textwidth]{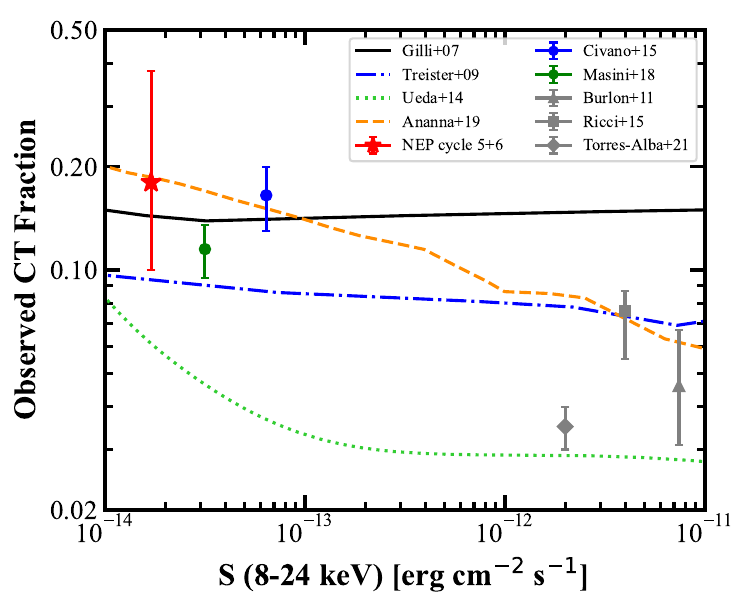}
\caption{CT fraction in different surveys as a function of survey sensitivity limit. The CT fraction measured here is plotted as a red star. Blue and green circles represent the \NuSTAR\ measurements in the COSMOS \citep{Civano_2015} and UDS \citep{Masini_2018} fields. The grey triangle, square, and diamond show the {{Swift}}-BAT measurements \citep{Burlon11,Ricci15,Nuria}. Lines show the CT fractions predicted by CXB synthesis models: \citet{gilli07} (black solid line), \citet{Treister09} (blue dash-dot line), \citet{Ueda14} (green dot line), and \citet{Tasnim_Ananna_2019} (orange dashed line).}
\label{fig:CT}
\end{figure}   

The CT fraction measured in the NEP field is consistent with the CT fraction measured in other surveys as shown in Figure~\ref{fig:CT}. The most directly comparable values are for the \NuSTAR\ COSMOS field \citep[13\%--20\%;][]{Civano_2015} and the \NuSTAR\ UDS field \citep[11.5\%$\pm$2.0\%;][]{Masini_2018}. For the Swift-BAT all-sky survey, which samples the bright end of the nearby AGN population, \citet{Burlon11} and \citet{Ricci15} measured a CT fraction of $\sim$4.6\%--7.6\%. However, a recent analysis \citep{Nuria} of the CT-AGN candidates in the BAT sample using high-quality \NuSTAR\ observations found that many candidates are less than CT-obscured. That brought the CT fraction of the entire BAT sample down to 3.5\%$\pm$0.5\%. However, \citet{Nuria} also found that the CT fraction of the BAT sample depends on the redshift range. A CT fraction of 20\% was found for the $z\le0.01$ sample and 8\% for $z\le0.05$ sample. This discrepancy was explained by BAT being biased against the detection of CT sources at higher redshift.
Figure~\ref{fig:CT} compares the measured CT fractions with population-synthesis model predictions \citep{gilli07, Treister09, Ueda14, Tasnim_Ananna_2019}. The recent \citet{Tasnim_Ananna_2019} model is in good agreement with the hard X-ray observed CT-AGN fraction at both bright and faint fluxes. 

\subsection{An obscured and variable Seyfert galaxy}
In addition to the bright blazar mentioned in Section~\ref{s:lz}, there is another prominent Seyfert galaxy (and radio source; \citealt{Willner2023}) in the TDF field.
The source XMM ID 17 (shown in Figure~\ref{Fig:visual_check}, \NuSTAR\ ID 58) has \XMM\ $\rm HR\ge0.99$ suggesting $N_{\rm H}>10^{23}$\,cm$^{-2}$ (Figure~\ref{fig:HR_XMM}). The \NuSTAR\ $\rm HR = 0.00_{-0.07}^{+0.05}$ also suggests $N_{\rm H}\sim10^{23}$\,cm$^{-2}$ (Figure~\ref{fig:HR_NuSTAR}). The bright core seen in the JWST long-wave imaging but not at shorter wavelengths supports the interpretation of high but not CT obscuration. The source's 2--10\,keV luminosity $L(\hbox{2--10\,keV}) = 4.5\pm0.5 \times 10^{42}$ erg s$^{-1}$) and 10--40\,keV luminosity $L(\hbox{10--40\,keV}) = 1.36\pm0.08 \times 10^{43}$ erg s$^{-1}$ suggest a type~2 AGN\null. 
More intriguing, this source is variable at 3$\sigma$ in 3--24\,keV band and at 2.9$\sigma$ in the 3--8 keV band but is not significantly variable in the 8--24\,keV band, suggesting a variable spectral shape. Its count rates increased by 115\% in the 3--24\,keV and 230\% in the 3--8\,keV band from 2019 Oct to 2022 Jan (Figure~\ref{fig:varia}). We did not find obvious variability of this source in either visible \citep[Zwicky Transient Facility, ZTF;][]{Bellm2019,Masci2019} or IR \citep[NEOWISE;][]{Mainzer2014}. This suggests that the X-ray variability of the source might be caused by the decreasing of the line-of-sight obscuration rather than the variability of the intrinsic accretion rate. However, further investigation is needed.

\section{Conclusions}
The \NuSTAR\ extragalactic survey of the JWST NEP-TDF attained a total of 1.5 Ms exposure and covered an area of $\sim$0.16~deg$^2$. This makes it the deepest \NuSTAR\ extragalactic survey to date. The survey consists of 21 observations in \NuSTAR\ cycles 5 and 6 across seven epochs from 2019 Sep to 2022 Jan, enabling a multi-year, multi-epoch study of this field in hard X-rays. Principal results are:
\begin{enumerate}
\item The \NuSTAR\ cycle 6 survey was taken from Oct.~2020 to Jan.~2022 with a total exposure of 880~ks acquired in 12 observations over four epochs covering an area of $\sim$0.11~deg$^2$. 
A total of 35 sources were detected above the 95\% reliability threshold in cycle 6. In the merged cycle 5 and 6 observations, which reach the deepest sensitivity, 60 sources were detected above the 95\% reliability threshold.

\item The survey's 8--24\,keV sensitivities at 20\%-area are $1.98 \times 10^{-14}$\,erg\,cm$^{-2}$\,s$^{-1}$ for \NuSTAR\ cycle 6 and $1.70 \times 10^{-14}$\,erg\,cm$^{-2}$\,s$^{-1}$ for \NuSTAR\ cycles 5+6. A ${\sim}1\sigma$ excess of faint 8--24\,keV sources compared to the population-synthesis models hints that more faint, heavily obscured sources might exist than predicted by the models.

\item To enable broadband (0.3--24~keV) X-ray spectral fitting and more reliable multiwavelength counterpart matching of the \NuSTAR\ detected sources, a total of 60~ks \XMM\ observations were taken simultaneously with \NuSTAR\ in cycle 6. A total of 286 \XMM\ sources were detected including more 3--8\,keV sources at the bright end compared to previous number counts.

\item Of the 60 \NuSTAR\ sources, 37 have \XMM\ counterparts. Of the 23 \NuSTAR\ sources without \XMM\ counterparts, 17 appear to be heavily obscured. 

\item 
The NEP-TDF has extensive multiwavelength coverage, including Subaru/HSC, J-PAS, and SDSS in optical and MMT/MMIRS and {WISE} in IR. 
A total of 214 \XMM\ sources have secure counterparts in multiwavelength catalogs, and 19 more have ambiguous counterparts. Deeper optical and IR observations covering the entire FoV of the \XMM\ NEP-TDF survey are needed to identify counterparts of the remaining 53 \XMM\ sources. In addition, VLA, HST, and JWST surveyed a fraction of the \XMM\ NEP-TDF and a total of 55, 102, and 32 \XMM\ sources have VLA, HST, and JWST counterparts, respectively.

\item 
Optical spectra of \XMM\ counterparts produced 82 high-confidence redshifts. 
Two additional sources have photometric redshifts measured in SDSS DR17. The 84 \XMM\ sources with redshifts include 22 \NuSTAR\ sources. In addition, spectroscopic redshifts of 40 VLA and Chandra sources in the NEP-TDF are reported in Table~\ref{Table:Hectospec_catalog}.

\item Half (48\%) of the \XMM\ sources are obscured with $N_{\rm H}>10^{22}$\,cm$^{-2}$, and 47\% of the \NuSTAR\ sources are heavily obscured with $N_{\rm H}>10^{23}$\,cm$^{-2}$). 18$_{-10}^{+20}$\% of the \NuSTAR\ sources are Compton-thick. Broadband spectral analysis is needed to accurately measure the column densities of the sources (S.~Creech et al., in prep).

\item A type 2 AGN at $z = 0.1791$ has X-ray obscuration $N_{\rm H}\sim10^{23}$\,cm$^{-2}$, and significant obscuration is supported by JWST and HST images. The source is significantly variable with its 3--8\,keV band flux having increased by 230\% in 26 months. 

\item The prime goal of the \NuSTAR\ NEP-TDF observations is to study hard X-ray variability. 
Preliminary results for the 60 sources detected in cycles 5+6 
show four sources varying with $p<0.05$ (${\sim}2\sigma$) in at least one energy band in the 26 months of observations. A detailed study of the source variability is in preparation.

\item Subsequent to the work reported here, an additional 855\,ks of \NuSTAR\ observations and 30\,ks of \XMM\ observations have been obtained in \NuSTAR\ cycle 8 (PI: Civano, pid 8180, Silver et al., in prep.). These targeted the NEP-TDF simultaneously with JWST\null. A further 900\,ks of \NuSTAR\ observations and 40\,ks of \XMM\ observations were approved for \NuSTAR\ cycle 9 (PI: Civano, ID: 9267). Thus the NEP-TDF will acquire a total of 3.25~Ms \NuSTAR\ observations, making it the newest and deepest \NuSTAR\ extragalactic survey. The data will constitute five years of continuous hard X-ray monitoring of the field, making it the first long-term monitoring, contiguous survey of hard X-ray variability.

\item The rich multiwavelength coverage and multi-year \NuSTAR\ monitoring of the NEP-TDF make it an ideal field for the next generation of hard X-ray surveys. The High-Energy X-ray Probe ({HEX-P}) concept\footnote{https://hexp.org} \citep[][]{Madsen2019} has a larger effective area, much better PSF, and lower background compared with \NuSTAR\null. This will allow $\sim$20 times deeper sensitivity in the 8--24\,keV band compared with the current deepest \NuSTAR\ NEP-TDF survey and detect $\sim$40 times more hard X-ray sources. {HEX-P} will be able to resolve more than 80\% of the CXB into individual sources up to 40~keV \citep{Civano2023} and better constrain current population-synthesis models. Its broadband coverage (0.2--80\,keV) will allow X-ray spectral analysis of both obscured and unobscured sources and thus more accurately constrain the CT fraction down to 10$^{-15}$~erg~cm$^{-2}$~s$^{-1}$ \citep{Civano2023,Boorman2023}.

\end{enumerate}
\bigskip

The authors thank the anonymous referee for their helpful comments.
XZ acknowledges NASA funding under contract numbers 80NSSC20K0043 and 80NSSC22K0012. The authors thank Jinguo Liu for helping improve the code efficiency when calculating the variability of the \NuSTAR\ and \XMM\ sources, thereby shortening the calculation time by more than a factor of 10; Cheng Cheng for helping reduce the Hectospec spectra; Alberto Masini for help with the proposal preparation and for sharing the data of his previous papers; Nelson Caldwell and the MMT observing team for the help in generating the observing catalog and scheduling the Hectospec observations; Brian Grefenstette for helpful discussion of \NuSTAR\ and its data analysis; Karl Foster and the \NuSTAR\ observation planning team for their help in designing the observation plan and scheduling the observations. 

This work has made use of data from the \NuSTAR\ mission, a project led by the California Institute of Technology, managed by the Jet Propulsion Laboratory, and funded by NASA. 
This research has made use of the \NuSTAR\ Data Analysis Software (NuSTARDAS) jointly developed by the ASI Science Data Center (ASDC, Italy) and the California Institute of Technology (USA). 
This research made use of data and software provided by the High Energy Astrophysics Science Archive Research Center (HEASARC), which is a service of the Astrophysics Science Division at NASA/GSFC and the High Energy Astrophysics Division of the Smithsonian Astrophysical Observatory. 
This work is based on observations obtained with \XMM, an ESA science mission with instruments and contributions directly funded by ESA Member States and NASA. 
The MMIRS, Hectospec, and Binospec observations reported were obtained at the MMT Observatory, a joint facility of the Smithsonian Institution and the University of Arizona. 
This paper uses data products produced by the OIR Telescope Data Center, supported by the Smithsonian Astrophysical Observatory. 
This work makes use of the data from SDSS IV. Funding for the Sloan Digital Sky Survey IV has been provided by the Alfred P. Sloan Foundation, the U.S. Department of Energy Office of Science, and the Participating Institutions. 
This work is partly based on the data from WISE, which is a joint project of the University of California, Los Angeles, and the Jet Propulsion Laboratory/California Institute of Technology, and NEOWISE, which is a project of the Jet Propulsion Laboratory/California Institute of Technology.

\facilities{\NuSTAR, \XMM, \cha, MMT (Binospec, Hectospec, and MMIRS), JWST, HST, Subaru/HSC, SDSS, J-PAS, WISE}

\section*{Data Availability}
Electronic versions of the generated \NuSTAR\ and \XMM\ catalogs as described in Appendix~B is available at CDS via \url{https://cdsarc.cds.unistra.fr/viz-bin/cat/J/ApJ/965/188} or via \url{http://vizier.cds.unistra.fr/viz-bin/VizieR?-source=J/ApJ/965/188}.
\bibliography{referencezxr}{}
\bibliographystyle{aasjournal}

\clearpage
\appendix
\renewcommand{\thesection}{APPENDIX~\Alph{section}}

\section{Measured to input flux and Signal to noise ratio} \label{Sec:flux_SN}
\begin{figure} 
\centering
\includegraphics[width=.49\textwidth]{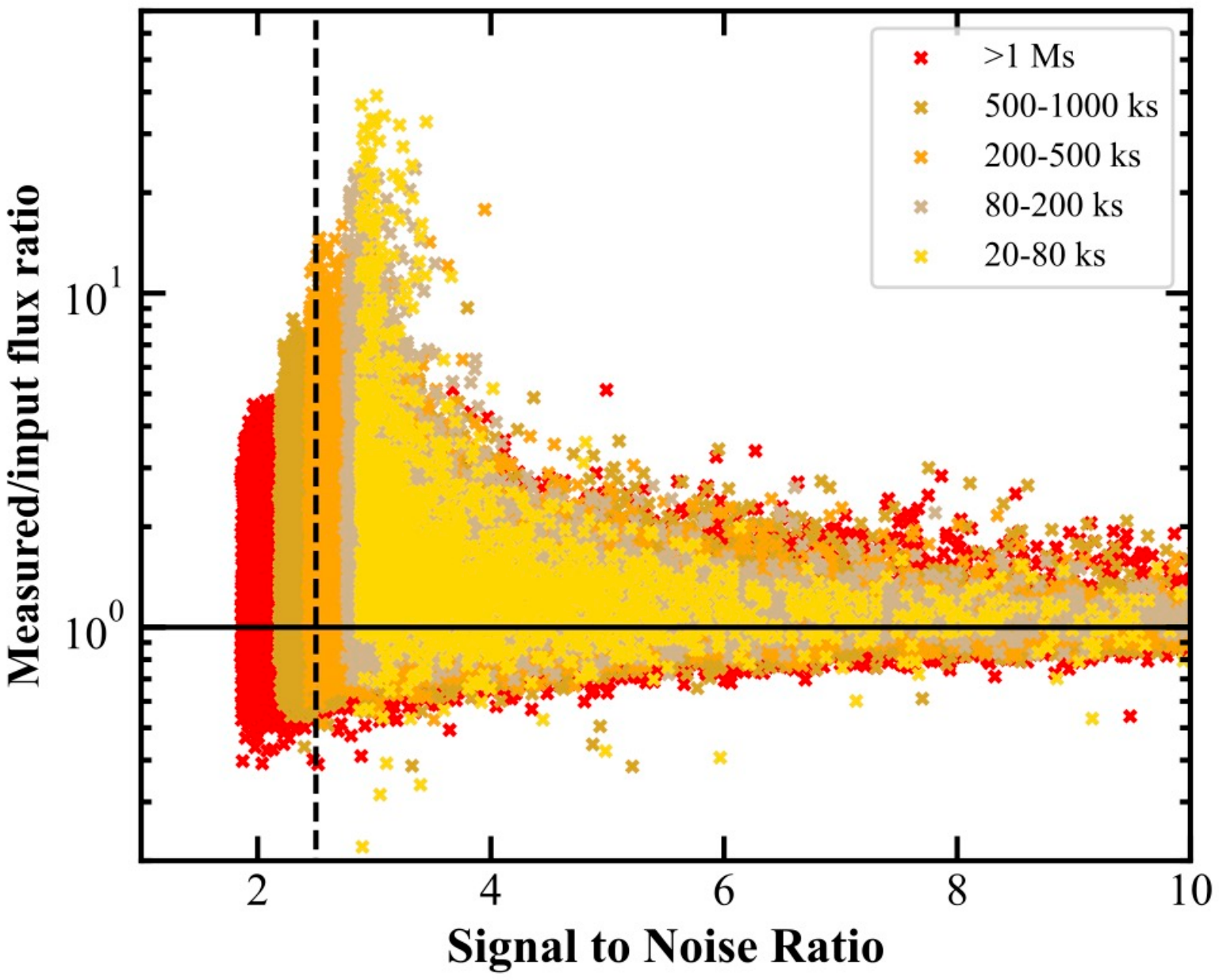}
\caption{Ratio of measured (3--24~keV) flux to true flux as a function of measured signal to noise ratio. Points show the ratio for simulated (Section~\ref{sec:simul}) individual sources with different exposure times indicated as shown in the legend. Points include only (simulated) sources detected with $>$95\% reliability. The vertical dashed line indicates $\rm S/N = 2.5$, which was the cut for calculating \lns\ (Section~\ref{sec:logN_logS}). }
\label{fig:SN_flux}
\end{figure}

The accuracy of the \NuSTAR\ source flux measurement is strongly correlated with the S/N (as defined in Section~\ref{sec:logN_logS}). Figure~\ref{fig:SN_flux} shows the simulation results.
The dispersion of the measured to input flux ratio is quite high at low S/N, and there is a strong bias for measured fluxes to be higher than the true flux, especially at low S/N.

\section{NuSTAR and XMM-Newton Catalog Description}\label{sec:catalog_description}
The description of each column of the 95\% reliability level catalog of \NuSTAR\ detected sources in cycle 6 and cycles 5+6 survey is in Table~\ref{Table:catalog}.
The description of each column of the catalog of \XMM\ detected sources is in Table~\ref{Table:XMM_catalog}.

In future work, the \NuSTAR\ sources can be referred to as ``NuSTAR JHHMMSS+DDMM.m'' where the sequence (JHHMMSS+DDMM.m) is the contents of column 2 of either of the \NuSTAR\ data tables. Similarly, the XMM sources can be referred to as ``TDFXMM JHHMMSS+DDMM.m'' with the sequence given in column 2 of the \XMM\ data table. 

\section{Hectospec observations of non-XMM targets in NEP-TDF} \label{sec:hectospec}
Table~\ref{Table:Hectospec_catalog} reports the coordinates, redshifts, and spectral types of the 40 VLA and Chandra sources in the NEP-TDF.

\section{Variability Calculation}\label{sec:variability}

\begin{figure} 
\centering
\includegraphics[width=\linewidth]{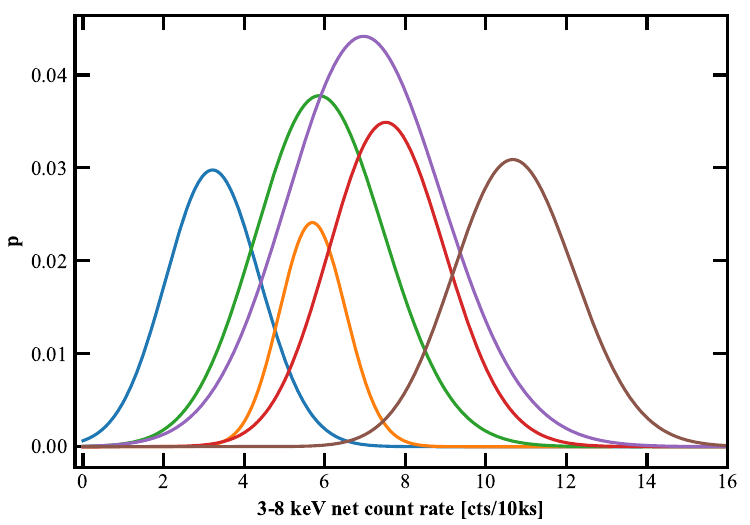}
\caption{3--8~keV net count-rate posterior probability distribution of \NuSTAR\ source ID 58 in each epoch. Epochs are distinguished by color in the order blue, orange, green, red, violet, brown, magenta (the same as Figure~\ref{fig:varia2}, which shows dates for each epoch). Count-rate PPDs are count PPDs divided by exposure time.}
\label{fig:ppd58}
\end{figure}   

\begin{figure} 
\centering
\includegraphics[width=\linewidth]{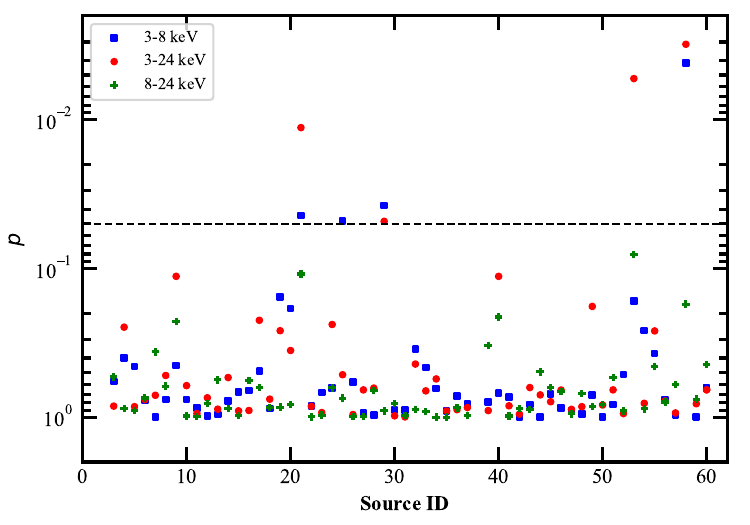}
\caption{``False-alarm'' probability $p$ of each source. Colors and point shapes show different energy bands as indicated in the legend. The horizontal dashed line shows $p = 0.05 \approx2\sigma$. Points above this line have higher probability of being true variables.}
\label{fig:varia}
\end{figure}   

Studying source variability is the prime goal of the \NuSTAR\ NEP-TDF. Thanks to the multi-year and multi-epoch observations in the field, NEP-TDF became the first \NuSTAR\ contiguous survey to study hard-X-ray (3--24~keV) variability. 

The traditional method for analyzing X-ray-source variability \citep[e.g.,][]{Yang_2016} is not suitable for \NuSTAR\ contiguous surveys because that method requires good counting statistics and negligible backgrounds. After adding all seven epochs in cycles 5 and 6, the median \NuSTAR-detected (3--24~keV) source has 120 net source counts on top of 700 background counts. This gives low S/N and an uncertain net count rate for individual epochs. Therefore, we developed a dedicated pipeline to analyze source variability in the low-count regime. 
This paper describes the pipeline and briefly summarizes the source variability results. A future paper will provide a systematic discussion of source variability in the \NuSTAR\ and \XMM\ NEP-TDF.

The pipeline follows the Bayesian approach developed by \citet{Primini2014} and used to generate the Chandra Source Catalog\footnote{\url{https://cxc.cfa.harvard.edu/csc/}} (CSC)\null. The key calculation is the probability distribution of the expected net source counts in each epoch. This approach is able to deal with the Poisson (not Gaussian) statistical noise and is valid in low-counts regimes because it uses Poisson likelihoods. The net count posterior probability distribution (PPD) was calculated using Equation~16 of \citet{Primini2014} assuming non-informative prior distributions. We used a circle with 20\arcsec\ radius to extract the total and background counts from each epoch's image and background map. Thus we analyzed only the inter-epoch variability rather than the intra-epoch's variability. As the sources are not detectable in every epoch, we used a fixed source position (the one reported in the catalog) for all epochs and energy bands. Figure~\ref{fig:ppd58} shows the net count-rate PPDs for the most-likely variable source (ID~58).

\begin{figure*} 
\begin{minipage}[b]{.35\linewidth}
\centering
\includegraphics[width=\linewidth]{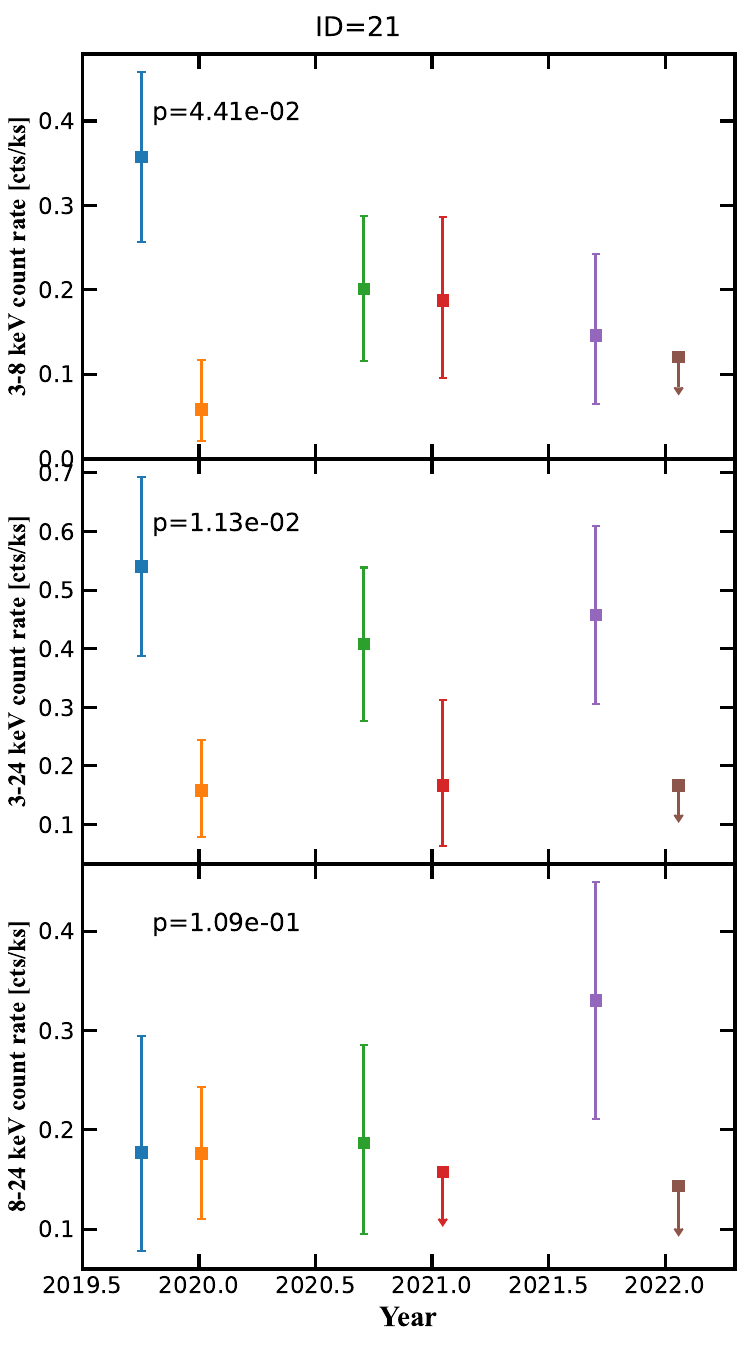}
\includegraphics[width=\linewidth]{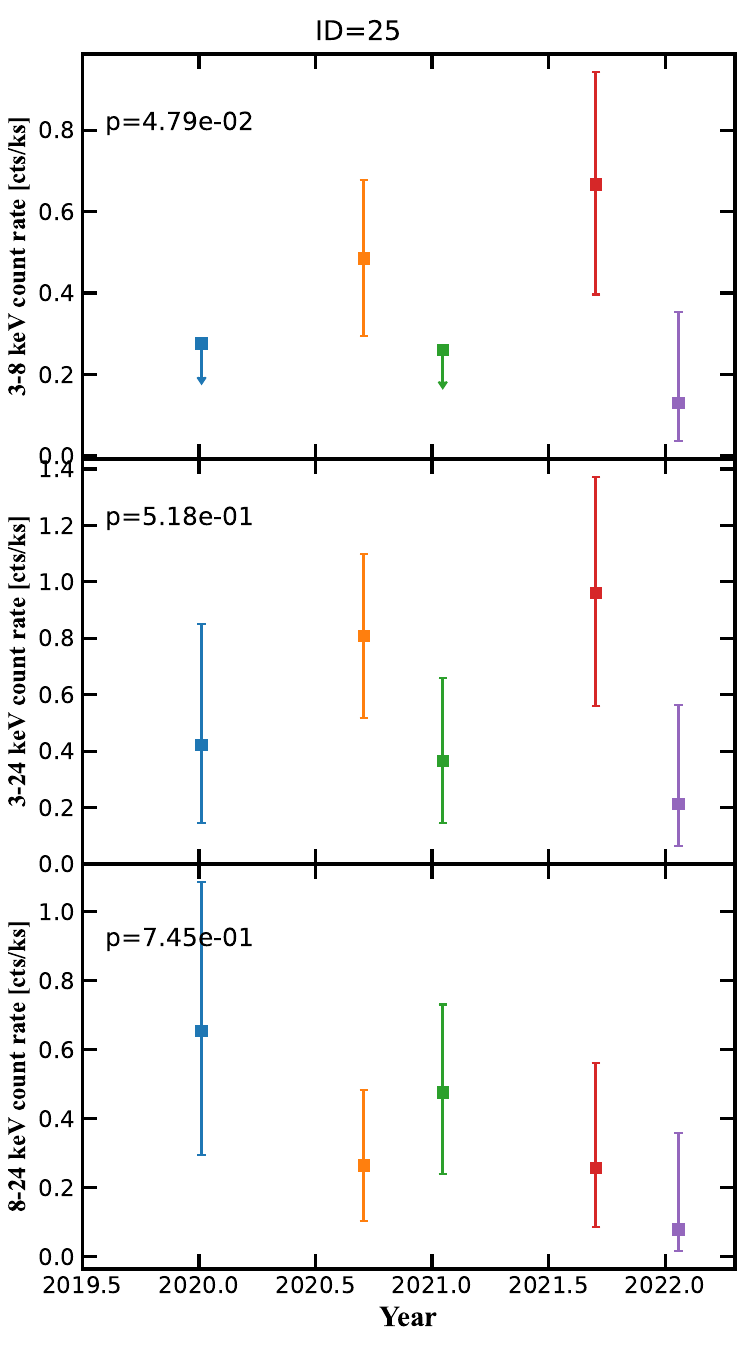}\\
\end{minipage}
\begin{minipage}[b]{.35\linewidth}
\centering
\includegraphics[width=\linewidth]{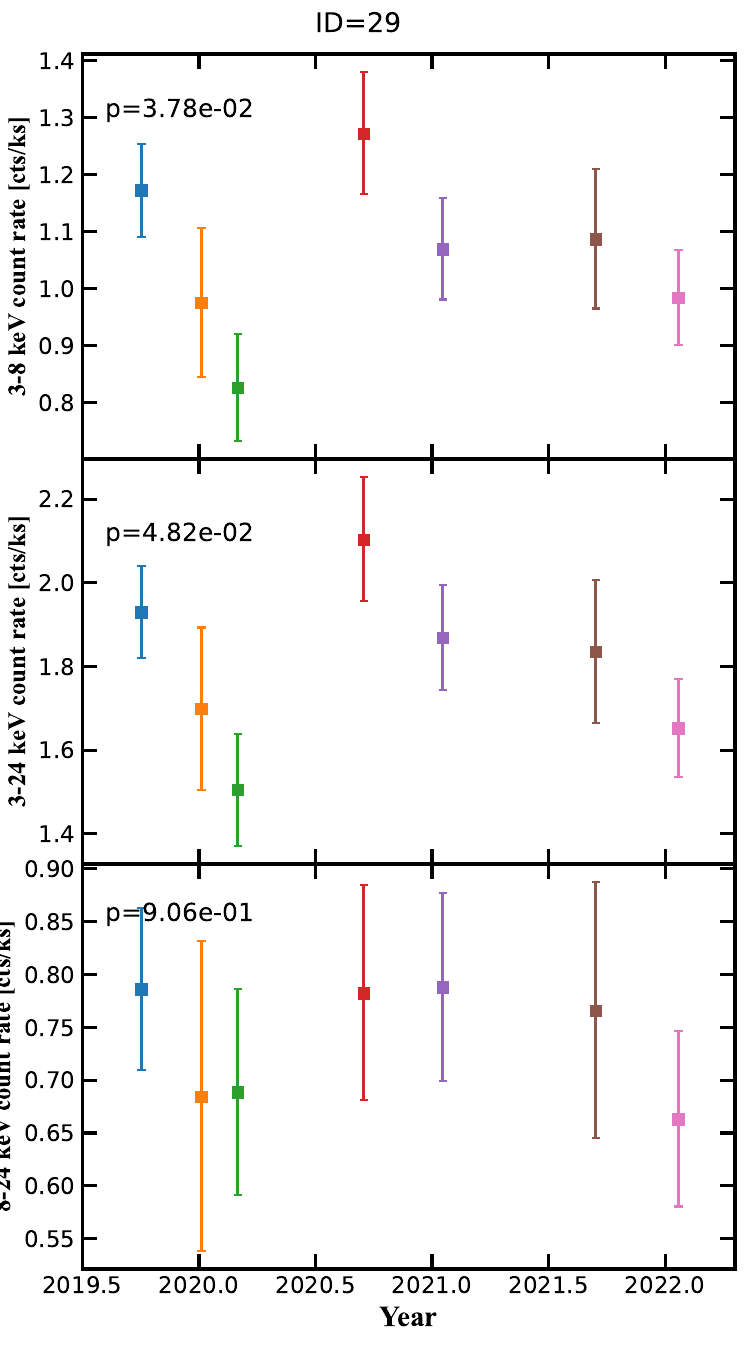}
\includegraphics[width=\linewidth]{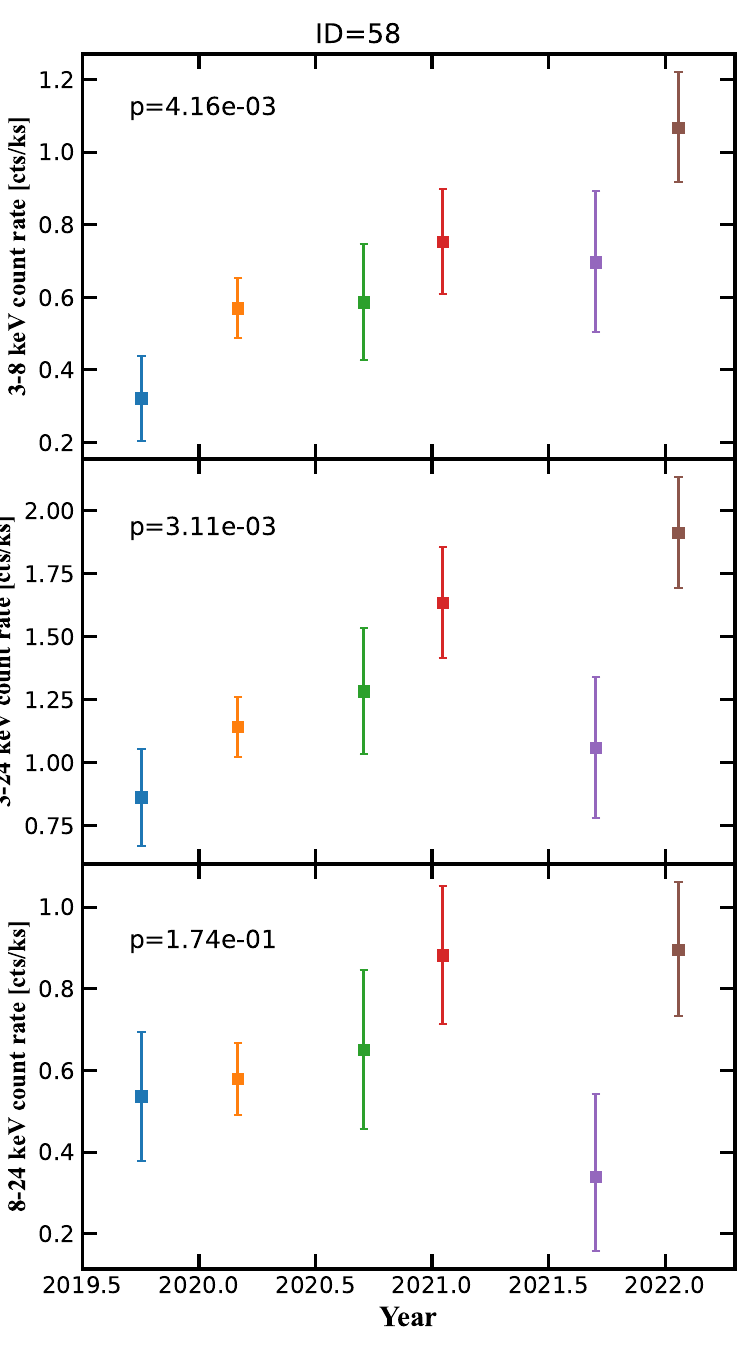}
\end{minipage}
\caption{Light curves of the four sources that show the most variability. Sources are labeled at the top of each panel, and sections top to bottom show different energy ranges as labeled. Point colors for each epoch are the same as in Figure~\ref{fig:ppd58}.}
\label{fig:varia2}
\end{figure*}

The probability of source variability came from applying the $\chi^2$ test following the CSC method \citep{Nowak2016}. The method computes the deviation between the most probable flux in each epoch and the most probable flux in the entire survey. The null hypothesis is that there is no variability, and the ``false-alarm'' probability $p$ was calculated from equation~10 of \citet{Nowak2016}. A smaller $p$ suggests a larger probability that variability exists. Figure~\ref{fig:varia} shows $p$ for the \NuSTAR\ sources. Of the 60 sources, 44 were observed in all seven epochs, and 9, 4, and 3 sources were observed in six, five, and one epoch(s), respectively. Four sources (ID 21/25/29/58) show variability at $p<0.05 (\sim2\sigma)$ in at least one band, and Figure~\ref{fig:varia2} shows their light curves. The same pipeline can be used for \XMM\ data, and those results will be presented in a future paper.

\begingroup
\renewcommand*{\arraystretch}{1.}
\begin{table*}
\centering
\caption{\NuSTAR\ 95\% reliability source catalog description.}
\label{Table:catalog}
  \begin{tabular}{ll}
       \hline
       \hline     
 	Col.&Description\\
	\hline 
	1&\NuSTAR\ source ID used in this paper.\\
	2&Source name (use ``NuSTAR JHHMMSS+DDMM.m'').\\
	3--4&X-ray coordinates (J2000) of the source in whichever energy band has the highest DET\_ML.\\
	5&3--24\,keV band deblended DET\underline{\;\;}ML ($-99$ if the source is not detected in a given band). \\
	6&3--24\,keV band vignetting-corrected exposure time in ks at the position of the source.\\
	7&3--24\,keV band total counts (source + background) in a 20\arcsec\ radius aperture.\\
	8&3--24\,keV band deblended background counts in a 20\arcsec\ radius aperture ($-99$ if the source is not detected in a given band).\\
	9&3--24\,keV band not deblended background counts in a 20\arcsec\ radius aperture.\\
	10&3--24\,keV band net counts (deblended if detected \& above DET\underline{\;\;}ML threshold or 90\% confidence upper limit if \\ 
	&undetected or detected but below DET\underline{\;\;}ML threshold) in a 20\arcsec\ radius aperture.\\
	11--12&3--24\,keV band 1\,$\sigma$ positive/negative net counts uncertainty ($-99$ for upper limits).\\
	13&3--24\,keV band count rate (90\% confidence upper limit if not detected or detected but below the threshold) in a 20\arcsec \\
	&radius aperture.\\
	14&3--24\,keV band aperture corrected flux (erg\,cm$^{-2}$\,s$^{-1}$; 90\% confidence upper limit if below 95\% confidence threshold).\\
	15--16&3--24\,keV band positive/negative flux uncertainties (erg\,cm$^{-2}$\,s$^{-1}$; $-99$ for upper limits).\\
	17--28&same as columns 5--16 but for 3--8\,keV.\\
	29--40&same as columns 5--16 but for 8--24\,keV.\\
	41--52&same as columns 5--16 but for 8--16\,keV.\\
	53--64&same as columns 5--16 but for 16--24\,keV.\\
	65&Hardness ratio computed using BEHR.\\
	66--67&Lower/upper limit of hardness ratio.\\
	68&\XMM\ source ID from the \XMM\ catalog ($-1$ if non-detection).\\
	69,70&Soft X-ray coordinates of the associated source ($-1$ if no \XMM\ counterpart).\\
	71&\NuSTAR\ to soft X-ray counterpart position separation in arcsec.\\
	72&3--8\,keV flux converted from \XMM\ 2--10\,keV flux (90\% confidence upper limit if $\tt mlmin<6$).\\
	73&3--8\,keV \XMM\ flux 1$\sigma$ uncertainty ($-99$ for upper limit). \\
	74&Flag for \NuSTAR\ counterparts (S, P, Sec, C if the XMM source is the single, primary, secondary, or\\
	&confusing counterpart of the \NuSTAR\ source, respectively).\\
	75&Flag for ancillary class (S for secure, A for ambiguous, and U for unidentified)\\
    76,77&Ancillary coordinates of the associated source ($-99$ if no detection).\\
	78&Optical (HSC) $i$ band AB magnitude ($-99$ if no detection).\\
    79,80&MMIRS $J$ and $K$ band AB magnitude ($-99$ if no detection).\\
	81,82&WISE $W1$ and $W2$ band AB magnitude ($-99$ if no detection).\\
	83&VLA 3 GHz counterpart ID from \citet{Hyun2023}.\\
	84&VLA 3~GHz flux density in $\mu$Jy \citep{Hyun2023}.\\
    85&HST F606W AB magnitude ($-99$ if no detection).\\
    86&JWST F444W AB magnitude ($-99$ if no detection).\\
	87&Spectroscopic redshift of the associated source.\\
	88&Photometric redshift of the associated source.\\
	89&Spectroscopic classification (Q for quasars, G for galaxies, S for stars, N/A if no measurement). Galaxies are defined\\ & as objects without broad 
	emission lines and therefore include type 2 AGN.\\
	90&Luminosity distance in Mpc (70/0.3/0.7 cosmology, $-99$ if no redshift measurement).\\
	91&10--40\,keV band rest-frame luminosity ($-99$ if no redshift measurement).\\
	92--93&10--40\,keV band 1$\sigma$ positive/negative rest-frame luminosity uncertainty ($-99$ if no redshift measurement).\\
	94&Source ID in the \citet{Zhao2021} \NuSTAR\ cycle 5 catalog ($-99$ for non-detection in the cycle 5 catalog).\\
	\hline
\end{tabular}
\raggedright
\tablecomments{Multiple rows are used for the \NuSTAR\ sources with multiple \XMM\ or optical/IR counterpart candidates.}
\end{table*}
\endgroup

\begingroup
\renewcommand*{\arraystretch}{1.}
\begin{table*}
\centering
\caption{\XMM\ source catalog description.}
\label{Table:XMM_catalog}
  \begin{tabular}{ll}
       \hline
       \hline     
 	Col.&Description\\
	\hline 
	1&\XMM\ source ID used in this paper.\\
	2&\XMM\ source name (use ``TDFXMM JHHMMSS+DDMM.m'').\\
	3--4&X-ray coordinates (J2000) of the source.\\
	5&0.5--2\,keV band DET\underline{\;\;}ML. \\
	6&0.5--2\,keV band vignetting-corrected exposure time (in ks) at the position of the source.\\
	7&0.5--2\,keV band net counts of the source (90\% confidence upper limit if $\tt mlmin<6$).\\
	8&0.5--2\,keV band net counts 1$\sigma$ uncertainty ($-99$ for upper limits).\\
	9&0.5--2\,keV band flux (erg\,cm$^{-2}$\,s$^{-1}$; 90\% confidence upper limit if $\tt mlmin<6$).\\
	10&0.5--2\,keV band flux 1\,$\sigma$ error ($-99$ for upper limits).\\
	11-16&same as columns 5--10 but for 2--10\,keV.\\
	17&Hardness ratio (90\% confidence upper or lower limits if not constrained).\\
	18&Hardness ratio 1$\sigma$ uncertainty ($-99$ for upper limits and 99 for lower limits). \\
	19&\NuSTAR\ source ID from the \NuSTAR\ cycles 5+6 catalog ($-1$ if non-detection).\\
	20&Flag for \NuSTAR\ cycle 56 counterparts (S, P, Sec, C if the XMM source is the single, primary, secondary,\\
	&or confused counterpart of the \NuSTAR\ source, respectively).\\
	21&\NuSTAR\ source ID from the \NuSTAR\ cycle 6 catalog ($-1$ if non-detection).\\
	22&Flag for \NuSTAR\ cycle 6 counterparts (S, P, Sec, C if the XMM source is the single, primary, secondary,\\
	&or confused counterpart of the \NuSTAR\ source, respectively).\\
	23&Flag for ancillary class (S for secure, A for ambiguous, or U for unidentified)\\
    24,25&Ancillary coordinates of the associated source ($-99$ if no detection).\\
	26&Optical (HSC) $i$ band AB magnitude ($-99$ if no detection).\\
	27&Flag for SDSS detection (1 if SDSS has detection, $-1$ if SDSS has no detection)\\
	28&Flag for J-PAS detection (1 if J-PAS has detection, $-1$ if J-PAS has no detection)\\
     	29,30&MMIRS $J$ and $K$ band AB magnitude ($-99$ if no detection).\\
	31,32&WISE $W1$ and $W2$ band AB magnitude ($-99$ if no detection).\\
	33&VLA 3 GHz counterpart ID from \citet{Hyun2023}.\\
    34&VLA 3~GHz flux in $\mu$Jy \citep{Hyun2023}.\\
    35&HST F606W AB magnitude ($-99$ if no detection).\\
    36&JWST F444W AB magnitude ($-99$ if no detection).\\
	37&Spectroscopic redshift of the associated source ($-99$ if no measurement).\\
	38&Photometric redshift of the associated source ($-99$ if no redshift measurement).\\
	39&Spectroscopic classification (Q for quasars, G for galaxies, S for stars, N/A if no measurement). Galaxies are defined\\ & as objects without broad 
	emission lines and therefore include type 2 AGN.\\
	40&Luminosity distance in Mpc (70/0.3/0.7 cosmology, $-99$ if no redshift measurement).\\
	41&0.5--2\,keV band rest-frame luminosity before correcting for absorption assuming a photon index of $\Gamma$ = 1.40\\
	& ($-99$ if not detected in the 0.5--2\,keV band).\\
	42&0.5--2\,keV band rest-frame luminosity 1$\sigma$ uncertainty.\\
	43&2--10\,keV band rest-frame luminosity before correcting for absorption assuming a photon index of $\Gamma$ = 1.80 \\
	&($-99$ if not detected in the 2--10\,keV band).\\
	44&2--10\,keV band rest-frame luminosity 1$\sigma$ uncertainty.\\
	\hline
\end{tabular}
\raggedright
\tablecomments{Multiple rows are used for the \XMM\ sources with multiple optical/IR counterpart candidates.}
\end{table*}
\endgroup

\begingroup
\renewcommand*{\arraystretch}{1.}
\begin{table}
\centering
\caption{Hectospec observations of 40 VLA and Chandra targets in NEP-TDF. }
\label{Table:Hectospec_catalog}
  \begin{tabular}{lcccc}
       \hline
       \hline     
 	Name&RA&DEC&$z$&Class\\
	\hline 
	VLA~3&260.351313&65.814148&0.2923&G\\
	VLA~48&260.507012&65.740059&0.6297&G\\
	VLA~52&260.522058&65.680923&0.5010&G\\
	VLA~62&260.530900&65.924820&0.2720&G\\
	VLA~74&260.562717&65.810898&0.0806&G\\
	VLA~82&260.573729&65.660584&0.6312&G\\
	VLA~140&260.638762&65.653336&0.1785&G\\
	VLA~159&260.648546&65.797379&0.2953&G\\
	VLA~164&260.652038&65.93177&0.0415&G\\
	VLA~173&260.658704&65.849815&0.4972&G\\	VLA~185&260.666254&65.671364&2.80&Q\rlap{\tablenotemark{a}}\\
	VLA~198&260.674983&65.976944&0.0746&G\\	
	VLA~200&260.678471&65.83725&0.5658&G\\	
	VLA~222&260.692892&65.861908&0.2946&G\\
	VLA~246&260.71455&65.753357&0.5397&G\\
	VLA~260&260.721475&65.813164&0.545&G\\
	VLA~382&260.799721&65.837288&0.3759&G\\
	VLA~386&260.805617&65.730148&1.0415&G\\
	VLA~429&260.838487&65.837677&0.8905&G\\
	VLA~439&260.846929&65.744003&0.3774&G\\
	VLA~477&260.881892&65.722382&0.3748&G\\
	VLA~514&260.910617&65.905113&0.3579&G\\
	VLA~528&260.919458&65.831345&0.3762&G\\
	VLA~561&260.946979&65.874069&0.3820&G\\
	VLA~574&260.961083&65.761307&0.0964&G\\
	VLA~592&260.978937&65.782906&0.2923&G\\
	VLA~628&261.009637&65.638786&0.5643&G\\
	VLA~656&261.038017&65.746704&0.5567&G\\
	VLA~675&261.057729&65.784676&0.2946&G\\
	VLA~688&261.073267&65.855888&0.1055&G\\
	VLA~705&261.116296&65.777801&0.4136&G\\
	VLA~721&261.151712&65.851639&0.4464&G\\
	VLA~752&261.251333&65.815597&0.5010&G\\
	VLA~755&261.289646&65.827973&0.1816&G\\
	Cha~11&260.729742&65.926109&0.008&G\\
	Cha~38&260.404500&65.799156&0&S\\
	Cha~43&260.538454&65.827545&0.776&G\\
	Cha~47&260.397692&65.838997&0.6487&G\\
	Cha~50&260.804071&65.886581&0.8347&Q\\
	Cha~76&260.363158&65.852623&0.2762&G\\
	\hline
\end{tabular}
\raggedright
\tablecomments{The source name is composed of catalog \cite[VLA;][]{Hyun2023} or Chandra (P.\ Maksym et al., in prep.)\ and the ID of the source in the corresponding catalog. {G}, {Q}, and {S} in spectral class represent galaxy, quasar, and star, respectively. Narrow emission-line (type 2) quasars are categorized as galaxies. }
\tablenotetext{a}{This broad absorption-line quasar also has a JCMT detection.}
\end{table}
\endgroup

\end{document}